\documentclass[12pt]{article}
\usepackage{eqsection,subeqnarray,indent,amsfonts,amssymb}
\usepackage{amsfonts}
\usepackage{graphicx}

\begin{document}

\bibliographystyle{plain}


\title{\vspace*{-2.5cm}
 Exact Potts Model Partition Functions for Strips of the 
 Honeycomb Lattice} 

\author{
  \\
  {\small    Shu-Chiuan Chang$^{a,b}$}                              \\[-0.2cm]
  {\small\it $(a)$ Department of Physics}  \\[-0.2cm]
  {\small\it National Cheng Kung University}                  \\[-0.2cm]
  {\small\it Tainan 70101}                 \\[-0.2cm]
  {\small\it Taiwan }                                          \\[-0.2cm]
  {\small\it $(b)$ Physics Division}  \\[-0.2cm]
  {\small\it National Center for Theoretical Science}                  \\[-0.2cm]
  {\small\it National Taiwan University}                  \\[-0.2cm]
  {\small\it Taipei 10617}                 \\[-0.2cm]
  {\small\it Taiwan }                                          \\[-0.2cm]
  {\small\tt scchang@mail.ncku.edu.tw}                       \\[5mm]
  {\small Robert Shrock$^{c}$}                  \\[-0.2cm]
  {\small\it $(c)$ C.~N.~Yang Institute for Theoretical Physics}  \\[-0.2cm]
  {\small\it State University of New York}                  \\[-0.2cm]
  {\small\it Stony Brook, N.~Y.~11794-3840}                 \\[-0.2cm]
  {\small\it USA }                                          \\[-0.2cm]
  {\small\tt robert.shrock@sunysb.edu}                      \\[-0.2cm]
  {\protect\makebox[5in]{\quad}}  
  \\
}

\maketitle

\thispagestyle{empty}   

\begin{abstract}

We present exact calculations of the Potts model partition function $Z(G,q,v)$
for arbitrary $q$ and temperature-like variable $v$ on $n$-vertex strip graphs
$G$ of the honeycomb lattice for a variety of transverse widths equal to $L_y$
vertices and for arbitrarily great length, with free longitudinal boundary
conditions and free and periodic transverse boundary conditions.  These
partition functions have the form $Z(G,q,v)=\sum_{j=1}^{N_{Z,G,\lambda}}
c_{Z,G,j}(\lambda_{Z,G,j})^m$, where $m$ denotes the number of repeated
subgraphs in the longitudinal direction.  We give general formulas for
$N_{Z,G,j}$ for arbitrary $L_y$. We also present plots of zeros of the
partition function in the $q$ plane for various values of $v$ and in the $v$
plane for various values of $q$. Explicit results for partition functions are
given in the text for $L_y=2,3$ (free) and $L_y=4$ (cylindrical), and plots of
partition function zeros are given for $L_y$ up to 5 (free) and $L_y=6$
(cylindrical).  Plots of the internal energy and specific heat per site for
infinite-length strips are also presented.

\end{abstract}


\clearpage

\newcommand{\beq}{\begin{equation}}
\newcommand{\eeq}{\end{equation}}
\newcommand{\beqs}{\begin{eqnarray}}
\newcommand{\eeqs}{\end{eqnarray}}
\newcommand{\lsim}{\mathrel{\raisebox{-.6ex}{$\stackrel{\textstyle<}{\sim}$}}}
\newcommand{\gsim}{\mathrel{\raisebox{-.6ex}{$\stackrel{\textstyle>}{\sim}$}}}
\newtheorem{theorem}{Theorem}[section]
\newtheorem{corollary}{Corollary}[section]
\newtheorem{conjecture}{Conjecture}[section]
\newtheorem{lemma}{Lemma}[section]
\newcommand{\be}{\begin{equation}}
\newcommand{\ee}{\end{equation}}
\newcommand{\widebar}{\overline}
\def\reff#1{(\protect\ref{#1})}
\def\spose#1{\hbox to 0pt{#1\hss}}
\def\ltapprox{\mathrel{\spose{\lower 3pt\hbox{$\mathchar"218$}}
 \raise 2.0pt\hbox{$\mathchar"13C$}}}
\def\gtapprox{\mathrel{\spose{\lower 3pt\hbox{$\mathchar"218$}}
 \raise 2.0pt\hbox{$\mathchar"13E$}}}
\def\textprime{${}^\prime$}
\def\proof{\par\medskip\noindent{\sc Proof.\ }}
\def\qed{\hbox{\hskip 6pt\vrule width6pt height7pt depth1pt \hskip1pt}\bigskip}
\def\proofof#1{\bigskip\noindent{\sc Proof of #1.\ }}
\def\half{ {1 \over 2} }
\def\third{ {1 \over 3} }
\def\twothird{ {2 \over 3} }
\def\smfrac#1#2{\textstyle{#1\over #2}}
\def\smhalf{ \smfrac{1}{2} }
\newcommand{\real}{\mathop{\rm Re}\nolimits}
\renewcommand{\Re}{\mathop{\rm Re}\nolimits}
\newcommand{\imag}{\mathop{\rm Im}\nolimits}
\renewcommand{\Im}{\mathop{\rm Im}\nolimits}
\newcommand{\sgn}{\mathop{\rm sgn}\nolimits}
\newcommand{\tr}{\mathop{\rm tr}\nolimits}
\newcommand{\diag}{\mathop{\rm diag}\nolimits}
\newcommand{\Gal}{\mathop{\rm Gal}\nolimits}
\newcommand{\mycup}{\mathop{\cup}}
\newcommand{\Arg}{\mathop{\rm Arg}\nolimits}
\def\hboxscript#1{ {\hbox{\scriptsize\em #1}} }
\def\zhat{ {\widehat{Z}} }
\def\phat{ {\widehat{P}} }
\def\qtilde{ {\widetilde{q}} }
\newcommand{\mod}{\mathop{\rm mod}\nolimits}
\renewcommand{\emptyset}{\varnothing}

\def\scra{\mathcal{A}}
\def\scrb{\mathcal{B}}
\def\scrc{\mathcal{C}}
\def\scrd{\mathcal{D}}
\def\scrf{\mathcal{F}}
\def\scrg{\mathcal{G}}
\def\scrl{\mathcal{L}}
\def\scro{\mathcal{O}}
\def\scrp{\mathcal{P}}
\def\scrq{\mathcal{Q}}
\def\scrr{\mathcal{R}}
\def\scrs{\mathcal{S}}
\def\scrt{\mathcal{T}}
\def\scrv{\mathcal{V}}
\def\scrz{\mathcal{Z}}

\def\q{{\sf q}}

\def\Z{{\mathbb Z}}
\def\R{{\mathbb R}}
\def\C{{\mathbb C}}
\def\Q{{\mathbb Q}}

\def\T{{\mathsf T}}
\def\H{{\mathsf H}}
\def\V{{\mathsf V}}
\def\D{{\mathsf D}}
\def\J{{\mathsf J}}
\def\P{{\mathsf P}}
\def\QQ{{\mathsf Q}}
\def\RR{{\mathsf R}}

\def\bsigma{\mbox{\protect\boldmath $\sigma$}}
\def\bone{{\mathbf 1}}
\def\vv{{\bf v}}
\def\uu{{\bf u}}
\def\w{{\bf w}}

\section{Introduction} 
\label{sec.intro}

In this paper we present some theorems on structural properties of Potts model
partition functions on strips of the honeycomb ($hc$) lattice of arbitrary
width equal to $L_y$ vertices and arbitrarily great length.  We also report
exact calculations of these partition functions for a number of
honeycomb-lattice strips of various widths and arbitrarily great lengths.
Using these results, we consider the limit of infinite length.  For this limit
we calculate thermodynamic functions and determine the loci in the complex $q$
and temperature planes where the free energy is non-analytic. This is an
extension of our previous study for this lattice in \cite{hca}.

We briefly review the definition of the model and relevant notation.  Consider
a graph $G=(V,E)$, defined by its vertex set $V$ and edge set $E$.  (Here we
keep the discussion general; shortly, we will specialize to strips of the
honeycomb lattice.) Denote the number of vertices and edges as $|V| \equiv n$
and $|E|$, respectively. On the graph $G$, at temperature $T$, the Potts model
is defined by the partition function $Z(G,q,v) = \sum_{ \{ \sigma_n \} }
e^{-\beta {\cal H}}$ with the (zero-field) Hamiltonian ${\cal H} = -J
\sum_{\langle i j \rangle} \delta_{\sigma_i \sigma_j}$ where
$\sigma_i=1,\ldots,q$ are the spin variables on each vertex $i \in V$; $\beta =
(k_BT)^{-1}$; $\langle i j \rangle \in E$ denotes pairs of adjacent vertices,
and $J$ is the spin-spin interaction constant. We use the notation $K = \beta
J$, $a = e^K$, and $v = e^K-1$.  The physical ranges are thus (i) $a \ge 1$,
i.e., $v \ge 0$ corresponding to $\infty \ge T \ge 0$ for the Potts ferromagnet
with $J > 0$, and (ii) $0 \le a \le 1$, i.e., $-1 \le v \le 0$, corresponding
to $0 \le T \le \infty$ for the Potts antiferromagnet with $J < 0$.  Let
$G^\prime=(V,E^\prime)$ with $E^\prime \subseteq E$.  Then $Z(G,q,v)$ can be
defined for arbitrary $q$ and $v$ by the formula \cite{fk}
\beq
Z(G,q,v) = \sum_{G^\prime \subseteq G} q^{k(G^\prime)}v^{|E^\prime|}
\label{cluster}
\eeq
where $k(G^\prime)$ denotes the number of connected components of
$G^\prime$. We define a (reduced) free energy per site $f=-\beta F$,
where $F$ is the free energy, via
\beq
f(\{G\},q,v) = \lim_{n \to \infty} \ln [Z(G,q,v)^{1/n} ] \ , 
\label{ef}
\eeq
where the symbol $\{G\}$ denotes $\lim_{n \to \infty}G$ for a given family of
graphs $G$. 

Our exact results on $Z(G,q,v)$ apply for arbitrary $q$ and $v$. We consider
free and cylindrical strip graphs $G$ of the honeycomb lattice of width $L_y$
vertices and of arbitrarily great length $L_x$ vertices.  Here, free boundary
conditions (sometimes denoted FF), mean free in both the transverse and
longitudinal directions (the latter being the one that is varied for a fixed
width), while cylindrical boundary conditions (sometimes denoted PF) mean
periodic in the transverse direction and free in the longitudinal direction.
We represent the strip of the honeycomb lattice in the form of bricks oriented
horizontally. For the honeycomb lattice with cylindrical boundary conditions,
the number of vertices in the transverse direction, $L_y$, must be an even
number, and the smallest value without degeneracy (multiple edges) is
$L_y=4$. Exact partition functions for arbitrary $q$ and $v$ have previously
been presented for strips of the honeycomb lattice with free boundary
conditions for width $L_y=2$ in \cite{hca}.  Our new results include theorems
that describe the structure of the Potts model partition function for strips
with free and cylindrical boundary conditions, of arbitrary width and length
and explicit calculations using the transfer matrix method (in the
Fortuin--Kasteleyn representation \cite{bn}) for strips with free and
cylindrical boundary conditions with width $ L_y=3$ (free) and $L_y=4$
(cylindrical).  We have carried out similar calculations for $L_y \leq 7$
(free) and $L_y = 6$ (cylindrical); these are too lengthy to include here. We
shall also present plots of partition function zeros in the limit of infinite
length, for widths $2 \le L_y \le 5$ (free) and $L_y = 4, 6$ (cylindrical).
Related calculations of Potts model partition functions for arbitrary $q$ and
$v$ on fixed-width, arbitrary-length strips of the square and triangular
lattices are \cite{a}-\cite{jrs05} and \cite{jrs05,ta,tt}, respectively.
Analogous partition function calculations for arbitrary $q$ and $v$ on finite
sections of 2D lattices with fixed width and length include \cite{kl,kc}. The
special case $v=-1$ is the zero-temperature limit of the Potts antiferromagnet,
for which $Z(G,q,-1)=P(G,q)$, where $P(G,q)$ is the chromatic polynomial
expressing the number of ways of coloring the vertices of the graph $G$ with
$q$ colors such that no two adjacent vertices have the same color.

As part of our work, we calculate zeros of the partition function in the $q$
plane for fixed $v$ and in the $v$ plane for fixed $q$. In the limit of
infinite strip length, $L_x \to \infty$, there is a merging of such zeros to
form continuous loci of points where the free energy is nonanalytic, which we
denote generically as ${\cal B}$.  For the limit $L_x \to \infty$ of a given
family of strip graphs, this locus is determined as the solution to an
algebraic equation and is hence an algebraic curve.

There are several motivations for this work.  Clearly, exact calculations of
Potts model partition functions with arbitrary $q$ and $v$ are of value in
their own right.  This is especially true since there are no exact calculations
of the free energy $f(\{G\},q,v)$ for arbitrary $q$ and $v$ on an infinite
lattice of dimension two or higher.  Exact calculations on lattice strips of
fixed width and arbitrarily great length thus provide a useful set of results
complementing other methods of analysis such as series expansions and Monte
Carlo simulations of the Potts model.  Our structural theorems elucidate the
form of the partition function on these strips for arbitrarily great widths as
well as lengths.  The honeycomb lattice is of interest since, together with the
square and triangular lattices, it comprises the third and last regular tiling
of the plane which is homopolygonal, i.e. composed of of a single type of
regular polygon.  While critical properties describing the second-order phase
transition of the Potts ferromagnet are universal and independent of lattice
type, the behavior of the Potts antiferromagnet is sensitively dependent on
lattice type, so that studies of this model on different lattices and lattice
strips are valuable.

   There is a particular motivation for carrying out exact calculations of
Potts model partition functions with arbitrary $q$ and $v$, because this allows
one to investigate more deeply a unique feature of the model, which is
qualitatively different from the behavior on either the square or triangular
lattice, namely the property that the critical temperature of the Potts
antiferromagnet on the honeycomb lattice decreases to zero, i.e., the critical
$v$ decreases to $-1$, at a non-integral value, $q=(3+\sqrt{5})/2$ (as reviewed
below in connection with the criticality condition, eq. (\ref{hc_eq})).  This
is formal, since for $q \not\in {\mathbb Z}_+$, the model (with either sign of
$J$) is only defined via the representation (\ref{cluster}), and in the
antiferromagnet case (i.e., for $-1 \le v < 0$), this formula can yield a
negative, and hence unphysical, result for the partition function.  One
obviously cannot investigate this formal criticality using the Hamiltonian
formulation, which requires $q \in {\mathbb Z}_+$.

Moreover, calculations of complex-temperature zeros of the partition function
show how the physical phases can be generalized to regions in the plane of a
complex-temperature variable, as was discussed for the 2D Ising model on the
square lattice \cite{fisher,earlyct}. Calculations of partition function zeros
on long finite lattice strips, and the loci ${\cal B}$ in the infinite-length
limit, also yield interesting insights into properties of the corresponding
phase diagrams in the complex-temperature and complex-$q$ planes.

\section{General Structural Theorems} 
\label{sec.2} 

\subsection{Preliminaries}

In this section we prove several general theorems that describe the structure
of the partition function for the honeycomb-lattice strips under consideration.
Let $m$ denote the number of bricks in the longitudinal direction for such a
strip.  Then the length (number of vertices in the longitudinal direction) is 
\beq
L_x=\cases{2m+1 & for odd $L_x$ \ , \cr\cr
           2m+2 & for even $L_x$ \ .}
\label{Lxm}           
\eeq
For this type of strip graph, $Z(G,q,v)$ has the form 
\beq
Z(G,q,v) = \sum_{j=1}^{N_{Z,G,\lambda}} c_{G,j}(\lambda_{Z,G,j})^m
\label{zgsum}
\eeq
where the coefficients $c_{G,j}$ and corresponding terms $\lambda_{G,j}$, as
well as the total number $N_{Z,G,\lambda}$ of these terms, depend on the type
of strip (width and boundary conditions) but not on its length. In the special
case $v=-1$, the numbers $N_{Z,G,\lambda}$ will be denoted $N_{P,G,\lambda}$.
We define $N_{Z,hc,BC_y \ BC_x,L_y,\lambda}$ as the total number of $\lambda$'s
for the honeycomb-lattice strip with the transverse and longitudinal boundary
conditions $BC_y$ and $BC_x$ of width $L_y$.  Henceforth where no confusion
will result, we shall suppress the $\lambda$ subscript.  The explicit labels
are $N_{Z,hc,FF,L_y}$ and $N_{Z,hc,PF,L_y}$ for the strips of the honeycomb
lattices with free and cylindrical boundary conditions.

\subsection{Case of Free Boundary Conditions}

\begin{theorem} \label{theorem1} For arbitrary $L_y$,
\beq 
N_{Z,hc,FF,L_y} = \cases{ C_{L_y} & for odd $L_y$ \ , \cr\cr \frac{1}{2}
\left [ C_{L_y} + {L_y \choose L_y/2} \right ] & for even $L_y$ \ .}
\label{nzhcff}
\eeq
\end{theorem}

{\sl Proof} \quad A honeycomb-lattice strip with free boundary conditions is
symmetric under reflection about the longitudinal axis if and only if $L_y$ is
even.  Therefore, for odd $L_y$, the total number of $\lambda$'s in the Potts
model partition function, $N_{Z,hc,FF,L_y}$, is the same as the number for the
triangular lattice strips $N_{Z,tri,FF,L_y}$, namely, the number of
non-crossing partitions of the set $\{1,2,...,L_y\}$. This is the Catalan
number \cite{ss00,cf}, $C_{L} = (L+1)^{-1} {2L \choose L}$.  For even $L_x$,
the reflection symmetry reduces $N_{Z,hc,FF,L_y}$ to the number for the square
lattice strips $N_{Z,sq,FF,L_y}$, which was given in Theorem 5 of \cite{ts}. \
$\Box$ We list the first few values of $N_{Z,hc,FF,L_y}$ in Table
\ref{nfreetable}.

\bigskip

\begin{table}[htbp]
\caption{\footnotesize{Numbers of $\lambda$'s for the Potts model partition
function and chromatic polynomials for the strips of the honeycomb lattices
having free boundary conditions and various widths $L_y$.}}
\begin{center}
\begin{tabular}{|c|c|c|}
\hline\hline $L_y$ & $N_{Z,hc,FF,L_y}$ & $N_{P,hc,FF,L_y}$ \\ 
\hline\hline  
 2  &      2 &     1   \\ \hline
 3  &      5 &     3   \\ \hline
 4  &     10 &     5   \\ \hline
 5  &     42 &    19   \\ \hline
 6  &     76 &    25   \\ \hline
 7  &    429 &   145   \\ \hline
 8  &    750 &   194   \\ \hline
 9  &   4862 &  1230   \\ \hline
10  &   8524 &  1590   \\ \hline
11  &  58786 & 11139   \\ \hline
12  & 104468 & 14681   \\ \hline \hline
\end{tabular}
\end{center}
\label{nfreetable}
\end{table}

\bigskip

\begin{figure}[htbp]
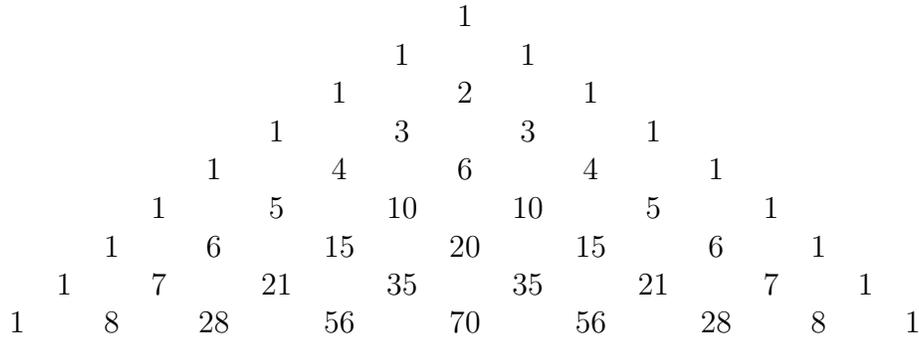

\begin{center}
\begin{tabular}{ccccccccccccccccc}
  &   &   &   &    &    &    &    & 1  &    &    &    &    &   &   &   &   \\
  &   &   &   &    &    &    & 1  &    & 1  &    &    &    &   &   &   &   \\
  &   &   &   &    &    & 1  &    & 2  &    & 1  &    &    &   &   &   &   \\
  &   &   &   &    & 1  &    & 3  &    & 3  &    & 1  &    &   &   &   &   \\
  &   &   &   & 1  &    & 4  &    & 6  &    & 4  &    & 1  &   &   &   &   \\
  &   &   & 1 &    & 5  &    & 10 &    & 10 &    & 5  &    & 1 &   &   &   \\
  &   & 1 &   & 6  &    & 15 &    & 20 &    & 15 &    & 6  &   & 1 &   &   \\
  & 1 &   & 7 &    & 21 &    & 35 &    & 35 &    & 21 &    & 7 &   & 1 &   \\
1 &   & 8 &   & 28 &    & 56 &    & 70 &    & 56 &    & 28 &   & 8 &   & 1
\end{tabular}
\end{center}
\caption{\footnotesize{Pascal's triangle.}}
\label{pascaltriangle}
\end{figure}

\begin{figure}[htbp]
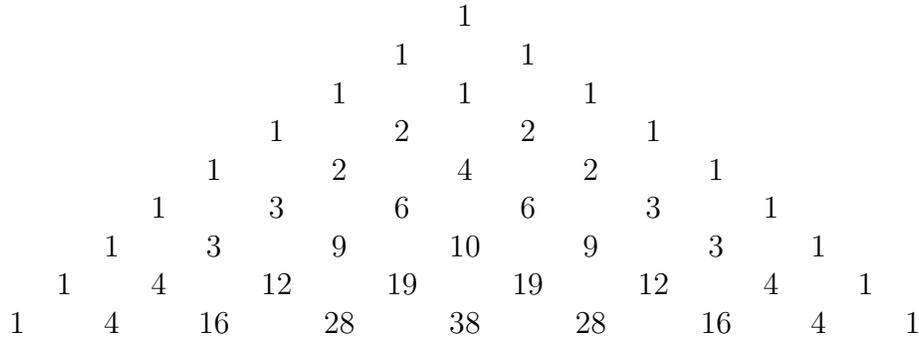

\begin{center}
\begin{tabular}{ccccccccccccccccc}
  &   &   &   &    &    &    &    & 1  &    &    &    &    &   &   &   &   \\
  &   &   &   &    &    &    & 1  &    & 1  &    &    &    &   &   &   &   \\
  &   &   &   &    &    & 1  &    & 1  &    & 1  &    &    &   &   &   &   \\
  &   &   &   &    & 1  &    & 2  &    & 2  &    & 1  &    &   &   &   &   \\
  &   &   &   & 1  &    & 2  &    & 4  &    & 2  &    & 1  &   &   &   &   \\
  &   &   & 1 &    & 3  &    & 6  &    & 6  &    & 3  &    & 1 &   &   &   \\
  &   & 1 &   & 3  &    & 9  &    & 10 &    & 9  &    & 3  &   & 1 &   &   \\
  & 1 &   & 4 &    & 12 &    & 19 &    & 19 &    & 12 &    & 4 &   & 1 &   \\
1 &   & 4 &   & 16 &    & 28 &    & 38 &    & 28 &    & 16 &   & 4 &   & 1
\end{tabular}
\end{center}
\caption{\footnotesize{Losanitsch's triangle.}}
\label{Losanitschtriangle}
\end{figure}

\begin{figure}[htbp]
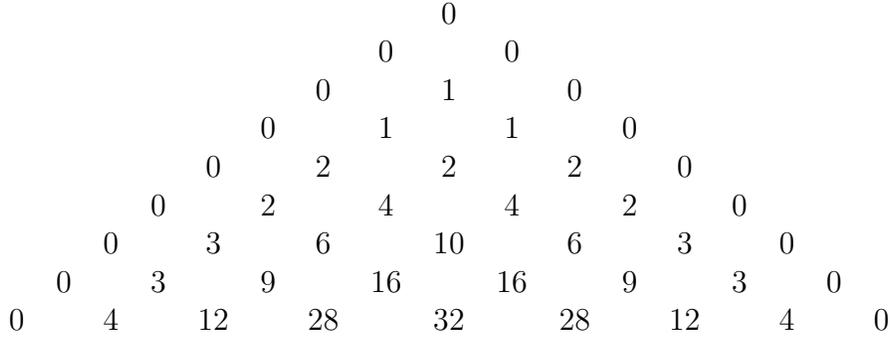

\begin{center}
\begin{tabular}{ccccccccccccccccc}
  &   &   &   &    &    &    &    & 0  &    &    &    &    &   &   &   &   \\
  &   &   &   &    &    &    & 0  &    & 0  &    &    &    &   &   &   &   \\
  &   &   &   &    &    & 0  &    & 1  &    & 0  &    &    &   &   &   &   \\
  &   &   &   &    & 0  &    & 1  &    & 1  &    & 0  &    &   &   &   &   \\
  &   &   &   & 0  &    & 2  &    & 2  &    & 2  &    & 0  &   &   &   &   \\
  &   &   & 0 &    & 2  &    & 4  &    & 4  &    & 2  &    & 0 &   &   &   \\
  &   & 0 &   & 3  &    & 6  &    & 10 &    & 6  &    & 3  &   & 0 &   &   \\
  & 0 &   & 3 &    & 9  &    & 16 &    & 16 &    & 9  &    & 3 &   & 0 &   \\
0 &   & 4 &   & 12 &    & 28 &    & 32 &    & 28 &    & 12 &   & 4 &   & 0
\end{tabular}
\end{center}
\caption{\footnotesize{Triangle formed by subtraction of elements of 
Losanitsch's triangle from corresponding elements of Pascal's triangle.}}
\label{Pascal-Losanitsch}
\end{figure}

We next discuss some combinatorics which will be used in our next theorem.  Let
us denote the elements in the $k$'th column of the $j$'th row of Pascal's
triangle as $P(j,k)$, with the value $P(j,k)={j \choose k}$ \ (the binomial
coefficient). The relation between these elements is
$P(j,k)=P(j-1,k-1)+P(j-1,k)$ with $P(0,0)=1$, $P(j,-1)=0$, and $P(j,k)=0$ for
$j<k$. That is, each element is the sum of the two numbers immediately above
it. For the reader's convenience, we display the first few rows of Pascal's
triangle in Fig. \ref{pascaltriangle}. Now, $P(j,k)$ is the number of ways to
place $k$ black beads and $j-k$ white beads in a line. Similarly, what is known
as Losanitsch's triangle \cite{sl} is given by the number of ways to put $j$
beads of two colors in a line, modulo reflection symmetry. We denote the entry
in the $k$'th column of the $j$'th row of this triangle as $L(j,k)$.  For
reference, the sequence formed by reading the entries in this triangle (from
left to right) by rows is listed as sequence $A034851$ in \cite{sl}.  The
relation between these entries is essentially the same as for Pascal's
triangle, except when $j$ is even and $k$ is odd:
\beq
L(j,k)=L(j-1,k-1)+L(j-1,k)- \delta _{j \ {\rm even}, k \ {\rm odd}} 
{j/2-1 \choose (k-1)/2}
\eeq
where $\delta _{j \ {\rm even}, k \ {\rm odd}}=1$ if $j$ is even and $k$ is odd
and zero otherwise. For reference, the first few rows of Losanitsch's triangle
are shown in Fig. \ref{Losanitschtriangle}. Now form a new triangle by
subtracting the entries in Losanitsch's triangle from the corresponding entries
in Pascal's triangle, and denote its elements as $PL(j,k)$.  The first few rows
of this triangle are displayed in Fig. \ref{Pascal-Losanitsch}.

The next theorem concerns the number of $\lambda$'s $N_{P,hc,FF,L_y}$ in the 
chromatic polynomial for the free $hc$ strip. 

\bigskip

\begin{theorem} \label{theorem2} For arbitrary $L_y$,
\beq
N_{P,hc,FF,L_y}=
\cases{ \sum_{i=0}^{(L_y-1)/2} M_{L_y-1-i} {(L_y-1)/2 \choose i} & 
for odd $L_y$ \ , \cr\cr
\sum_{i=0}^{L_y/2-1} N_{P,sq,FF,L_y-i} {L_y/2-1 \choose i} & \cr
- \frac12 \sum_{i=1}^{L_y/2-2} PL(L_y/2-1,i) N_{P,FP,[(L_y+1-i)/2]} & 
for even $L_y$ \ .} 
\label{nphcpf}
\eeq
where $M_L$ is the Motzkin number \cite{ds}, $N_{P,sq,FF,L_y}$ is the number of
$\lambda$'s for the square lattice with free boundary conditions given in
Theorem 2 of \cite{ts}, $N_{P,FP,L_y}$ is the total number of $\lambda$'s for
the square, triangular, or honeycomb lattices with cyclic boundary conditions
given in eq.(5.2) of \cite{cf}, and $PL(j,k)$ is the number given by the
subtraction of the elements of Losanitsch's triangle from the corresponding
elements of Pascal's triangle.  
\end{theorem}

{\sl Proof} \quad Consider odd strips widths $L_y$ first; for these, there is
no reflection symmetry. For each transverse slice containing $L_y$ vertices of
the honeycomb lattice with free boundary conditions, there are $(L_y-1)/2$
edges. Compared with the calculation of partitions for strips of the square
lattice, the same numbers of edges are removed, such that the end vertices of
each of these missing edges are allowed to have the same color. Firstly, all of
the possible partitions for the triangular-lattice strip with free boundary
conditions $N_{P,tri,FF,L_y}=M_{L_y-1}$ \cite{ss00} are valid for the honeycomb
lattice. Among $(L_y-1)/2$ missing edges, if one pair of end vertices has the
same color, the number of partitions is given by $N_{P,tri,FF,L_y-1}$. As an
example, for the $L_y=3$ slice of the honeycomb-lattice strip, there is an edge
connecting vertices 1 and 2 and no edge connecting vertices 2 and 3. It follows
that the set of partitions is comprised of $\{ 1, \delta_{1,3}, \delta_{2,3}\}$
in the shorthand notation used in \cite{ts,tt}. Similarly, if two pairs of end
vertices of missing edges separately have the same color, there are ${(L_y-1)/2
\choose 2}$ choices for the locations of missing edges and the number of
partitions for each choice is $N_{P,tri,FF,L_y-2}$. By including all the
possible missing edges with the same color on their end vertices, the first
line in eq. (\ref{nphcpf}) is established for the honeycomb lattice with odd
$L_y$.

For the honeycomb lattice with even $L_y$, reflection symmetry must be taken
into account. We thus consider partitions with the transverse slice composed of
edges connecting vertices 1 and 2, vertices 3 and 4, ..., vertices $L_y-1$ and
$L_y$. That is, there are $L_y/2$ edges in each slice and $L_y/2-1$ missing
edges. Naively, one would expect that the number of partitions is
$\sum_{i=0}^{L_y/2-1} N_{P,sq,FF,L_y-i} {L_y/2-1 \choose i}$ by the above
argument. However, this would actually involve an overcounting, because certain
partitions with the same color on end vertices of missing edge(s) are
equivalent under the reflection symmetry. For example, the $L_y=6$ strip of the
honeycomb lattice has two missing edges, i.e., there is no edge between
vertices 2 and 3, and between vertices 4 and 5. There are $N_{P,sq,FF,5}=7$
partitions if vertices 2 and 3 have the same color, and separately seven
partitions if vertices 4 and 5 have the same color.  Five partitions for the
first case $\delta_{2,3}$, $\delta_{2,3} \delta_{1,6}$, $\delta_{2,3,5}$,
$\delta_{2,3} \delta_{1,4,6}$ and $\delta_{2,3,5} \delta_{1,6}$ are equivalent
under reflection to the following partitions for the second case:
$\delta_{4,5}$, $\delta_{4,5} \delta_{1,6}$, $\delta_{2,4,5}$, $\delta_{4,5}
\delta_{1,3,6}$, and $\delta_{2,4,5} \delta_{1,6}$. In general, the
double-counted partitions are those symmetric partitions in $N_{P,sq,FF,L_y-i}$
such that $i$ pairs of end vertices of missing edges have the same color. This
number of symmetric partitions is $\frac12 N_{P,FP,[(L_y+1-i)/2]}$, given as
Theorem 1 in \cite{ts}. The double counting occurs when two choices of $i$
missing edges are reflection-symmetric to each other. This number of asymmetric
choices of missing edges is given by $PL(j,k)$.  \ $\Box$ We list the first few
values of $N_{P,hc,FF,L_y}$ for the honeycomb lattice with free boundary
conditions in Table \ref{nfreetable}.

\subsection{Case of Cylindrical Boundary Conditions}

For the honeycomb lattice with cylindrical boundary condition, only strips with
even $L_y$ can be defined. The number of $\lambda$'s can be reduced from the
number of non-crossing partitions $N_{Z,tri,FF,L_y}=C_{L_y}$ with length-two
rotational symmetry, and this first reduction will be denoted as
$N_{Z,hc,PF,L_y}^\prime$. This number can be further reduced with reflection
symmetry, and this further-reduced number will be denoted as
$N_{Z,hc,PF,L_y}$. We list the first few relevant numbers in the Potts model
partition function for the strips of the honeycomb lattices with cylindrical
boundary conditions in Table \ref{nztable}.

\begin{table}[htbp]
\caption{\footnotesize{Numbers of $\lambda$'s in the Potts model partition
function for the strips of the honeycomb lattices having cylindrical boundary
conditions and various even $L_y$.}}
\begin{center}
\begin{tabular}{|c|c|c|c|c|c|}
\hline\hline $L_y$ & $N_{Z,tri,FF,L_y}$ & $N_{Z,hc,PF,L_y}^\prime$ &
$N_{Z,hc,PF,L_y}$ & $2N_{Z,hc,PF,L_y}$ & $\frac{L_y}{2}N_{Z,hc,PF,L_y}^\prime$
 \\ 
 & & & & $-N_{Z,hc,PF,L_y}^\prime$ & $-N_{Z,tri,FF,L_y}$ \\
\hline\hline  
 2  &      2 &     2 &     2 &   2 &    0  \\ \hline
 4  &     14 &    10 &     8 &   6 &    6  \\ \hline
 6  &    132 &    48 &    34 &  20 &   12  \\ \hline
 8  &   1430 &   378 &   224 &  70 &   82  \\ \hline
10  &  16796 &  3364 &  1808 & 252 &   24  \\ \hline
12  & 208012 & 34848 & 17886 & 924 & 1076  \\ \hline \hline
\end{tabular}
\end{center}
\label{nztable}
\end{table}

\bigskip

\begin{lemma} \label{lemma1} For arbitrary even $L_y$,
\beq
\frac{L_y}{2} N_{Z,hc,PF,L_y}^\prime - N_{Z,tri,FF,L_y} = \sum_{{\rm even} \ 
d|L_y; \ 1 \le d < L_y} \phi \left (\frac{L_y}{d} \right ) {2d \choose d} 
\label{nzhcpfprime-nztriff}
\eeq
where $d|L_y$ means that $d$ divides $L_y$, and $\phi(n)$ is the Euler 
totient function, equal to the number of positive integers not exceeding the
positive integer $n$ and relatively prime to $n$.
\end{lemma}

{\sl Proof} \quad Because the honeycomb-lattice strip has length-two rotational
symmetry, the value $\frac{L_y}{2} N_{Z,hc,PF,L_y}^\prime$ must be larger than
$N_{Z,tri,FF,L_y}$. For strips with cylindrical boundary conditions, partitions
can be classified according to the periodicity $d$ modulo rotations. That is, a
partition could be transformed back to itself when the $L_y$ vertices are
rotated by length $d$, where $d$ denotes any of the positive integers that
divide $L_y$. The number of partitions which have periodicity $d$ modulo these
rotations was denoted as $2\alpha_d$ in the proof of Theorem 2.2 of \cite{tt},
and is given by
\beq
2\alpha_d = \frac{1}{d} \sum_{d^\prime|d} \mu (d/d^\prime) {2d^\prime 
\choose d^\prime} 
\label{mobiusalpha3}
\eeq
where $\mu(n)$ is the M\"obius function, defined as $\mu(n)=-1$ if $n$ is
prime, $\mu(n)=0$ if $n$ has a square factor, and $\mu(n)=1$ for other $n$.
Now the periodicity $d$ can be either odd or even. For a partition with odd
periodicity $d$, the contribution to the excess of $(L_y/2)
N_{Z,hc,PF,L_y}^\prime$ relative to $N_{Z,tri,FF,L_y}$ is
$2\alpha_d(L_y/2-d)$. On the honeycomb-lattice strip with length-two rotational
symmetry, a partition with even periodicity $d$ and its length-one rotation are
not equivalent. Thus for a partition with even periodicity $d$ and its
length-one rotated counterpart, the contribution is $2\alpha_d(L_y-d)$. We have
\beqs 
\frac{L_y}{2} N_{Z,hc,PF,L_y}^\prime-N_{Z,tri,FF,L_y} & = & \sum_{{\rm
even} \ d|L_y} 2\alpha_d(L_y-d) + \sum_{{\rm odd} \ d|L_y}
2\alpha_d(\frac{L_y}{2}-d) \cr\cr & = & \sum_{d|L_y} 2\alpha_d(L_y-d) -
\sum_{{\rm odd} \ d|L_y} \alpha_d L_y \cr\cr & = & \sum_{d|L_y; \ 1 \le d <
L_y} \phi (L_y/d) {2d \choose d} - \sum_{{\rm odd} \ d|L_y} \alpha_d L_y
\label{nzhcpfprime}
\eeqs
where the last line follows from the proof of Theorem 2.2 in \cite{tt}. Using
eq. (\ref{mobiusalpha3}) for the second summation in eq. (\ref{nzhcpfprime}) is
\beqs 
\sum_{{\rm odd} \ d|L_y} \alpha_d L_y & = & \sum_{{\rm odd} \ d|L_y}
\frac{L_y}{2d} \sum_{d^\prime|d} \mu (d/d^\prime) {2d^\prime \choose d^\prime}
\cr\cr & = & \sum_{{\rm odd} \ d^\prime|L_y} \sum_{{\rm odd} \ d|L_y;
d^\prime|d} \frac{L_y}{2d} \mu (d/d^\prime) {2d^\prime \choose d^\prime} \cr\cr
& = & \sum_{{\rm odd} \ d^\prime|L_y} \frac12 {2d^\prime \choose d^\prime}
\sum_{{\rm odd} \ d^{\prime\prime}|L_y^\prime}
\frac{L_y^\prime}{d^{\prime\prime}} \mu (d^{\prime\prime}) \ , 
\eeqs
where we change the variables to $d^{\prime\prime}=d/d^\prime$ and
$L_y^\prime=L_y/d^\prime$. (This is given as sequence A062570 in \cite{sl}.) 
The summation of $d^{\prime\prime}$ is
\beq 
\sum_{{\rm odd} \ d^{\prime\prime}|L_y^\prime}
\frac{L_y^\prime}{d^{\prime\prime}} \mu (d^{\prime\prime}) = \phi (2L_y^\prime)
= 2\phi (L_y/d^\prime) \ , \eeq
where that second identity follows because $L_y^\prime$ is even by the formula
for the Euler totient function, $\phi (n)=n \prod _{{\rm prime} \
p|n}(1-1/p)$. Therefore, we find 
\beq
\sum_{{\rm odd} \ d|L_y} \alpha_d L_y  = \sum_{{\rm odd} \ d|L_y} \phi (L_y/d) {2d \choose d} \ ,
\eeq
and the proof is completed. \ $\Box$

\bigskip
\begin{lemma} \label{lemma2} For arbitrary even $L_y$,
\beq
2N_{Z,hc,PF,L_y} - N_{Z,hc,PF,L_y}^\prime = N_{Z,FP,\frac{L_y}{2}}
\label{2nzhcpf-nzhcpfprime}
\eeq
where $N_{Z,FP,L_y}={2L_y \choose L_y}$ is the total number of $\lambda$'s for
the square or triangular or honeycomb lattices with cyclic boundary conditions,
given in eq.(5.6) of \cite{cf}.
\end{lemma}

{\sl Proof} \quad We use the result in Theorem 2.1 of \cite{tt} that
$2N_{Z,sq,PF,L_y}-N_{Z,tri,PF,L_y} =N_{Z,FP,\frac{L_y}{2}}$ for even $L_y$ is
the number of partitions with both rotation and reflection symmetries. The
partitions that do not have reflection symmetry appear once in
$N_{Z,sq,PF,L_y}$ but twice (itself plus its mirror image) in
$N_{Z,tri,PF,L_y}$, so that they do not contribute to
$2N_{Z,sq,PF,L_y}-N_{Z,tri,PF,L_y}$. It was shown there that for even $L_y$,
there are two classes of reflection symmetries, denoted as type I and type
II. For type I partitions, the reflection axis does not go through any vertex;
while for type II partitions, the reflection axis goes through two
vertices. There are at least two partitions that belong to both type I and II
classes, namely, the partitions 1 (identity) and $\delta_{1,2,...,L_y}$ (unique
block). Denote the set of partitions of $N_{Z,tri,PF,L_y}$ as ${\bf
P}_{Z,tri,PF,L_y}$, that of $N_{Z,hc,PF,L_y}^\prime$ as ${\bf
P}_{Z,hc,PF,L_y}^\prime$, and that of $N_{Z,hc,PF,L_y}$ as ${\bf
P}_{Z,hc,PF,L_y}$. For example, ${\bf P}_{Z,tri,PF,4} = \{ 1, \delta_{1,2,3,4},
\delta_{1,2}, \delta_{1,3}, \delta_{1,2}\delta_{3,4}, \delta_{1,3,4} \}$. As
the cylindrical strip of the honeycomb lattice only has length-two rotation
symmetry, $N_{Z,hc,PF,L_y}^\prime$ is larger than $N_{Z,tri,PF,L_y}$. The
excess partitions in ${\bf P}_{Z,hc,PF,L_y}^\prime$ relative to ${\bf
P}_{Z,tri,PF,L_y}$ can be obtained by making length-one rotations for certain
partitions in ${\bf P}_{Z,tri,PF,L_y}$. For example, in addition to those in
${\bf P}_{Z,tri,PF,L_y}$, ${\bf P}_{Z,hc,PF,4}^\prime$ contains $\delta_{2,3},
\delta_{2,4}, \delta_{2,3}\delta_{1,4}, \delta_{1,2,4}$. Notice that the
partitions that belong to both type I and II classes remain the same after the
length-one rotation (modulo length-two rotational symmetry), so that they
should be excluded in the doubling. For the honeycomb lattice, the reduction
from ${\bf P}_{Z,hc,PF,L_y}^\prime$ to ${\bf P}_{Z,hc,PF,L_y}$ occurs only for
type I partitions. Considering the $L_y=4$ strip again as an example, we
observe that $\delta_{1,3}$ and $\delta_{2,4}$ are equivalent, so that only one
of them should be kept for ${\bf P}_{Z,hc,PF,L_y}$, and similarly for
$\delta_{1,3,4}$ and $\delta_{1,2,4}$. Since the numbers of partitions in type
I and II classes are the same, the right hand side of
eq. (\ref{2nzhcpf-nzhcpfprime}) is the same as that in Theorem 2.1 of \cite{tt}
for even $L_y$.  \ $\Box$

The exact formula for $N_{Z,hc,PF,L_y}$ follows from Lemmas~\ref{lemma1}
and~\ref{lemma2}:

\begin{theorem} \label{theorem3} For arbitrary even $L_y$,
\beq
N_{Z,hc,PF,L_y} = \frac{1}{2} {L_y \choose L_y/2} + \frac{1}{L_y} 
\left[ C_{L_y} + \sum_{{\rm even} \ d|L_y; \ 1 \le d < L_y} \phi 
\left (\frac{L_y}{d} \right ) {2d \choose d} \right] \ .
\label{nzhcpf}
\eeq
\end{theorem}

\bigskip

We list the first few relevant numbers in the chromatic polynomial for the
strips of the honeycomb lattices with cylindrical boundary conditions in Table
\ref{nptable}. Analogously to Lemmas~\ref{lemma1} and~\ref{lemma2}, we state
the following conjectures,

\begin{table}[htbp]
\caption{\footnotesize{Numbers of $\lambda$'s for the chromatic polynomial for
the strips of the honeycomb lattices having cylindrical boundary conditions and
various even $L_y$.}}
\begin{center}
\begin{tabular}{|c|c|c|c|c|c|}
\hline\hline $L_y$ & $n_P(hc,L_y,0)$ & $N_{P,hc,PF,L_y}^\prime$ &
$N_{P,hc,PF,L_y}$ & $2N_{P,hc,PF,L_y}$ & $\frac{L_y}{2}N_{P,hc,PF,L_y}^\prime$
 \\ 
 & & & & $-N_{P,hc,PF,L_y}^\prime$ & $-n_P(hc,L_y,0)$ \\
\hline\hline  
 2  &     1 &    1 &    1 &   1 &   0  \\ \hline
 4  &     6 &    5 &    4 &   3 &   4  \\ \hline
 6  &    43 &   17 &   12 &   7 &   8  \\ \hline
 8  &   352 &   99 &   62 &  25 &  44  \\ \hline
10  &  3114 &  626 &  346 &  66 &  16  \\ \hline
12  & 29004 & 4907 & 2576 & 245 & 438  \\ \hline \hline
\end{tabular}
\end{center}
\label{nptable}
\end{table}

\bigskip

\begin{conjecture} \label{conjecture1} For arbitrary even $L_y$,
\beq
\frac{L_y}{2} N_{P,hc,PF,L_y}^\prime - n_P(hc,L_y,0) = \sum_{{\rm even} \ 
d|L_y; \ 1 \le d < L_y} \phi \left (\frac{L_y}{d} \right ) N_{P,hc,FP,d}
\label{nphcpfprime-nphcff}
\eeq
where $n_P(hc,L_y,0)$ is the number of $\lambda$'s for the cyclic strips of the
honeycomb lattice with level $d=0$, and $N_{P,hc,FP,L_y}$ is the total number
of $\lambda$'s for these strips given in \cite{hca}.
\end{conjecture}
Compared with Lemma \ref{lemma1}, the number of non-crossing partitions
$N_{Z,tri,FF,L_y}=C_{L_y}$ is now replaced by the corresponding number
$n_P(hc,L_y,0)$ for the honeycomb strip in the chromatic polynomial, and the
number ${2d \choose d} =N_{Z,hc,FP,d}$ is replaced by $N_{P,hc,FP,d}$.

\begin{conjecture} \label{conjecture2} For arbitrary even $L_y$,
\beq
2N_{P,hc,PF,L_y} - N_{P,hc,PF,L_y}^\prime =
\cases{ \frac{1}{2} N_{P,hc,FP,\frac{L_y}{2}} & for odd $L_y/2 > 1$ \ , \cr\cr
\frac{1}{2} \left [ N_{P,hc,FP,\frac{L_y}{2}} + N_{P,hc,FP,\frac{L_y-2}{2}} 
\right ] & for even $L_y/2$ \ .} 
\label{2nphcpf-nphcpfprime}
\eeq
\end{conjecture}

Conjectures~\ref{conjecture1} and~\ref{conjecture2} imply an exact formula for
$N_{P,hc,PF,L_y}$.

\section{Potts Model Partition Functions for Strips of the Honeycomb Lattice 
         with Free Boundary Conditions} \label{sec.free_bc}

The Potts model partition function for a strip of the honeycomb lattice of 
width $L_y$ and length $L_x$ vertices with free boundary conditions is given by
\beq
Z(L_y \times L_x,FF,q,v) 
        = \w^{\rm T} \cdot \T^m \cdot \uu_{\rm id} 
\label{def_Z_free}
\eeq
where $\T=\V \cdot \H_2 \cdot \V \cdot \H_1$ is the transfer matrix. $\H_1$ and
$\H_2$ are matrices corresponding respectively to adding two kinds of
transverse bonds in a slice, and $\V$ corresponding to adding longitudinal
bonds in each slice. The number $m$ is related to $L_x$ as defined in
eq. (\ref{Lxm}), and the vector $w$ is given by 
\beq
\w^{\rm T} = \cases{ \w^{\rm T}_{\rm odd} = \vv^{\rm T} \cdot \H_1 & for odd 
$L_x$ \ , \cr\cr
\w^{\rm T}_{\rm even} = \vv^{\rm T} \cdot \H_2 \cdot \V \cdot \H_1 & for even 
$L_x$ \ .} 
\eeq
Hereafter we shall follow the notation and the computational methods of 
\cite{ts,tt}. 

The matrices $\T$, $\V$, $\H_1$ and $\H_2$ act on the space of connectivities
of sites on the first slice, whose basis elements $\vv_{\cal P}$ are indexed by
partitions ${\cal P}$ of the vertex set $\{1,\ldots,L_y\}$. In particular,
$\uu_{\rm id} = \vv_{\{\{1\},\{2\},\ldots,\{L_y\}\}}$. We denote the set of
basis elements for a given strip as ${\bf P} = \{ \vv_{\cal P} \}$.  

An equivalent way to present a general formula for the partition function is
via a generating function.  Labelling a lattice strip of a given type and width
as $G_m$, with $m$ the length, one has
\beq
\Gamma(G,q,v,z) = \sum_{m=0}^\infty z^m Z(G_m,q,v)
\label{gamma}
\eeq
where $\Gamma(G,q,v,z)$ is a rational function
\beq
\Gamma(G,q,v,z) = \frac{{\cal N}(G,q,v,z)}{{\cal D}(G,q,v,z)}
\label{gammaform}
\eeq
with
\beq
{\cal N}(G,q,v,z)=\sum_{j=0}^{\deg_z({\cal N})} A_{G,j}z^j
\label{numgamma}
\eeq
\beq
{\cal D}(G,q,v,z) = 1+\sum_{j=1}^{N_{Z,hc,BC,L_y}} b_{G,j}z^j 
                  = \prod_{j=1}^{N_{Z,hc,BC,L_y}}
                        (1- \lambda_{Z,G,j}z)
\label{dengamma}
\eeq
where the subscript $BC$ denotes the boundary conditions.  In the
transfer-matrix formalism, the $\lambda_{Z,G,j}$'s in the denominator of the
generating function, eq.~(\ref{dengamma}), are the eigenvalues of $\T$.

Strips of the honeycomb lattice with free boundary conditions are well-defined
for widths $L_y \ge 2$.  The partition function $Z(G,q,v)$ was calculated, for
arbitrary $q$, $v$, and $m$, for the strip with $L_y=2$ and free boundary
conditions in \cite{hca}, using a systematic iterative application of the
deletion-contraction theorem. Here after re-expressing the results for $L_y=2$
in the present transfer matrix formalism, we shall report explicit results for
the partition function for strips with $L_y = 3$ and free boundary conditions.
For $4 \le L_y \le 7$, the expressions for $\T(L_y)$, $\w(L_y)$ and $\uu_{\rm
id}(L_y)$ are too lengthy to include here. They are available from the authors
on request.

%
%
\subsection{$L_y=2$} \label{sec.2F} 

For the strip with width $L_y=2$, we only have to consider odd $L_x$.  The
number of elements in the basis is equal to $C_2=2$: ${\bf P} = \{ 1,
\delta_{1,2} \}$. In this basis, the transfer matrix and the other relevant
quantities are given by
\begin{subeqnarray}
\T &=& \left( \begin{array}{cc}
              R_{11} &  D_1 F_2 R_{12} \\
              v^5 &  v^4 D_1 
           \end{array} \right) \\
\w_{\rm odd}^{\rm T} &=& q \left( F_1, D_1 \right)
\label{Tt2ffcompact}
\end{subeqnarray}
where
\begin{subeqnarray}
\slabel{def_Dk}
D_k &=& v+k \\
\slabel{def_Fk}
F_k &=& q+kv \\
\slabel{def_Ek}
R_{11} &=& q^4 + 5q^3v + 10q^2v^2 + 10qv^3 + 5v^4 \\
R_{12} &=& q^2 + 2qv + 2v^2 \ . 
\end{subeqnarray}
In terms of this transfer matrix and these vectors, one calculates the
partition function $Z(G_m,q,v)$ for the strip with a given length $m$ via
eq.~(\ref{def_Z_free}).  Equivalently, one can calculate the partition
function\ using a generating function, and this was the way in which the
results were presented in \cite{hca}, with
${\cal D} = \prod_{j=1}^2(1-\lambda_{hcf2,j}z)$ and
\beq
\lambda_{hcf2,(1,2)} = \frac{1}{2} \biggl [ M_1 \pm \sqrt{M_2}
 \ \biggr ]
\label{lams}
\eeq
where
\beq
M_1=q^4+5q^3v+10q^2v^2+10qv^3+6v^4+v^5
\label{t12}
\eeq
and
\beqs
M_2 & = & q^8+10q^7v+45q^6v^2+120q^5v^3+208q^4v^4-2q^4v^5+244q^3v^5-6q^3v^6
\cr\cr
& & +196q^2v^6-4q^2v^7+104qv^7+4qv^8+32v^8+8v^9+v^{10} \ .
\label{rs12}
\eeqs
The product of these eigenvalues, i.e., the determinant of $\T$, is
\beq
{\rm det}(\T) = v^4(1+v)(v+q)^4 = v^4 D_1 F_1^4 \ .
\label{det_tff2}
\eeq
The vanishing of this determinant at $v=-1$ and $v=-q$ occurs because in each
case one of the two eigenvalues is absent for, respectively, the chromatic and
flow polynomials \cite{hs,f}.  Analogous formulas can be given for ${\rm
det}(\T)$ for higher values of $L_y$; we omit these for brevity.
%
%
%
\subsection{$L_y=3$} \label{sec.3F} 

For the $hc$ strip of width $L_y=3$ the number of elements in the basis for
enumerating partitions is given by the Catalan number $C_3=5$.  This basis is
${\bf P} = \{ 1, \delta_{1,2}, \delta_{1,3}, \delta_{2,3}, \delta_{1,2,3}
\}$. In this basis, the transfer matrices and the other relevant quantities are
\begin{subeqnarray}
\T &=& \left( \begin{array}{ccccc}
F_1 F_2 S_{11} & D_1 F_1 S_{12} & S_{13} S_{13}^\prime & S_{14} & D_1 S_{15} \\
v^5 F_1 F_2 & v^4 D_1 F_1 F_2 & v^5 S_{23} & v^5 S_{25} & v^4 D_1 S_{25} \\
v^6 F_1 & v^5 D_1 F_1 & v^4 S_{33} & v^5 S_{25} & v^4 D_1 S_{25} \\
v^3 F_1 F_2 R_{12} & v^3 D_1 F_1 S_{42} & v^3 F_2 S_{43} & v^3 S_{44} & v^3 D_1 S_{45} \\ 
v^7 F_1 & v^6 D_1 F_1 & v^6 S_{25} & v^7 D_3 & v^6 D_1 D_3 \\
             \end{array} \right) \\
\w_{\rm odd}^{\rm T} &=&   
   q \left( q(q+v), q(1+v), q+v, q+v, 1+v \right) \\ 
\w_{\rm even}^{\rm T} &=&   
   q \left( F_1^5, D_1 F_1^4, F_1 X_1, F_1^2 X_2, D_1 F_1 X_2 \right) \\ 
\uu_{\rm id}^{\rm T} &=&  \left( 1, 0, 0, 0, 0 \right)  
\end{subeqnarray}
where the $S_{ij}$ and $X_k$ are defined in a shorthand notation as 
\begin{subeqnarray}
S_{11} &=& q^4 + 5q^3v + 11q^2v^2 + 12qv^3 + 7v^4  \\
S_{12} &=& q^4 + 6q^3v + 15q^2v^2 + 19qv^3 + 11v^4  \\
S_{13} &=& q^2 + 4qv + 5v^2  \\
S_{13}^\prime &=& q^3 + 4q^2v + 7qv^2 + 7v^3 + v^4  \\
S_{14} &=& q^5 + 8q^4v + 28q^3v^2 + q^3v^3 + 54q^2v^3 + 5q^2v^4 + 59qv^4 + 9qv^5 + 32v^5 + 7v^6  \cr
& & \\
S_{15} &=& q^4 + 7q^3v + 21q^2v^2 + q^2v^3 + 33qv^3 + 4qv^4 + 24v^4 + 5v^5  \\
S_{23} &=& 2q + 5v + v^2 \\
S_{25} &=& q + 4v + v^2 \\
S_{33} &=& q^2 + 4qv + 6v^2 + v^3 \\
S_{42} &=& q^2 + 3qv + 3v^2 \\
S_{43} &=& q^2 + 3qv + 5v^2 + v^3 \\
S_{44} &=& q^3 + 6q^2v + q^2v^2 + 12qv^2 + 3qv^3 + 10v^3 + 3v^4 \\
S_{45} &=& q^2 + 5qv + qv^2 + 7v^2 + 2v^3  
\end{subeqnarray}
\begin{subeqnarray}
X_1 &=& q^3 + 4q^2v + 6qv^2 + 4v^3 + v^4  \\
X_2 &=& q^2 + 3qv + 3v^2 +v^3 
\end{subeqnarray}
We will discuss properties of the resultant partition functions below.

\section{Potts Model Partition Functions for Honeycomb-lattice Strips 
         with Cylindrical Boundary Conditions}

For the honeycomb lattice with cylindrical boundary conditions, the width must
be even (and larger than two in order to avoid the degenerate situation of
vertical edges forming emanating from and returning to a given vertex). The
Potts model partition function $Z(G,q,v)$ for a honeycomb-lattice strip with
cylindrical boundary conditions can be written in the same form as in
eq.(\ref{def_Z_free}). Here either $\H_1$ or $\H_2$ should include the bond
connecting the boundary sites in the transverse direction.  The dimension of
the transfer matrix can be reduced by the two symmetries discussed in Section
\ref{sec.2}, namely the length-two translation symmetry along the transverse
direction and reflection symmetry. This number is $N_{Z,hc,PF,L_y}$ and is
given in terms of $L_y$ by eq.~\reff{nzhcpf} (see Table~\ref{nztable} for some
numerical values).  We consider the basis in the translation-invariant and
reflection-invariant subspace to construct the transfer matrix and the
corresponding vectors. To simplify the notation, we will still use $\T$, $\w$
and $\uu_{\rm id}$ as in eq.(\ref{def_Z_free}), so that the partition function
is given by the analogous equation, $Z(L_y \times L_x,PF,q,v)= \w^{\rm T} \cdot
\T^m \cdot \uu_{\rm id}$.  We have calculated the transfer matrix $\T(L_y)$ and
the vectors $\w(L_y)$ and $\uu_{\rm id}(L_y)$ for $L_y=4$ and $L_y=6$. The
explicit results for $L_y=4$ are given in the appendix; the results for $L_y =
6$ are too lengthy to present here, and are available upon request.

\section{Partition Function Zeros in the $\lowercase{q}$ Plane}

\begin{figure}[hbtp]
\vspace*{-1cm}
\centering
\begin{tabular}{cc}
   \includegraphics[width=170pt]{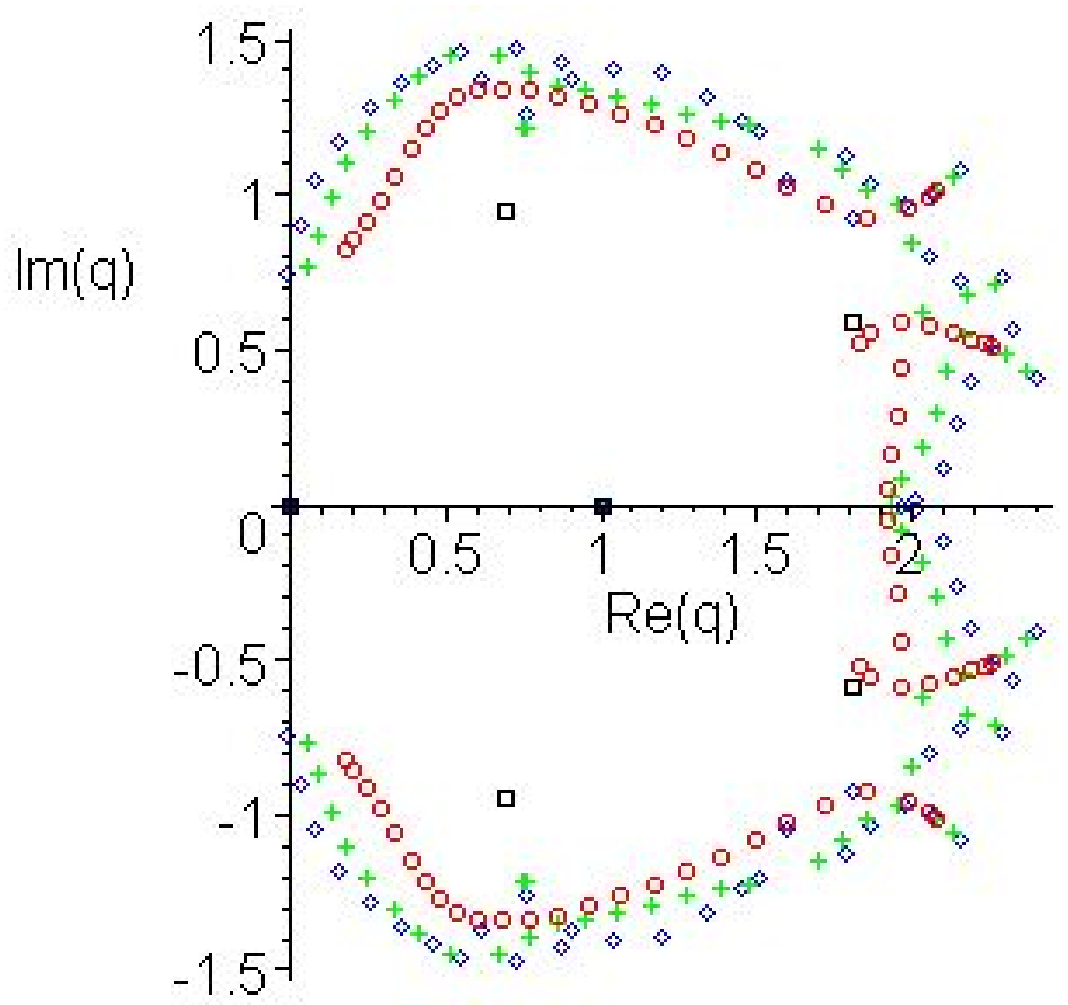} \qquad \qquad & \qquad \qquad 
   \includegraphics[width=170pt]{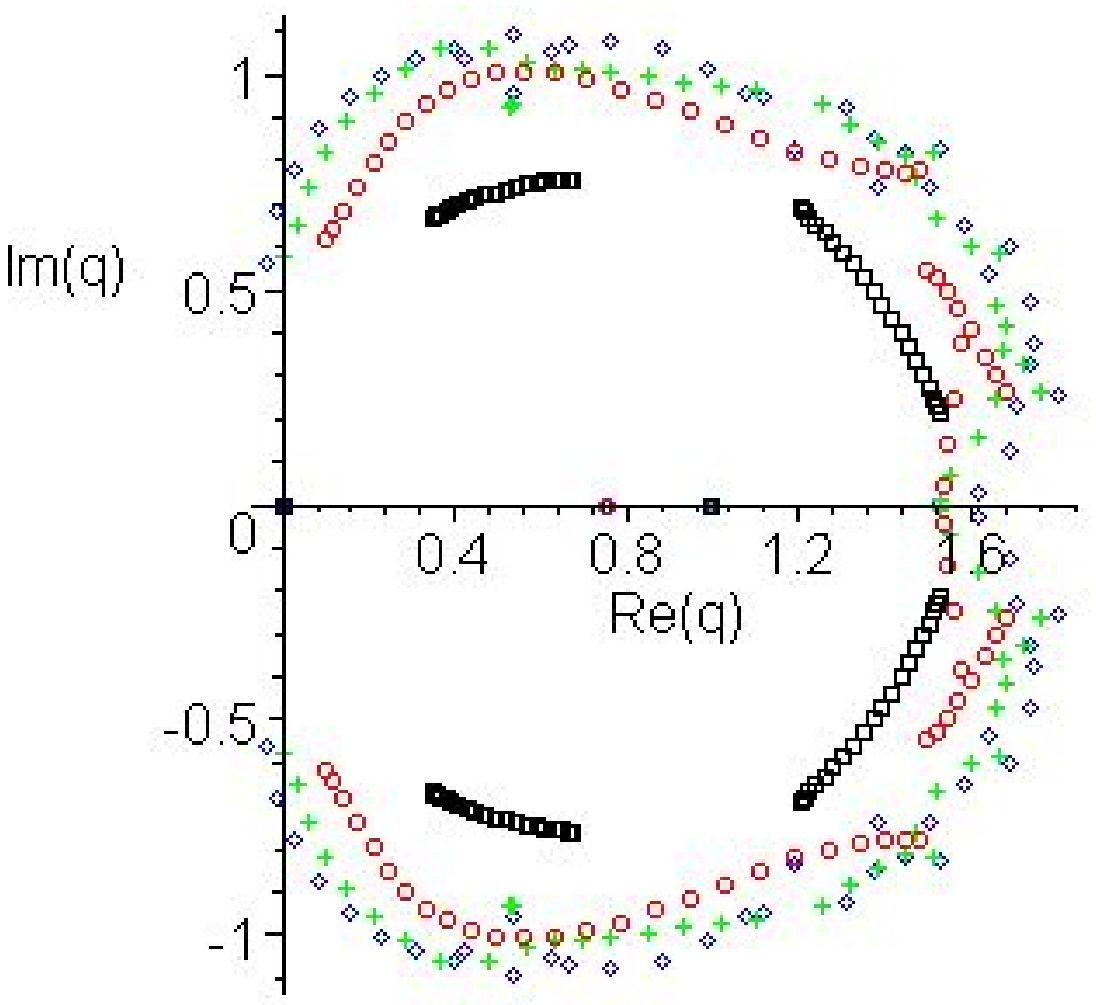} \\
   \phantom{(((a)}(a)    & \phantom{(((a)}(b) \\[5mm]
   \includegraphics[width=170pt]{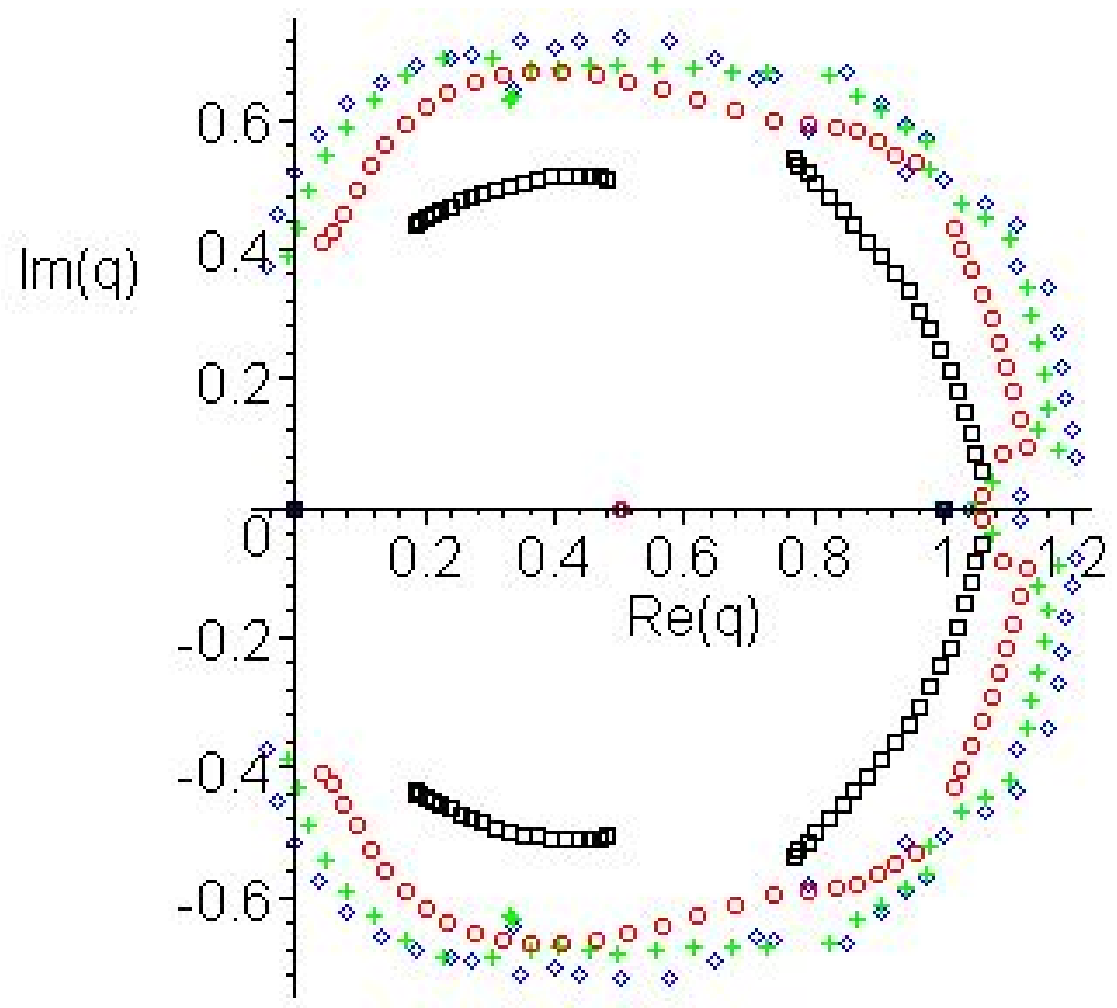} \qquad \qquad & \qquad \qquad 
   \includegraphics[width=170pt]{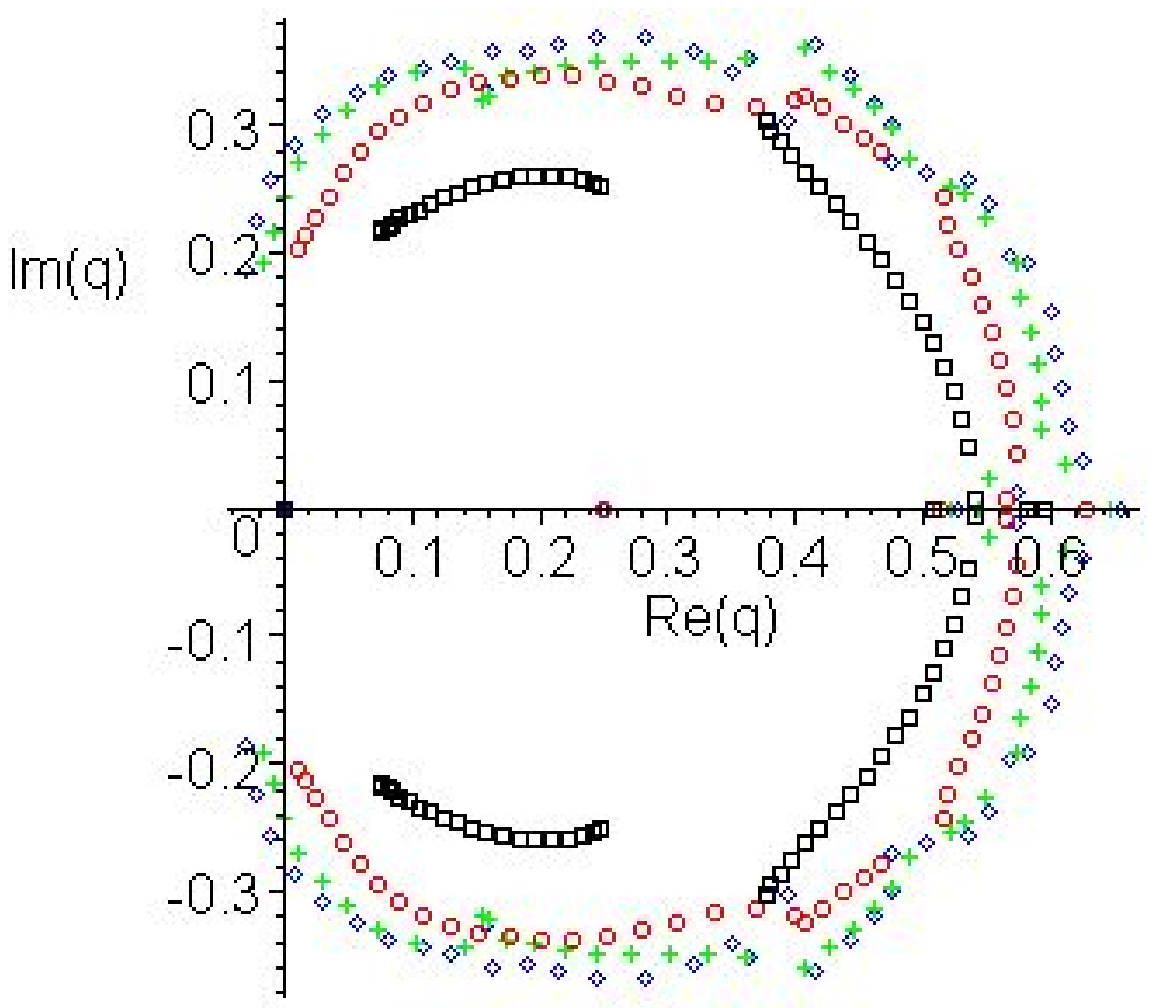} \\
   \phantom{(((a)}(c)    & \phantom{(((a)}(d) \\
\end{tabular}
\caption[a]{\protect\label{figures_qplane_F} Partition-function zeros in the
  $q$ plane for the Potts antiferromagnet with (a)
$v=-1.0$, (b) $v=-0.75$, (c) $v=-0.5$, and (d) $v=-0.25$ on strips with free
boundary conditions and several widths $L_y$: 2 ($\Box$, black), 3 ($\circ$,
red), 4 ($+$, green), and 5 ($\Diamond$, blue), where the colors refer to
  the online paper.}
\end{figure}

\begin{figure}[hbtp]
\vspace*{-1cm}
\centering
\begin{tabular}{cc}
   \includegraphics[width=170pt]{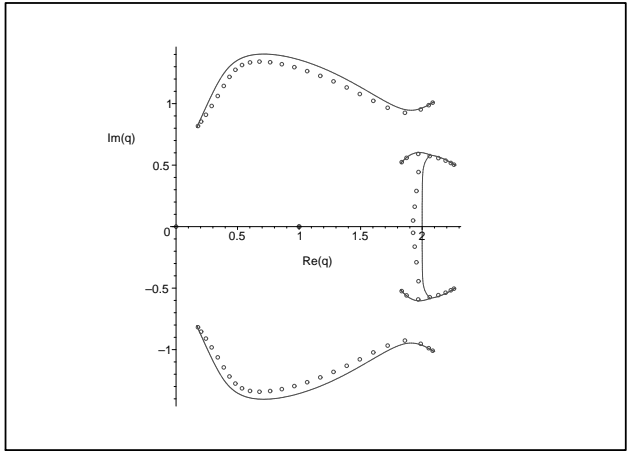} \qquad \qquad & \qquad \qquad 
   \includegraphics[width=170pt]{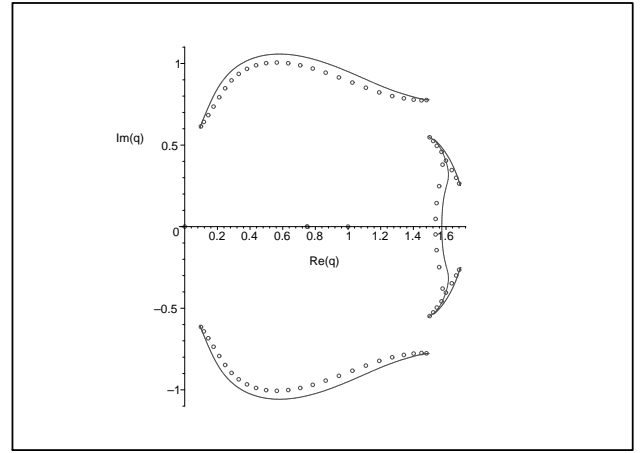} \\
   \phantom{(((a)}(a)    & \phantom{(((a)}(b) \\[5mm]
   \includegraphics[width=170pt]{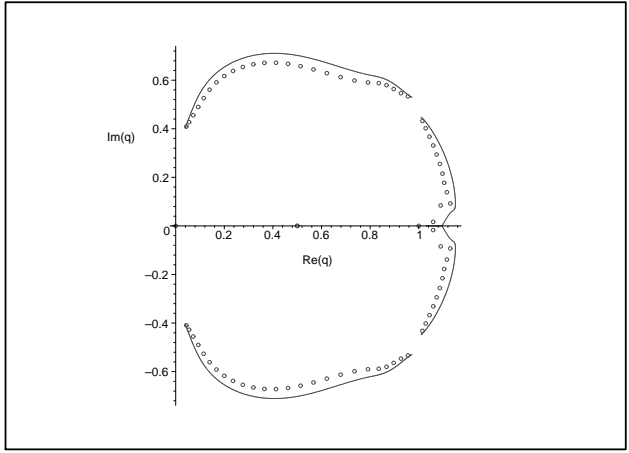} \qquad \qquad & \qquad \qquad 
   \includegraphics[width=170pt]{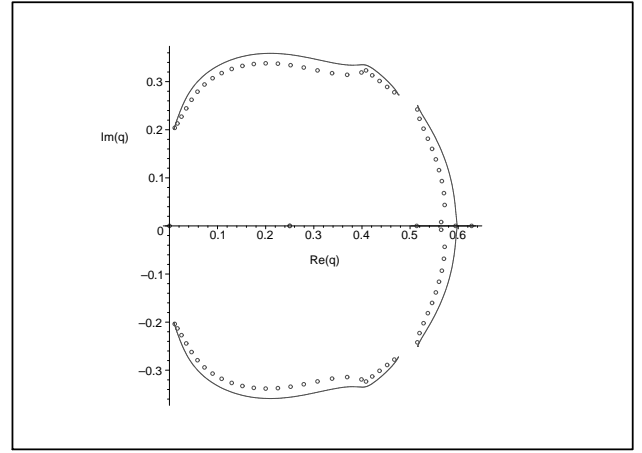} \\
   \phantom{(((a)}(c)    & \phantom{(((a)}(d) \\
\end{tabular}
\caption[a]{\protect\label{figures_qplane_F2} Partition-function zeros and
asymptotic loci ${\cal B}$ in the $q$ plane for the Potts antiferromagnet on
the $hc$ strip with width $L_y=3$ and free boundary conditions, for (a)
$v=-1.0$, (b) $v=-0.75$, (c) $v=-0.5$, (d) $v=-0.25$.}
\end{figure}

\begin{figure}[hbtp]
\vspace*{-1cm}
\centering
\begin{tabular}{cc}
   \includegraphics[width=170pt]{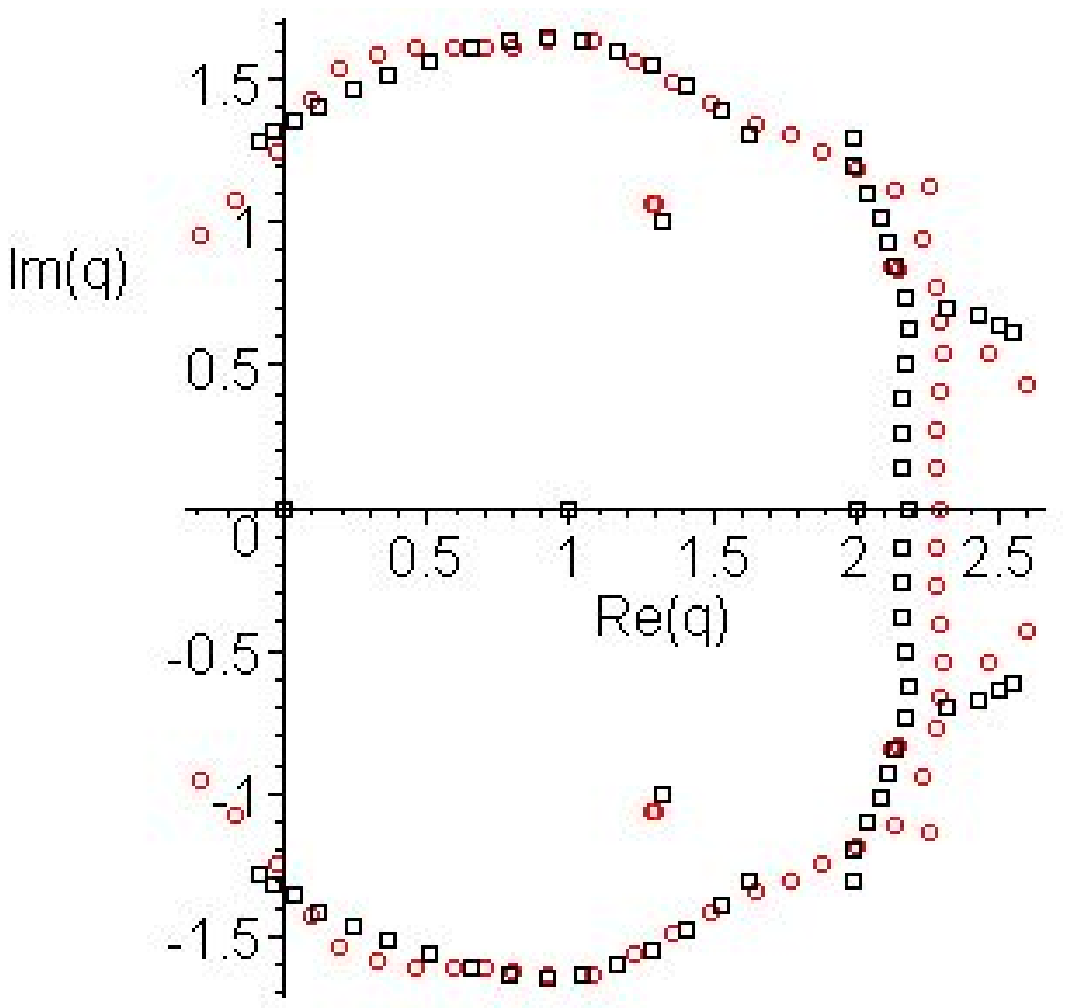} \qquad \qquad & \qquad \qquad 
   \includegraphics[width=170pt]{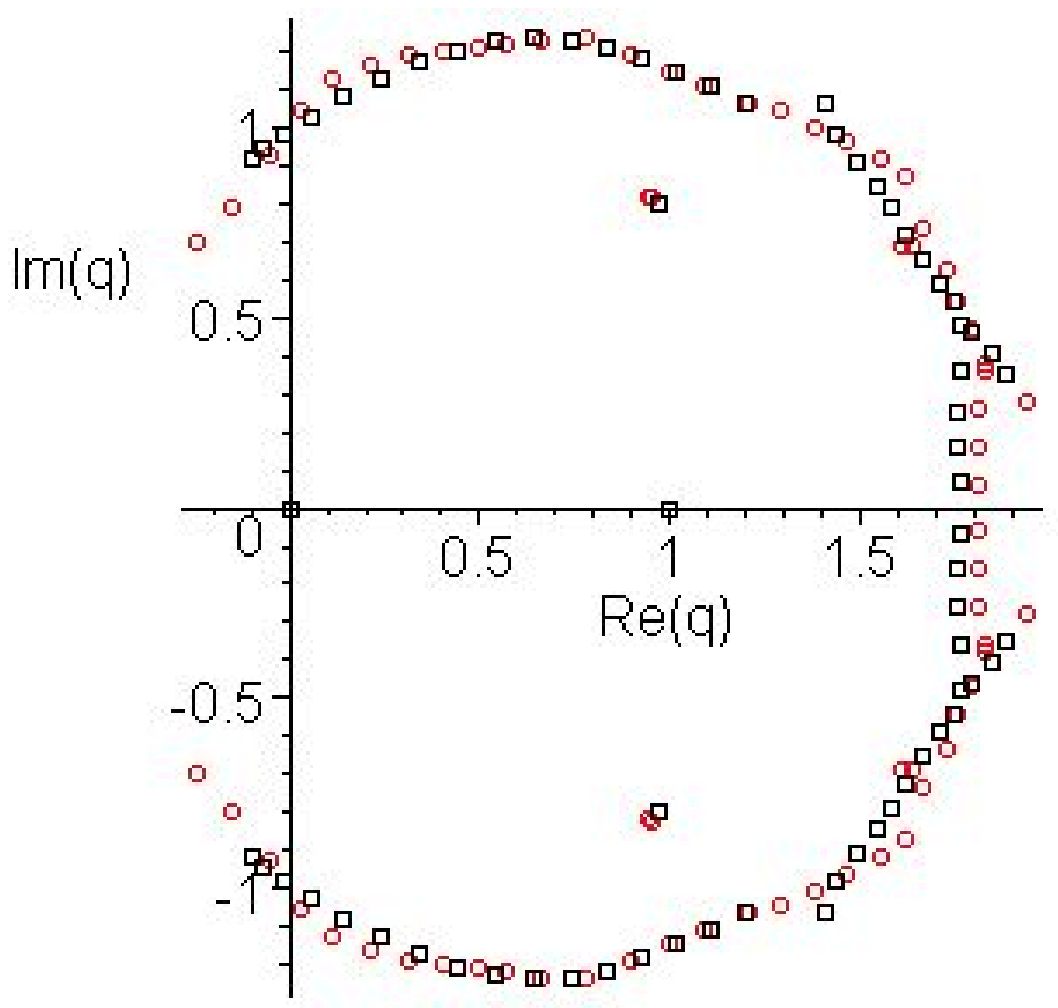} \\
   \phantom{(((a)}(a)    & \phantom{(((a)}(b) \\[5mm]
   \includegraphics[width=170pt]{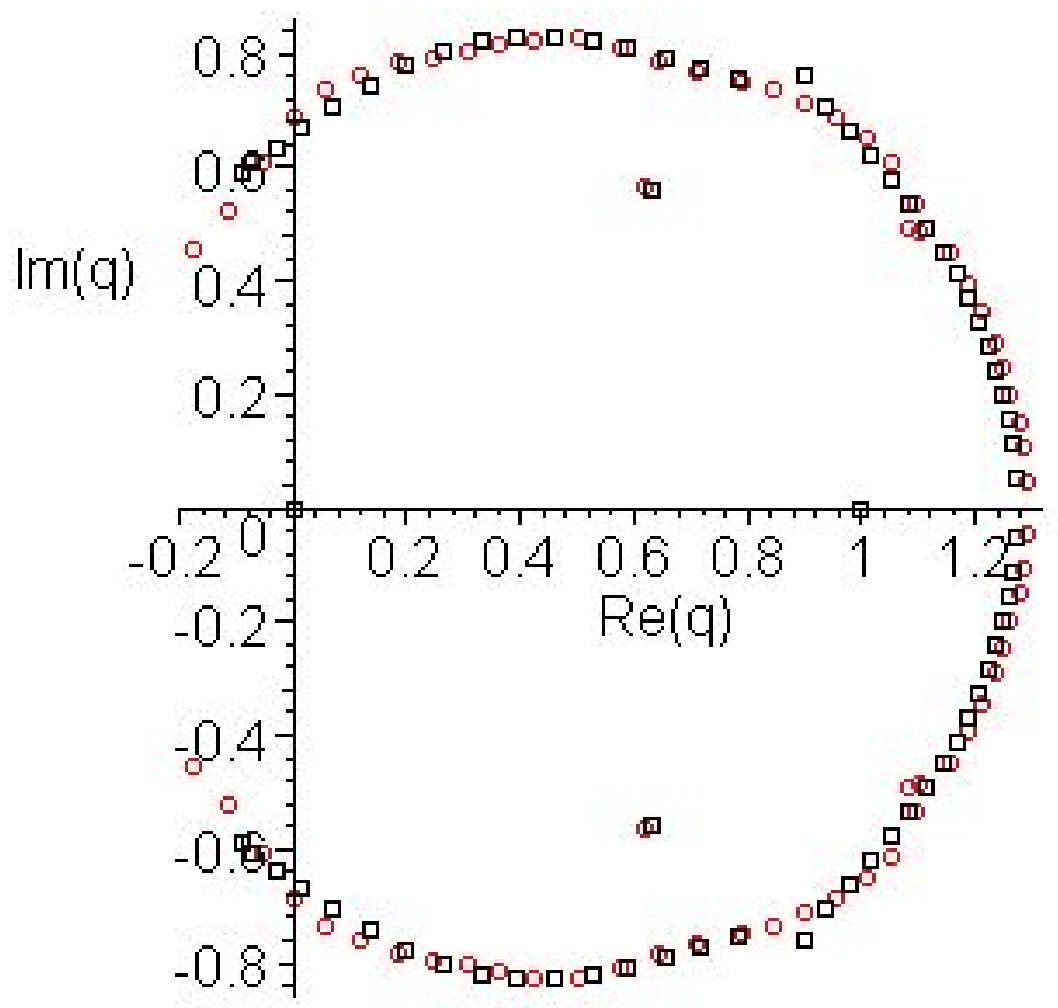} \qquad \qquad & \qquad \qquad 
   \includegraphics[width=170pt]{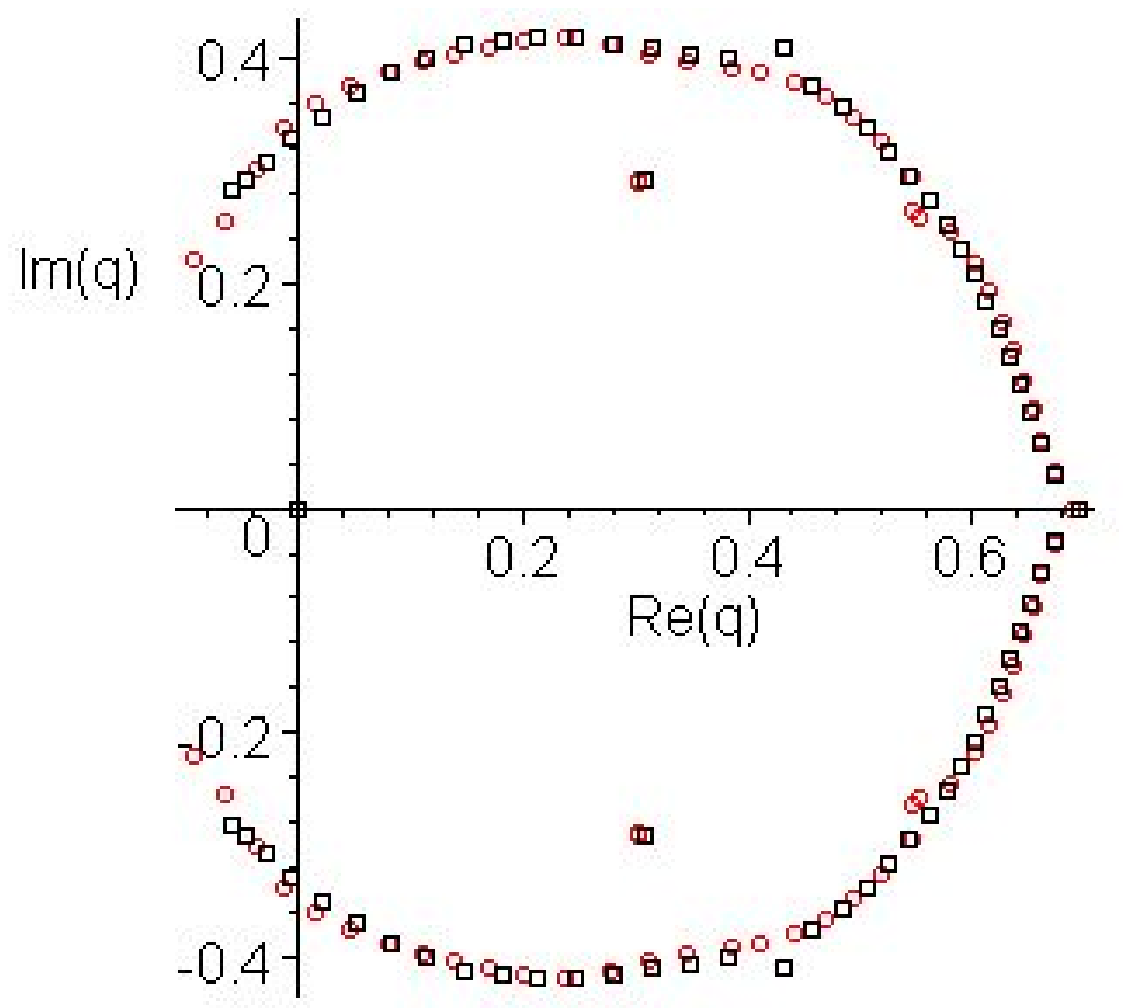} \\
   \phantom{(((a)}(c)    & \phantom{(((a)}(d) \\
\end{tabular}
\caption[a]{\protect\label{figures_qplane_P} Partition-function zeros for (a)
$v=-1.0$, (b) $v=-0.75$, (c) $v=-0.5$, and (d) $v=-0.25$ on strips with
cylindrical boundary conditions and several widths $L_y$: 4 ($\Box$, black), 6
($\circ$, red), where the colors refer to the online paper.} 
\end{figure}

In this section we shall present results for zeros in the $q$-plane for the
partition function of the Potts antiferromagnet on strips of the honeycomb
lattice with free and cylindrical boundary conditions, for various values of
the temperature-like variable $v$.  Fig. \ref{figures_qplane_F} shows these
zeros for strips of widths $2 \le L \leq 5$ and free boundary conditions.  In
the limit $L_x \to \infty$ the zeros merge to form sets of curves which,
together, comprise the locus ${\cal B}$.  As an illustration of this, in
Fig. \ref{figures_qplane_F2} we show these zeros for the free strips of width
$L_y=3$, together with the loci ${\cal B}$ for various values of $v$.  One sees
that the strip lengths that we use to calculate the zeros are sufficiently
great that most of these zeros lie rather close to the infinite-length
asymptotic loci.  This behavior - that partition function zeros calculated for
long strips generally lie close to the asymptotic loci ${\cal B}$ - is similar
to what we found in our previous work. Hence, one can draw a reasonably good
inference for many of the features of these loci ${\cal B}$ from the zeros.
The corresponding partition-function zeros for honeycomb-lattice strips with
$L_y=4,6$ and cylindrical boundary conditions are shown in
Figure~\ref{figures_qplane_P}.  Again, one could calculate the asymptotic loci
${\cal B}$, but since the zeros already give a reasonably good idea of the
structure of these loci, they will suffice for our present purposes.  Our
Figs. \ref{figures_qplane_F} and \ref{figures_qplane_P} include calculations up
to $L_y$=5 and 6, respectively. We previously studied the case $L_y=2$ in
\cite{hca} (again for arbitrary $q$ and temperature).  From our present
results, we see that, for general values of $v$, as the width $L_y$ increases,
the curve envelope moves outward somewhat and the arc endpoints on the left
move slowly toward $q=0$.  This behavior is consistent with the hypotheses that
for a given $v$, as $L_y \to \infty$, (i) ${\cal B}_q$ would approach a
limiting locus as $L_y \to \infty$ and (ii) this locus would separate the $q$
plane into different regions, with a curve passing through $q=0$ as well as a
maximal real value, $q_c(v)$.  This is qualitatively similar to the behavior
that was found earlier for the square-lattice strips \cite{a,s3a,ts} and the
triangular-lattice strips \cite{tt}.  As expected, the convergence to the limit
$L_y \to \infty$ seems to be faster with cylindrical boundary conditions, as
there are no surface effects when the length is made infinite.  In this limit
$L_y \to \infty$, one expects that the locus ${\cal B}$ will have the property
that the maximal point at which it crosses the real axis, $q_c(v)$, is equal to
the solution, eq. (\ref{qsol}) (with the plus sign for the square root), of the
criticality condition on the infinite honeycomb lattice, eq. (\ref{hc_eq})
below.

We recall some results for the special case $v=-1$ corresponding to the
zero-temperature Potts antiferromagnet. Chromatic zeros were calculated for a
finite patch of the honeycomb lattice with free and cylindrical boundary
conditions in (Fig. 8 of) \cite{baxter87}.  Chromatic zeros and asymptotic loci
for strips of this lattice were calculated for free longitudinal boundary
conditions in \cite{hca,hs,strip} and for periodic longitudinal boundary
conditions in \cite{hca,cf,pg,pt} (see, e.g., Fig. 6 of \cite{strip} for
$L_y=3$ and free b.c., Fig. 1 of \cite{pg} for $L_y=2$ and cyclic b.c., Fig. 17
of \cite{hca} and Fig. 7 of \cite{pt} for $L_y=3$ and cyclic b.c., and Figs. 8
and 9 of \cite{pt} for $L_y=4$ and $L_y=5$ with cyclic b.c.).  Comparisons of
the arcs on ${\cal B}$ for $v=-1$ obtained for strips with free longitudinal
boundary conditions led to the inference that in the limit $L_y \to \infty$
these loci would separate the $q$ plane into regions including a curve passing
through $q=0$ \cite{bcc}.  The property that, for $v=-1$, ${\cal B}_q$
separates the $q$ plane into regions with one of the curves on ${\cal B}_q$
passing through the origin is also observed for lattice strips with finite
width $L_y$ if one imposes periodic longitudinal boundary conditions
\cite{pg},\cite{wcyl}-\cite{bcc}.  In particular, for $v=-1$ and cyclic b.c.,
we found the following values of $q_c(-1)$ (i) 2 for $L_y=2$ \cite{pg}, (ii) 2
for $L_y=3$ \cite{hca}, (iii) $\simeq 2.1548$ for $L_y=4$ \cite{pt}, and (iv)
$\simeq 2.2641$ for $L_y=5$ \cite{pt}.  For $L_y \to \infty$, eq. (\ref{qsol})
formally yields $q_c(-1)=(3+\sqrt{5})/2 \simeq 2.6180$.  This is formal since
the Potts antiferromagnet is, in general, only defined for $q \in {\mathbb
Z}_+$, owing to the fact that the formula (\ref{cluster}) can yield a negative
and hence unphysical value for $Z(G,q,v)$ if $v$ is negative and $q$ is not a
non-negative integer.  There are at least two different ways in which, as $L_y
\to \infty$, the loci ${\cal B}$ could approach the limiting form containing
$q_c(-1)$: (i) the endpoints of the complex-conjugate prongs farthest to the
right could move down and pinch the real axis and/or (ii) the point at which
the locus ${\cal B}$ for finite $L_y$ crosses the real axis could increase to
this limiting value.  These are not necessarily mutually exclusive; a
combination of (i) and (ii) could presumably occur, so that the final locus
would have a structure similar to that of the $L_y=2$ cyclic case shown in
Fig. 1(a) of \cite{pg} or the $L_y=3$ case shown in Fig. 7 of \cite{pt}, in
which several curves on ${\cal B}$ meet at $q_c(-1)$. Alternatively, the
rightmost complex-conjugate prongs might bend back and intersect the rest of
the boundary ${\cal B}$ away from the real axis, forming ``bubble'' regions, as
we found for the cyclic $L_y=4,5$ strips (Figs. 8 and 9 in \cite{pt}).

As regards the general distribution of zeros, for a given $L_y$, as $v$
increases from $-1$ to 0 (i.e., $K=\beta J$ increases from $-\infty$ to 0,
these zeros contract to a point at $q=0$.  This is an elementary consequence of
the fact that $K \to 0$, the spin-spin interaction term in the Potts model
Hamiltonian, ${\cal H}$, vanishes, so that the sum over states just counts all
$q$ possible spin states independently at each vertex, and $Z(G,q,v)$
approaches the value $Z(G,q,0)=q^n$.  More generally, one can inquire about the
maximum modulus of a zero of $Z(G,q,v)$ in the $q$ plane as a function of $v$.
It has been proved \cite{sokalzero} that (for a graph $G$ without loops), for
the antiferromagnetic Potts model partition function with $|1+v| \le 1$, the
zeros of $Z(G,q,v)$ lie in the disk $|q| < C r |v|$, where $C \simeq 7.964$ and
$r$ is the maximal degree of a vertex, with $r=3$ for our honeycomb-lattice
strips.  Our zeros have moduli that lie considerably below this upper bound.
For example, for $v=-1$, the zeros that we have calculated and displayed in
Figs. \ref{figures_qplane_F}-\ref{figures_qplane_P} have moduli bounded above
by about 1.2 and 1.3, respectively, while the above-mentioned inequality would
give an upper bound of $|q| < 3C \simeq 23.9$.

One can also plot ${\cal B}_q$ for the ferromagnetic region $0 \le v \le
\infty$.  Although we have not included these plots here, we note that an
elementary Peierls argument shows that the Potts ferromagnet on
infinite-length, finite-width strips has no finite-temperature phase transition
and associated magnetic long range order. Hence, for this model ${\cal B}_q$
does not cross the positive real $q$ axis for $0 < v < \infty$.

\bigskip
\bigskip

%
%

%
%

%
%

\section{Partition Function Zeros in the $\lowercase{v}$ Plane}

\subsection{General}

In this section we shall present results for zeros in the $v$-plane, for
various values of $q$, for the partition function of the Potts model on
strips of the honeycomb lattice of widths $L_y \leq 5$ with free boundary
conditions and of widths $L_y=4,6$ with cylindrical boundary conditions.
We recall the possible noncommutativity in the
definition of the free energy for certain special integer values of $q$,
denoted $q_{sp}$, (see eqs.~(2.10), (2.11) of \cite{a}):
\beq
\lim_{n \to \infty} \lim_{q \to q_{sp}} Z(G,q,v)^{1/n} \ne 
\lim_{q \to q_{sp}} \lim_{n \to\infty} Z(G,q,v)^{1/n} \ . 
\label{fnoncom}
\eeq
As discussed in \cite{a}, because of this noncommutativity, the formal
definition (\ref{ef}) is, in general, insufficient to define the free energy
$f$ at these special points $q_{sp}$; it is necessary to specify the order of
the limits that one uses in the above equation. We denote the two definitions
using different orders of limits as $f_{qn}$ and $f_{nq}$: $f_{nq}(\{G\},q,v) =
\lim_{n \to \infty} \lim_{q \to q_{sp}} n^{-1} \ln Z(G,q,v)$ and
$f_{qn}(\{G\},q,v) = \lim_{q \to q_{sp}} \lim_{n \to \infty} n^{-1} \ln
Z(G,q,v)$.
 
As a consequence of this noncommutativity, it follows that for the special set
of points $q=q_{sp}$ one must distinguish between (i) $({\cal
B}_v(\{G\},q_{sp}))_{nq}$, the continuous accumulation set of the zeros of
$Z(G,q,v)$ obtained by first setting $q=q_{sp}$ and then taking $n \to \infty$,
and (ii) $({\cal B}_v(\{G\},q_{sp}))_{qn}$, the continuous accumulation set of
the zeros of $Z(G,q,v)$ obtained by first taking $n \to \infty$, and then
taking $q \to q_{sp}$.  For these special points (cf.~eq.~(2.12) of \cite{a}),
\beq
({\cal B}_v(\{G\},q_{sp}))_{nq} \ne ({\cal B}_v(\{G\},q_{sp}))_{qn} \ .
\label{bnoncom}
\eeq
Here this noncommutativity will be relevant for $q=0$ and $q=1$. 

\begin{figure}[hbtp]
\centering
\begin{tabular}{cc}
   \includegraphics[width=170pt]{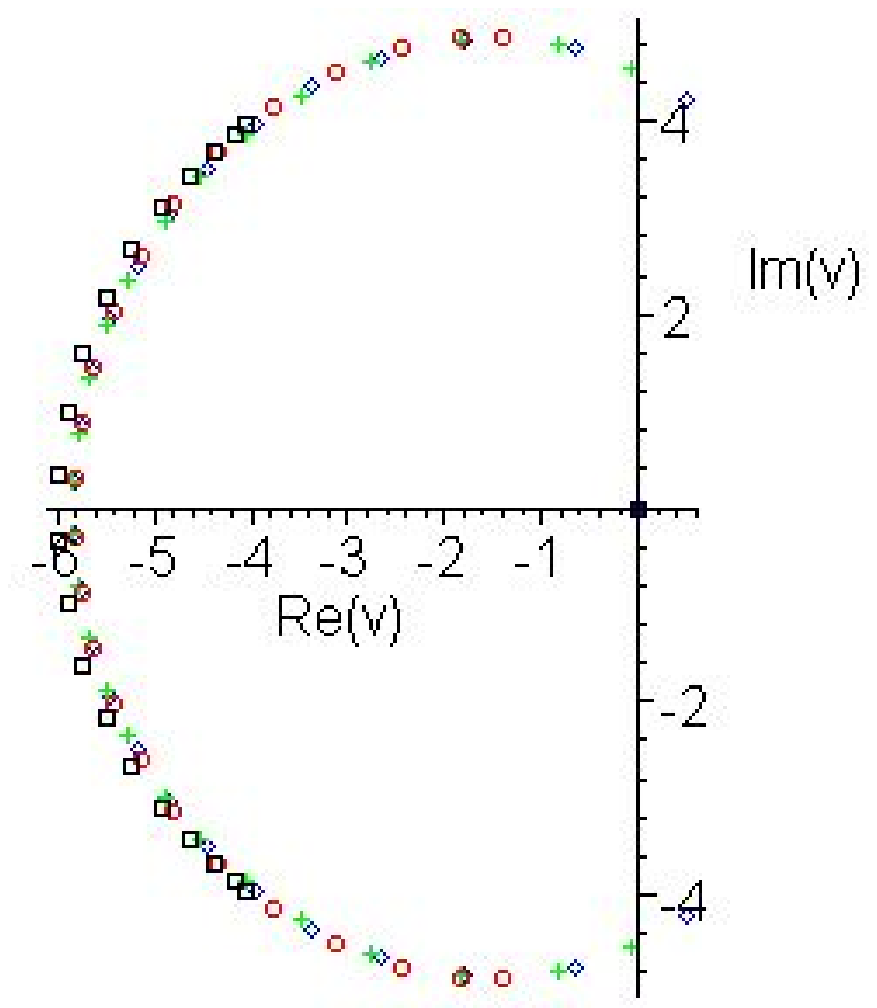} \qquad \qquad & \qquad \qquad 
   \includegraphics[width=170pt]{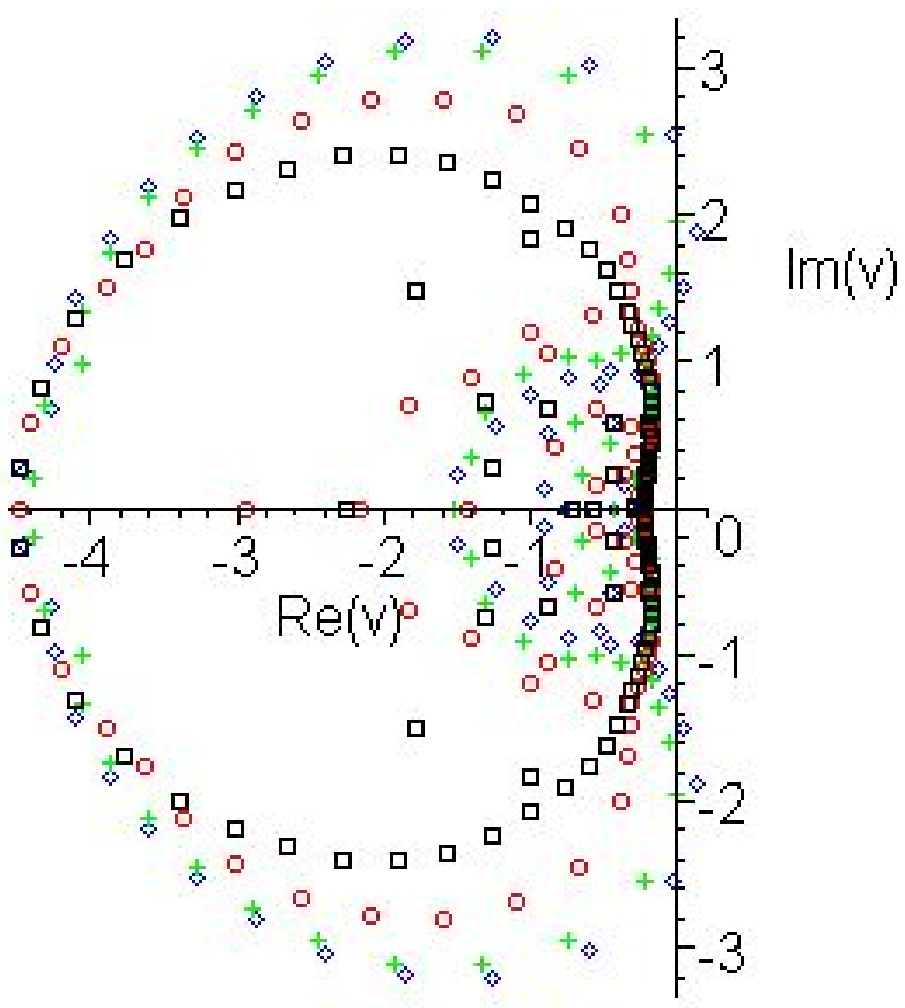} \\
   \phantom{(((a)}(a)    & \phantom{(((a)}(b) \\[5mm]
   \includegraphics[width=170pt]{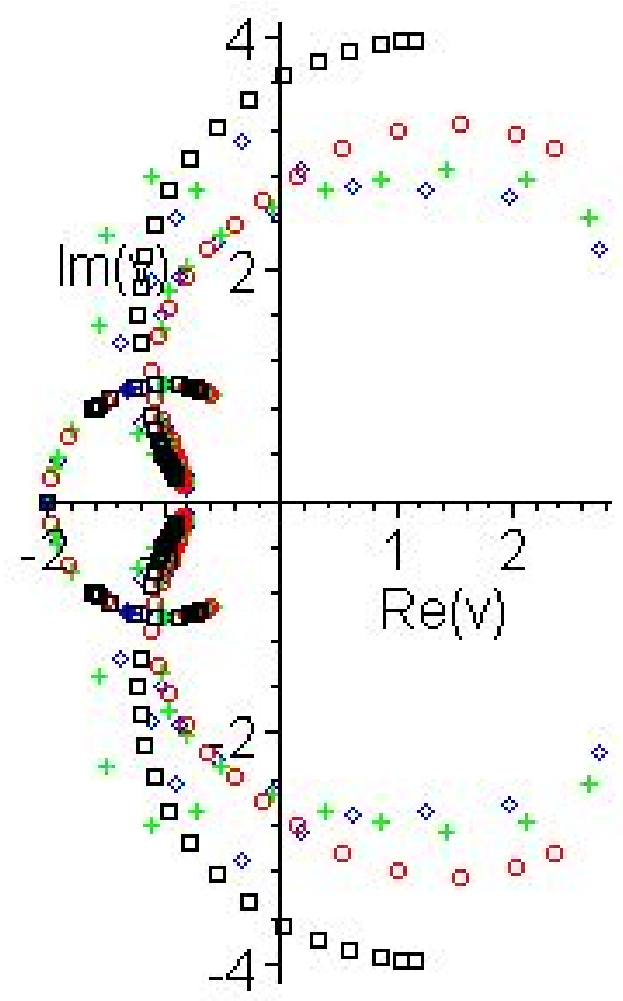} \qquad \qquad & \qquad \qquad 
   \includegraphics[width=170pt]{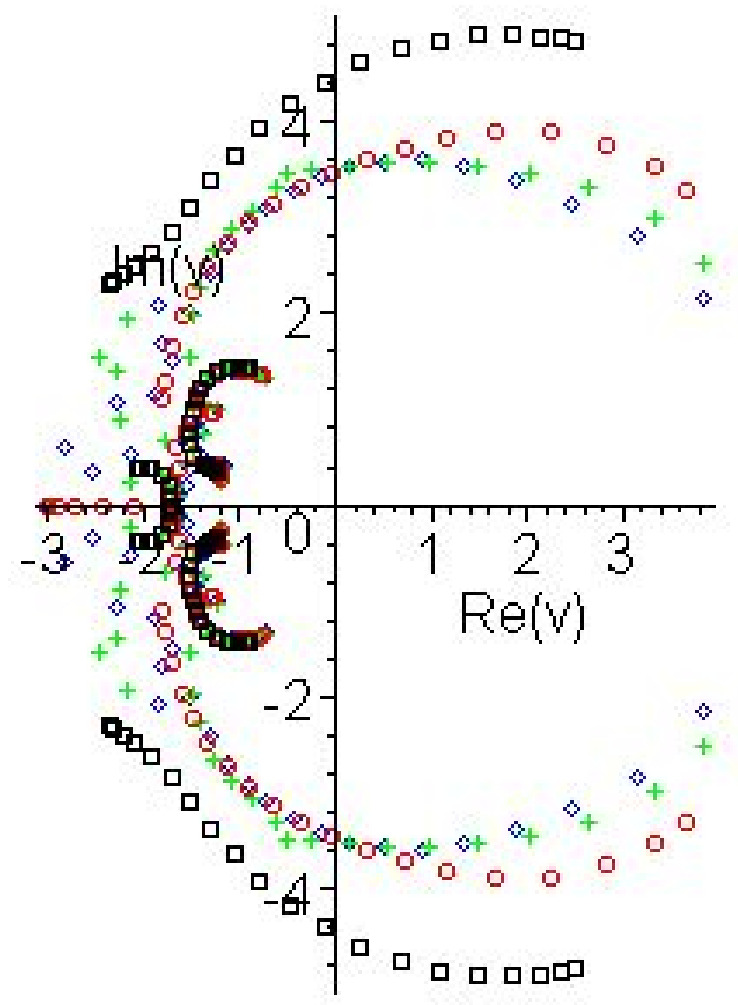} \\
   \phantom{(((a)}(c)    & \phantom{(((a)}(d) \\
\end{tabular}
\caption[a]{\protect\label{figures_vplane_F} Partition-function zeros, in the
  $v$ plane, of (a) $Z(G,q,v)/q$ for $q=0$, and $Z(G,q,v)$ for (b) $q=0.999$,
  (c) $q=2$, and (d) $q=3$ on strips with free boundary conditions and several
  widths $L_y$: 2 ($\Box$, black), 3 ($\circ$, red), 4 ($+$, green), and 5
  ($\Diamond$, blue), where the colors refer to the online paper.  }
\end{figure}

\begin{figure}[hbtp]
\centering
\begin{tabular}{cc}
   \includegraphics[width=170pt]{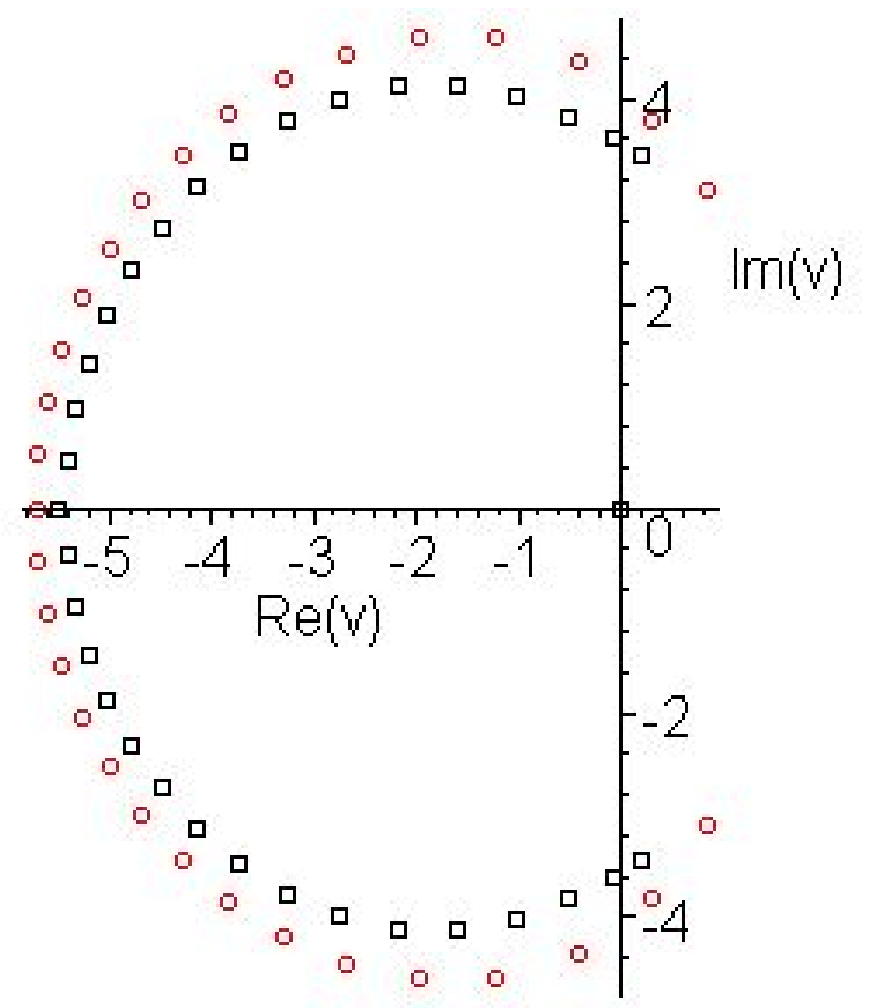} \qquad \qquad & \qquad \qquad 
   \includegraphics[width=170pt]{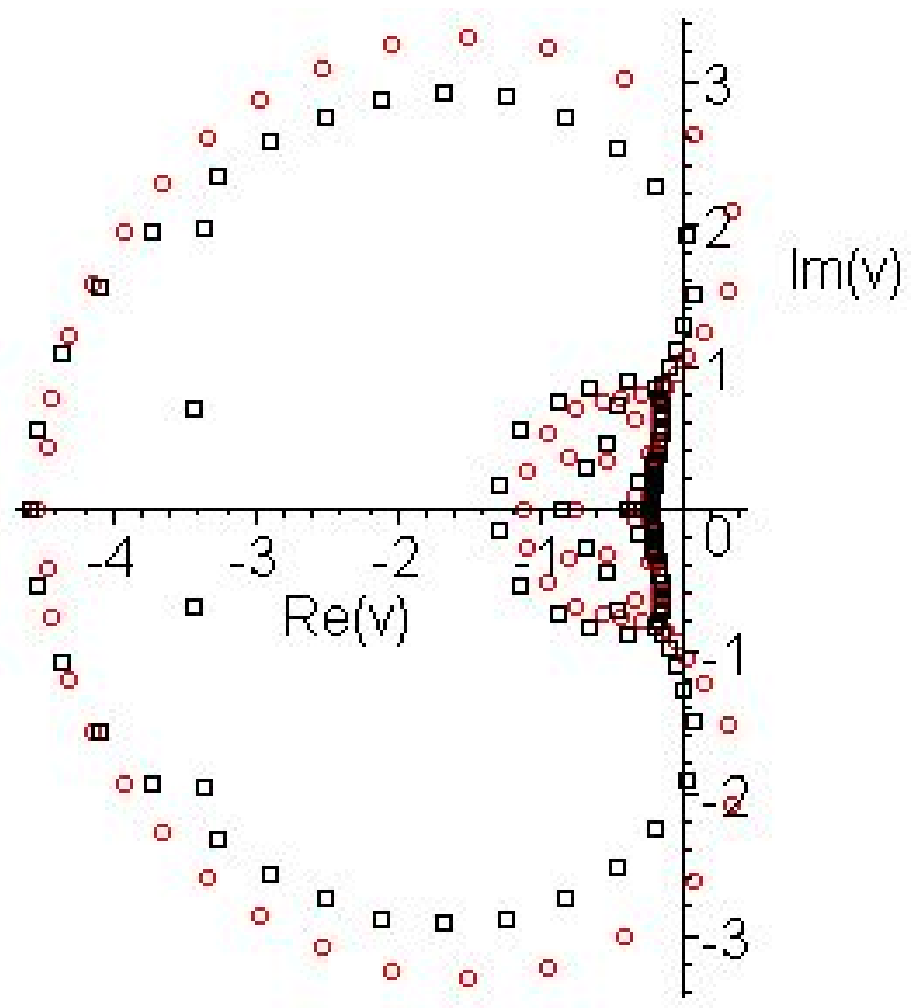} \\
   \phantom{(((a)}(a)    & \phantom{(((a)}(b) \\[5mm]
   \includegraphics[width=170pt]{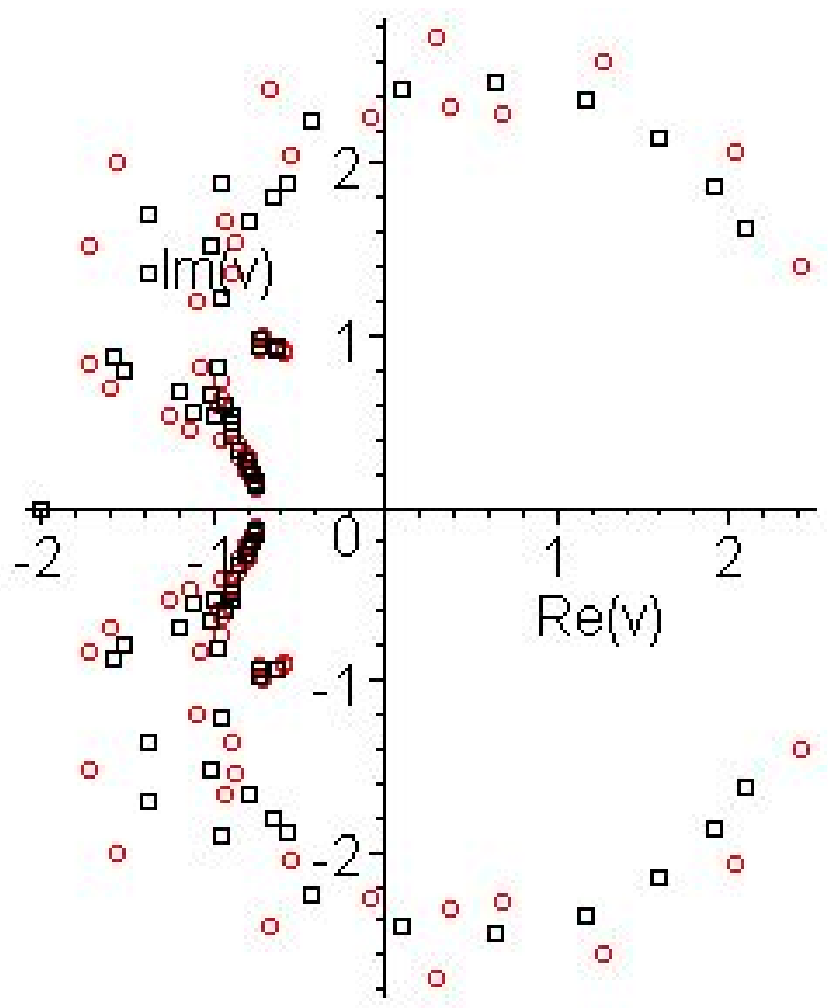} \qquad \qquad & \qquad \qquad 
   \includegraphics[width=170pt]{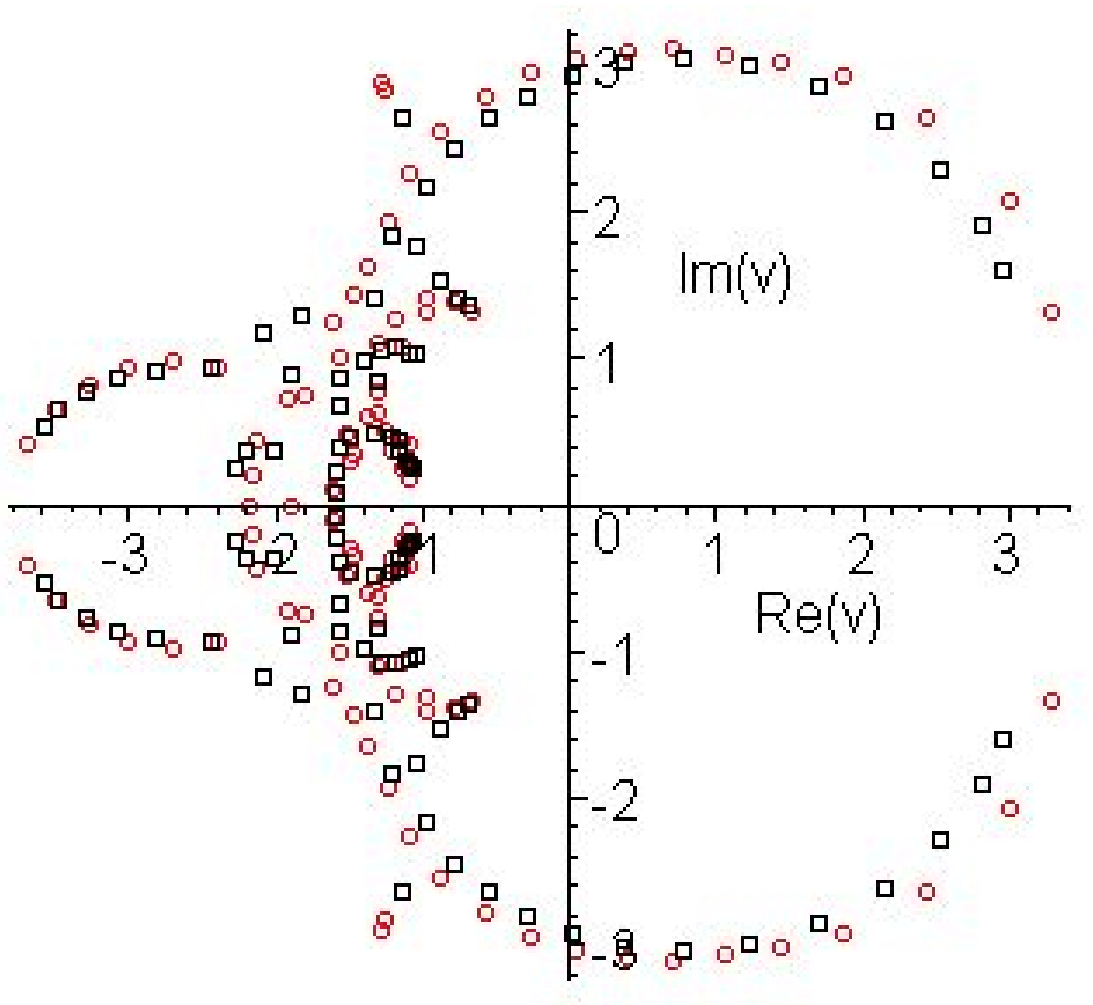} \\
   \phantom{(((a)}(c)    & \phantom{(((a)}(d) \\
\end{tabular}
\caption[a]{\protect\label{figures_vplane_P} Partition-function zeros, in the
  $v$ plane, of (a) $Z(G,q,v)/q$ for $q=0$, and $Z(G,q,v)$ for (b) $q=0.999$,
  (c) $q=2$, and (d) $q=3$ on strips with cylindrical boundary conditions and
  several widths $L_y$: 4 ($\Box$, black), 6 ($\circ$, red), where the colors
  refer to the online paper.  }
\end{figure}

In Figure~\ref{figures_vplane_F} we show the partition-functions zeros in the
$v$-plane, for fixed values of $q$, for strips with $2 \leq L_y \leq 5$ and
free boundary conditions.  We have displayed each value of $q$ on a different
plot: (a) $q=0$, (b) $q \simeq 1$, (c), $q=2$, and (d) $q=3$. The corresponding
partition-function zeros for honeycomb-lattice strips with $L_y$=4,6 and
cylindrical boundary conditions are shown in Figure~\ref{figures_vplane_P}.
Complex-temperature phase diagrams and associated partition function zeros were
given in \cite{hca} for $L_y=2$ for free longitudinal boundary conditions.  Our
present calculations extend that work to greater strip widths.

For our discussion we will need some results concerning the behavior of the
Potts model on the infinite honeycomb lattice (defined via the 2D thermodynamic
limit).  The criticality condition for the $q$-state Potts model on this
lattice is \cite{hccrit1}-\cite{wurev}
\beq
v^3-3qv-q^2=0 \ .
\label{hc_eq}
\eeq
Since eq. (\ref{hc_eq}) is cubic in $v$, it is cumbersome to write the general
solution for $v$ as a function of $q$.  The following information will suffice:
the equation has (i) one real root in $v$ for real $q < 0$ and $q > 4$ (ii)
three degenerate real roots, $v=0$, at $q=0$, (iii) three real roots, two of
which are degenerate, at $q=4$: $v=-2, \ -2, \ 4$; and (iv) three distinct real
roots for $0 < q < 4$.  A plot of these roots is given, e.g., as Fig. 4 of
\cite{p}. In the interval $0 \le q \le 4$, the maximal root, $v_{hc3}$, which
is the transition point between the paramagnetic (PM) and ferromagnetic (FM)
phases, increases monotonically from $v_{hc3}=0$ at $q=0$ to $v_{hc3}=4$ at
$q=4$, while the middle one, $v_{hc2}$, which is the transition point between
the paramagnetic and antiferromagnetic (AFM) phases, decrease monotonically
from $v_{hc2}=0$ at $q=0$, through $v_{hc2}=-1$ at $q=(3+\sqrt{5})/2$, and then
to $v_{hc2}=-2$ at $q=4$.  Only the interval $0 > v_{hc2} \ge -1$ corresponds
to a physical PM-AFM transition; for $q=(3+\sqrt{3})/2$, this transition occurs
at zero temperature, i.e., $v=-1$, and for larger values of $q$, the Potts
antiferromagnet has no physical PM-AFM transition and is disordered and
noncritical even at $T=0$ (e.g., \cite{p3afhc}).  A rigorous result is
that the $q$-state Potts antiferromagnet on the honeycomb lattice is disordered
and noncritical even at $T=0$ if $q \ge 4$ \cite{sstheorem}.  As noted above,
the critical point for the Potts antiferromagnet for $q \not\in {\mathbb Z}_+$
has a formal, rather than directly physical, significance.  The lowest root,
$v_{hc1}$, is unphysical; it is equal to 0 at $q=0$, decreases through $v=-2$
at $q=2$ to a minimum of $-9/4$ at $q=27/8 = 3.375$ and then increases slightly
to reach the value $-2$ again at $q=4$.  In the interval $0 \le q \le 4$ it is
convenient to write 
\beq
q = q(\theta) = 4\cos^2 \Big ( \frac{\theta}{2} \Big ) 
\label{qt}
\eeq
with $0 \le \theta \le \pi$.  The main cases of interest here are
contained within the discrete set of values $q=q_r \equiv q(\theta_r)$, where
$e^{i\theta_r}$ is a certain root of unity given by 
\beq
\theta_r = \frac{2 \pi}{r} \ , \quad r \in {\mathbb Z}_+ \ .
\label{qr}
\eeq
These special values $q_r$ in were discussed by Tutte and Beraha in connection
with zeros of chromatic polynomials \cite{t,bkw} and are also of interest since
they correspond to roots of unity for the deformation parameter in the
Temperley-Lieb algebra relevant for the Potts model \cite{saleur,mbook}.  Some
values are $q_r=4, \ 0, \ 1, \ 2, (3+\sqrt{5})/2, \ 3$ for $1 \le r \le 6$,
respectively.  For $q=q_r$, the solutions of eq. (\ref{hc_eq}) for $v$ have
simple expressions in terms of trigonometric functions, which will be of use
for our discussion below \cite{qv}:
\beq
v_{hc1}(r) = -4\cos \Bigl ( \frac{\pi}{r} \Bigr )
          \cos \biggl [ \frac{\pi}{3}\Bigl ( \frac{1}{r}-1 \Bigr )\biggr ]
\label{vhc1}
\eeq
\beq
v_{hc2}(r) = -4\cos \Bigl ( \frac{\pi}{r} \Bigr )
          \cos \biggl [ \frac{\pi}{3}\Bigl ( \frac{1}{r}+1 \Bigr )\biggr ]
\label{vhc2}
\eeq
and
\beq
v_{hc3}(r) = 4\cos \Bigl ( \frac{\pi}{r}  \Bigr )
              \cos \Bigl ( \frac{\pi}{3r} \Bigr ) \ .
\label{vhc3}
\eeq
This set is dual, via the map $v \to q/v$, to the set of solutions of the
corresponding criticality condition for the triangular lattice, $v^3+3v^2-q=0$,
viz., $v_{t,\eta}=-1+2\cos [2(1+\eta r)\pi/(3r)]$ for $\eta=1,0,-1$, given in
eqs. (27)-(29) of \cite{qv}.  In previous studies such as \cite{hca}, it has
been found that although infinite-length, finite-width strips are
quasi-one-dimensional systems, and hence the Potts model has no physical
finite-temperature transition for such systems, some aspects of the
complex-temperature phase diagram have close connections with those on the
(infinite) honeycomb lattice.  We shall discuss some of these connections
below.  One can also express the solutions of eq. (\ref{hc_eq}) as values of
$q$ for a given $v$; these are
\beq
q = \frac{1}{2}\Big [ -3v \pm \sqrt{v^2(4v+9)} \ \Big ]
\label{qsol}
\eeq
and are real for $v \ge -9/4$.

%
%

%
%

%
%
\begin{figure}[hbtp]
\centering
\begin{tabular}{cc}
   \includegraphics[width=170pt]{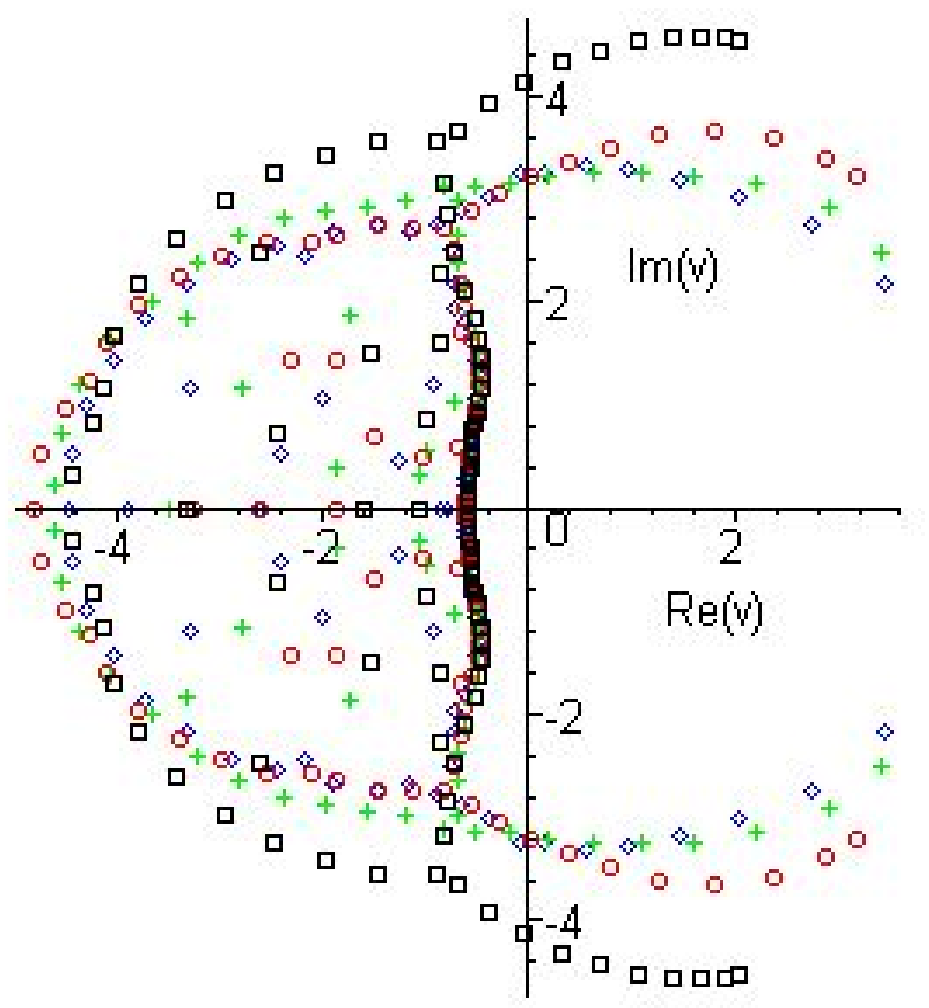} \qquad \qquad & \qquad \qquad 
   \includegraphics[width=170pt]{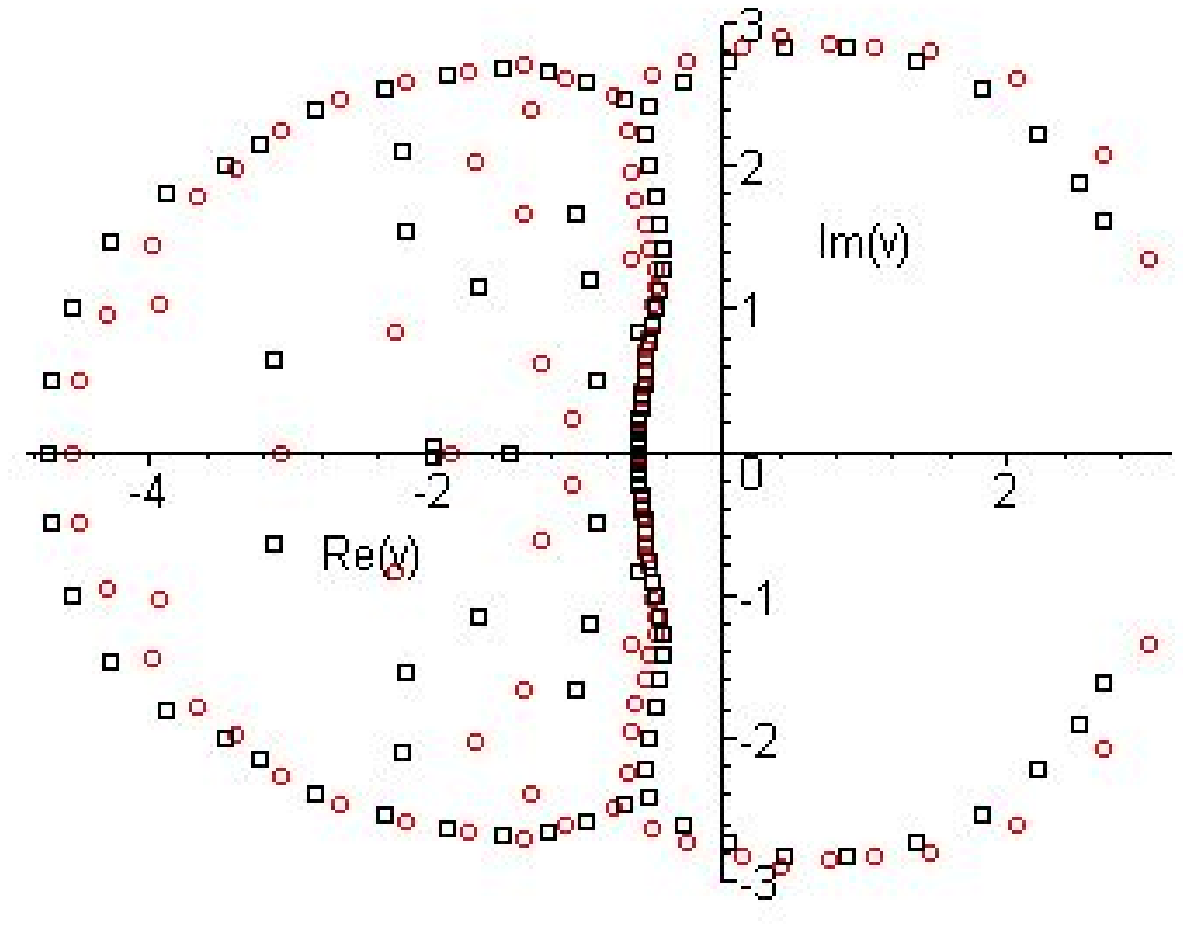} \\
   \phantom{(((a)}(a)    & \phantom{(((a)}(b) \\
\end{tabular}
\caption[a]{\protect\label{figures_vplane_q5} Partition-function zeros, in the
  $v$ plane for $q=q_5=(1/2)(3+\sqrt{5} \ )$ for (a) free boundary conditions 
with $L_y=$ 2 ($\Box$, black), 3 ($\circ$, red), 4 ($+$, green), and 5
  ($\Diamond$, blue); (b) cylindrical boundary conditions and
  $L_y=$: 4 ($\Box$, black), 6 ($\circ$, red), where the colors refer to the
  online paper.}
\end{figure}

\subsection{$q=0$} 

From the cluster representation of $Z(G,q,v)$, eq.~(\ref{cluster}), it follows
that this partition function has an overall factor of $q^{k(G)}$, where $k(G)$
denotes the number of components of $G$, i.e., an overall factor of $q$ for a
connected graph.  Hence, $Z(G,q=0,v)=0$.  In the transfer matrix formalism,
this is evident from the overall factor of $q$ coming from the vector $\w$.
However, if we first take the limit $n \to \infty$ to define ${\cal B}$ for $q
\ne 0$ and then let $q \to 0$ or, equivalently, extract the factor $q$ from the
left vector $\w$, we obtain a nontrivial locus, namely $({\cal
B}_v(\{G\},0)_{qn}$.  This is a consequence of the noncommutativity
(\ref{bnoncom}) for $q=0$.

With the second order of limits or the equivalent removal of the factor of $q$
in $Z$, we obtain the zeros for $q=0$ shown in
Figures~\ref{figures_vplane_F}(a) (free boundary conditions)
and~\ref{figures_vplane_P}(a) (cylindrical boundary conditions).  The zeros
appear to converge to a roughly circular curve. We see in these figures that
the limiting curves cross the real $v$-axis at $v\approx -6$. As $L_y$
increases, the arc endpoints on the upper and lower right move toward the real
axis.  It is possible that these could pinch this axis at $v=0$ as $L_y \to
\infty$, corresponding to the root of (\ref{hc_eq}) for $q=0$.

\subsection{$q=1$} 
 
For $q=1$, the spin-spin interaction in ${\cal H}$ always has the Kronecker
delta function equal to unity, and hence the Potts model partition function is
given by
\be 
Z(G,q=1,v) = e^{K|E|} = (1+v)^{|E|} 
\label{Z_q=1}
\ee
where $|E|$ is the number of edges in the graph $G$.  This has a single zero at
$v=-1$.  But again, one encounters the noncommutativity (\ref{bnoncom}) for
$q=1$.  It is interesting to analyze this in terms of the transfer matrix
formalism. At this value of $q$, both the transfer matrix and the left vector
$\w$ are non-trivial.  There is thus a cancellation of terms that yields the
result \reff{Z_q=1}. The strip with $L_y=2$ and free boundary conditions is the
simplest one to analyze: the eigenvalues and coefficients for $q=1$ are given
by
\begin{subeqnarray}
\lambda_1(1,v) &=& v^4 \ ;   \qquad \quad c_1(1,0) = 0 \\ 
\lambda_2(1,v) &=& (1+v)^5 \ ; \quad       c_2(1,0) = (1+v) 
\end{subeqnarray}
Thus, only the second eigenvalue contributes to the partition function, and it
gives the expected result $Z(2 \times m,FF,q=1,v) = (1+v)^{5m+1}$.  The strip
with $L_y=3$ and free boundary conditions is similar: there is a single
eigenvalue $\lambda_1(1,v)=(1+v)^8$ with a non-zero coefficient $c_1(1,v)$
equal to $(1+v)$ for odd $L_x$ and $(1+v)^5$ for even $L_x$. The other four
eigenvalues including $v^4(1+v)^2$ and the roots of
\beq
x^3 -2v^4(v^3+6v^2+4v+1)x^2 + v^8(1+v)^2(v^4+10v^3+15v^2+6v+1)x-v^{14}(1+v)^6
=0
\label{q1eq}
\eeq
have identically zero coefficients. Thus, the partition function takes the form
$Z(3\times m,FF,q=1,v) = (1+v)^{8m+k}=(1+v)^{|E|}$, where $k$=1 for odd $L_x$
and 5 for even $L_x$.

In general, we conclude that at $q=1$, only the eigenvalue
$\lambda=(1+v)^{3L_y}$ contributes to the partition function for cylindrical
boundary conditions, and only the eigenvalue $\lambda=(1+v)^{3L_y-1}$
contributes to the partition function for free b.c.  The other eigenvalues do
not contribute because they have zero coefficients in eq. (\ref{zgsum}).  
This is analogous to what we found in earlier work for cyclic strips
\cite{a,s3a,ta}. 

In our present case, in order to get insight into ${\cal B}_{qn}$, we have
computed the partition function zeros for a value of $q$ close to 1, namely,
$q=0.999$ (see Figures~\ref{figures_vplane_F}(b)
and~\ref{figures_vplane_P}(b)).  The patterns of zeros show less scatter for
cylindrical, as contrasted with free, boundary conditions.  Subsituting $r=3$,
i.e., $q=q_3=1$ in the solutions (\ref{vhc1})-(\ref{vhc3}) of the criticality
equation (\ref{hc_eq}), we have
\beq
v_{hc1} = -4\cos (\pi/3) \cos(2\pi/9) = -1.5320888... 
\label{vhc1_q1}
\eeq
\beq
v_{hc2} = -4\cos (\pi/3) \cos(4\pi/9) = -0.34729635... 
\label{vhc2_q1}
\eeq
\beq
v_{hc3} = 4\cos (\pi/3) \cos(\pi/9) = 1.8793852... 
\label{vhc3_q1}
\eeq
Our results are consistent with the inference that as $L_y \to \infty$, the
locus ${\cal B}_v $ crosses the real $v$ axis at the values of $v_{hc1}$ and
$v_{hc2}$ in eqs.  There is evidently no indication of any crossing near the
value $v_{hc3}$.  A noteworthy feature of the patterns of zeros, at least for
the case of cylindrical boundary conditions, is that they do not appear to
exhibit the prongs that tend to occur for some other cases discussed here.

\subsection{$q=2$}

The zeros for the $q=2$ Ising case are displayed in
Figures~\ref{figures_vplane_F}(c) and~\ref{figures_vplane_P}(c) for free and
cylindrical boundary conditions, respectively. (Fig. \ref{figures_vplane_F}(c)
extends our previous calculation presented in Fig. 3 of \cite{hca} to greater
widths.)  From our earlier work \cite{a,ta} one
knows that the loci ${\cal B}_v$ are different for strips with free or periodic
transverse boundary conditions and free longitudinal boundary conditions, on
the one hand, and free or periodic transverse boundary conditions and periodic
(or twisted periodic) longitudinal boundary conditions.  One anticipates,
however, that in the limit of infinite width, the subset of the
complex-temperature phase diagram that is relevant to real physical
thermodynamics will be independent of the boundary conditions used to obtain
the 2D thermodynamic limit.

In the 2D thermodynamic limit, one knows the complex-temperature phase diagram
exactly for the $q=2$ (Ising) case.  (This isomorphism involves the
redefinition of the spin-spin exchange constant $J_{\rm Potts} = 2J_{\rm
Ising}$ and hence $K_{\rm Potts} = 2K_{\rm Ising}$, where $K_{\rm Potts}$ is
denoted simply $K$ here.)  Since the honeycomb lattice is bipartite, the phase
boundary separating the paramagnetic and ferromagnetic phases maps into that
separating the paramagnetic and antiferromagnetic phases under the
transformation $K \to -K$, i.e., $a \to 1/a$, where $a=e^K=v+1$.  The total
boundary locus ${\cal B}$ is invariant under this inversion $a \to 1/a$.
Because of this symmetry, it is convenient to discuss the phase diagram first
in terms of the variable $a$; the features in the $v$ plane then follow in an
obvious manner.  Following the calculation of the zero-field free energy $f$ of
the Ising model on the square lattice \cite{on}, $f$ was calculated on the
triangular and honeycomb lattices in \cite{hc}.  The critical points separating
the PM and FM phases are given by $a_c=2+\sqrt{3}$ and $1/a_c=2-\sqrt{3}$,
respectively. In terms of $v$, these correspond to the values $v_{hc3}$ and
$v_{hc2}$ with $r=4$ in eqs. (\ref{vhc3}) and (\ref{vhc2}).  The
complex-temperature phase diagram, with boundaries comprised by the locus
${\cal B}$, was given in the plane of the variable ${\rm tanh}(K) =
(a-1)/(a+1)$ in (Fig. 3 of) \cite{abe} and in the variable $a$ in (Fig. 1(c)
of) \cite{chitri}.  The locus ${\cal B}$ separates the complex $a$ plane into
three phases: (i) the physical PM phase occupying the interval $2-\sqrt{3} \le
a \le 2+\sqrt{3}$, and its complex-temperature extension (CTE), where the $S_q$
symmetry is realized explicitly ($S_q$ being the symmetric group on $q$
numbers, the symmetry group of the Hamiltonian), (ii) the physical
ferromagnetic phase occupying the interval $2+\sqrt{3} \le a \le \infty$ (and
its CTE), and (iii) the physical antiferromagnetic phase occupying the interval
$0 \le a \le 2-\sqrt{3}$ and its CTE.  The boundary separating the CTE of the
FM and AFM phases is an arc of the unit circle $a=e^{i\theta}$ with $\pi/3 \le
\theta \le 5\pi/3$; it thus has endpoints at $a=e^{\pm i \pi/3}$ and crosses
the real $a$ axis at $a=-1$. The rest of ${\cal B}$ is a closed curve crossing
the real axis at $a=2 \pm \sqrt{3}$ and having intersection points with the
above-mentioned circular arc at the points at $a=\pm i$.  In \cite{hca} the
locus ${\cal B}$ for an infinite-length free and cyclic strip with width
$L_y=2$ were compared with this 2D phase diagram.

Using our exact results, we can compare the loci ${\cal B}_v$ for various strip
widths and either free or periodic transverse boundary conditions with the
known complex-temperature phase diagram for the Ising model on the infinite 2D
honeycomb lattice.  This comparison is simplest for the case of free boundary
conditions, so we concentrate on these results.  For the finite values of $L_y$
that we have considered, the loci ${\cal B}_v$ inferred from these zeros
clearly contain a circular arc crossing the real $v$ axis at $v=-2$ and have
intersection points at $v=-1 \pm i$, just as is the case with the exactly known
locus ${\cal B}$ for the infinite 2D honeycomb lattice.  Moreover, one sees
that as $L_y$ increases, the endpoints of the complex-conjugate arcs move down
toward the real axis.  As $L_y \to \infty$, we expect that these arc endpoints
will cross the real $v$ axis at the points $v=1 \pm \sqrt{3}$ that constitute
the intersections, with the real $v$ axis, of the complex-temperature phase
boundaries ${\cal B}$ for the Ising model on the infinite honeycomb lattice.
As in our earlier studies, this comparison shows that, although the behavior of
the asymptotic locus ${\cal B}$ for infinite-length lattice strips is 
qualitatively different from the locus for the thermodynamic limit of the
two-dimensional lattice as regards the physical phase transitions (owing to its
quasi-one-dimensional nature), its features for complex temperatures show many
similarities with the exactly known features for the 2D thermodynamic limit. 

\subsection{$q=(3+\sqrt{5})/2$} 

One of the useful features of exact solutions for $Z(G,q,v)$ for arbitrary $q$
and $v$ is that they allow one to analyze values of $q$ that are not positive
integers and hence cannot be represented in Hamiltonian form, but instead 
via the relation (\ref{cluster}).  Within the sequence of the $q_r$'s, the
first such example is provided by the case $r=5$, i.e., $q=q_5=(3+\sqrt{5})/2
\simeq 2.6180$. For this value, the criticality condition for the model on the
(thermodynamic limit of the) honeycomb lattice has the solutions
\beq
v_{hc3} = \frac{1}{2}\Big [ 1 + \sqrt{3(5+2\sqrt{5}) \ } \ \Big ] \simeq 3.165 
\label{vhc3q5}
\eeq
\beq
v_{hc2} = -1
\label{vhc2q5}
\eeq
\beq
v_{hc1} = \frac{1}{2}\Big [ 1 - \sqrt{3(5+2\sqrt{5}) \ } \ \Big ] \simeq -2.165
\label{vhc1q5}
\eeq
The first of these is the PM-FM critical point, while the second formally
corresponds to the critical temperature for the Potts antiferromagnet going to
zero.  The third is a complex-temperature singular point.  In Figs. 
\ref{figures_vplane_q5} we plot zeros of the partition function in the $v$
plane for $q=q_5$ and strips with free and cylindrical boundary conditions.
From these zeros, one can infer that as $L_y \to \infty$, the rightmost
complex-conjugate arc endpoints would move in and pinch the real axis at the
PM-FM value $v_{hc3} \simeq 3.165$ given in eq. (\ref{vhc3q5}).  The results
are also consistent with crossings on ${\cal B}$ at the other two points 
$v_{hc2}$ and $v_{hc1}$ as well as the values $v \simeq -0.6$ and $v \simeq
-4.4$ which are not roots of the criticality condition (\ref{hc_eq}).

\subsection{$q=3$} 

In contrast to the $q=2$ case, the free energy of the $q$-state Potts model has
not been calculated exactly for $q \ge 3$ on any 2D (or higher-dimensional)
lattice, and hence the corresponding complex-temperature phase diagrams are not
known exactly. For $q=3$, i.e., $r=6$, the solutions of the criticality
equation (\ref{hc_eq}) given by eqs. (\ref{vhc1})-(\ref{vhc3}) are 
\beq
v_{hc3}=2\sqrt{3}\cos(\pi/18) =3.4114741...
\label{vcq3}
\eeq
corresponding to the physical PM-FM phase transition point, and two other roots
at the complex-temperature values
\beq
v_{hc2}=-2\sqrt{3}\cos(7\pi/18)  = -1.1847925...
\label{vq3point2}
\eeq
and
\beq 
v_{hc1}=-2\sqrt{3}\cos(5\pi/18) = -2.2266815...
\label{vq3point3}
\eeq
Some discussions of the complex-temperature solutions of eq.~(\ref{hc_eq}) and
their connections with the complex-temperature phase diagram have been given in
\cite{mm}-\cite{p2}.

The partition-function zeros in the $v$ plane for $q=3$ are displayed in
Figures~\ref{figures_vplane_F}(d) and~\ref{figures_vplane_P}(d) for free and
cylindrical boundary conditions, respectively. We expect that the pair of
complex-conjugate endpoints in this regime will eventually converge to the
ferromagnetic critical point $v_{hc3}$ as $L_y \to \infty$.  However,
obviously, an infinite-length strip of finite width $L_y$ is a
quasi-one-dimensional system, so the Potts model has no physical
finite-temperature phase transition on such a strip for any finite $L_y$. The
$q=3$ Potts antiferromagnet is disordered on the honeycomb lattice for all
temperatures $T$ including $T=0$, so there is no finite-temperature PM-AFM
transition.

In the complex-temperature interval $v < -1$, there are considerable
finite-size and boundary condition effects.  Because of this, in previous work,
a combination of partition-function zeros and analyses of low-temperature
series expansions was used \cite{p2}; these enable one at least to locate some
points on the complex-temperature phase boundary.  As regards the infinite 2D
honeycomb lattice, because of a duality relation, the complete physical
temperature interval $0 \le T \le \infty$, i.e., $0 \le a \le 1$ of the
$q$-state Potts antiferromagnet on the triangular lattice is mapped to the
complex-temperature interval $-\infty \le v \le -q)$ on the honeycomb lattice
(and vice versa) \cite{hcl}.  Ref. \cite{hcl} found that there is a
complex-temperature singularity for the $q=3$ Potts model at
$v_{tri,PM-AFM,q=3} = -0.79691 \pm 0.00003$.  From duality, the corresponding
singularity on the honeycomb lattice is $v_{hc,q=3} = 3/v_{tri,PM-AFM,q=3} =
-3.76454 \pm 0.00015$. One anticipates that as $L_y \to \infty$ for the
infinite-length, width-$L_y$ strips of the honeycomb lattice, the left-most
arcs on ${\cal B}_v$ will cross the real $v$ axis at this point.  There are
several other cases of interest, such as $q=4$ and $q_r$ values with $r \ge 7$.
For brevity, we do not consider these here.

\section{Internal Energy and Specific Heat} 

It is of interest to display some of the physical thermodynamic functions for
the Potts model on the infinite-length limits of these strips.  
Having calculated the partition function, one obtains the free energy per
site, $f(G,q,v)$ as 
\be
f(G,q,v) = {1\over n} \log Z(G,q,v) 
\label{def_free_energy}
\ee
for finite $n$, with the $n \to \infty$ limit having been defined in
eq.~(\ref{ef}) above. The internal energy per site, $E$, is 
\beq
E(G,q,v)=-\frac{\partial f}{\partial \beta} = 
-J(v+1)\frac{\partial f}{\partial v}
\slabel{def_E}
\eeq
and the specific heat per site, $C$, is 
\beq
C = \frac{\partial E}{\partial T} = k_B K^2(v+1)\left [
\frac{\partial f}{\partial v} + (v+1)\frac{\partial^2 f}{\partial v ^2}
\right ] \ .
\label{def_C}
\eeq
As the strip width $L_y \to \infty$, these approach the internal energy and
specific heat for the infinite 2D honeycomb lattice. For convenience we define
a dimensionless internal energy
\beq
E_r = -\frac{E}{J} = (v+1)\frac{\partial f}{\partial v} \ .
\label{er}
\eeq
Note that $\sgn(E_r)$ is (i) opposite to $\sgn(E)$ in the ferromagnetic case
where $J > 0$ for which the physical region is $0 \le v \le \infty$ and (ii)
the same as $\sgn(E)$ in the antiferromagnet case $J<0$ for which the physical
region is $-1 \le v \le 0$.  Of course, the infinite-length limits of the
honeycomb-strips considered here are quasi-one-dimensional systems, so that
$f$, $E$, and $C$ are analytic functions of temperature for all finite
temperatures.

We recall the high-temperature (equivalently, small--$|K|$) expansion for an
infinite lattice of dimensionality $d \ge 2$ with coordination number $\Delta$:
\beq
-\frac{E}{J} = E_r = \frac{\Delta}{2}\left [ \frac{1}{q} + \frac{(q-1)K}{q^2} +
O(K^3) \right ] \ .
\label{ehightemp}
\eeq
Here, $\Delta=3$ for the infinite honeycomb lattice.  Again, we recall that in
papers on the $q=2$ Ising special case, the Hamiltonian is usually defined as
${\cal H}_I = -J_I\sum_{\langle i j \rangle} \sigma_i \sigma_j$ with $\sigma_i
= \pm 1$ rather than the Potts model definition of ${\cal H}$, so that one has
the rescaling $2K_I=K$, where $K_I = \beta J_I$.  Furthermore, $E_I = -J\langle
\sigma_i \sigma_j \rangle$ raher than the Potts definition $E = -J\langle
\delta_{\sigma_i\sigma_j} \rangle$, where $\langle i j \rangle$ are adjacent
vertices.  Hence, for example, for $q=2$, with the usual Ising model
definitions, $E_I(v=0)=0$ rather than $E=-J\Delta/(2q)$ and the
high-temperature expansion is $E_I = -J(\Delta/2)[K + O(K^3)]$ rather than the
$q=2$ form of (\ref{ehightemp}).  Similarly, for $T \to 0$, with the
conventional Ising definition, $E_I \to -|J|$ for both the ferromagnetic and
antiferromagnetic cases, while with our Potts-based definition, $E \to
-(\Delta/2)J = -3J/2$, i.e., $E_r \to 3/2$ for the ferromagnet and $E \to 0$
for the antiferromagnet.

\begin{figure}[hbtp]
\centering
\begin{tabular}{cc}
   \includegraphics[width=170pt]{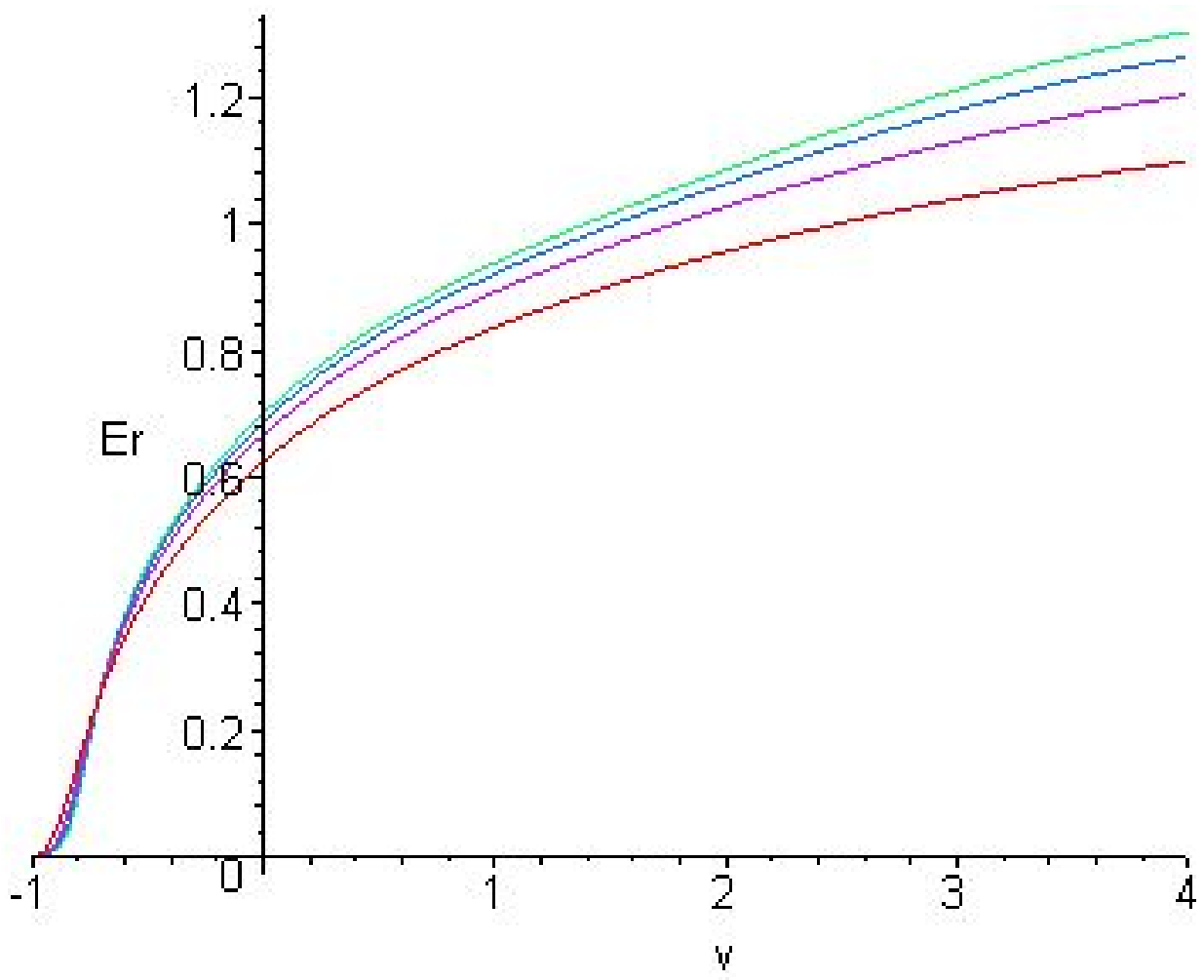} \qquad \qquad & \qquad \qquad 
   \includegraphics[width=170pt]{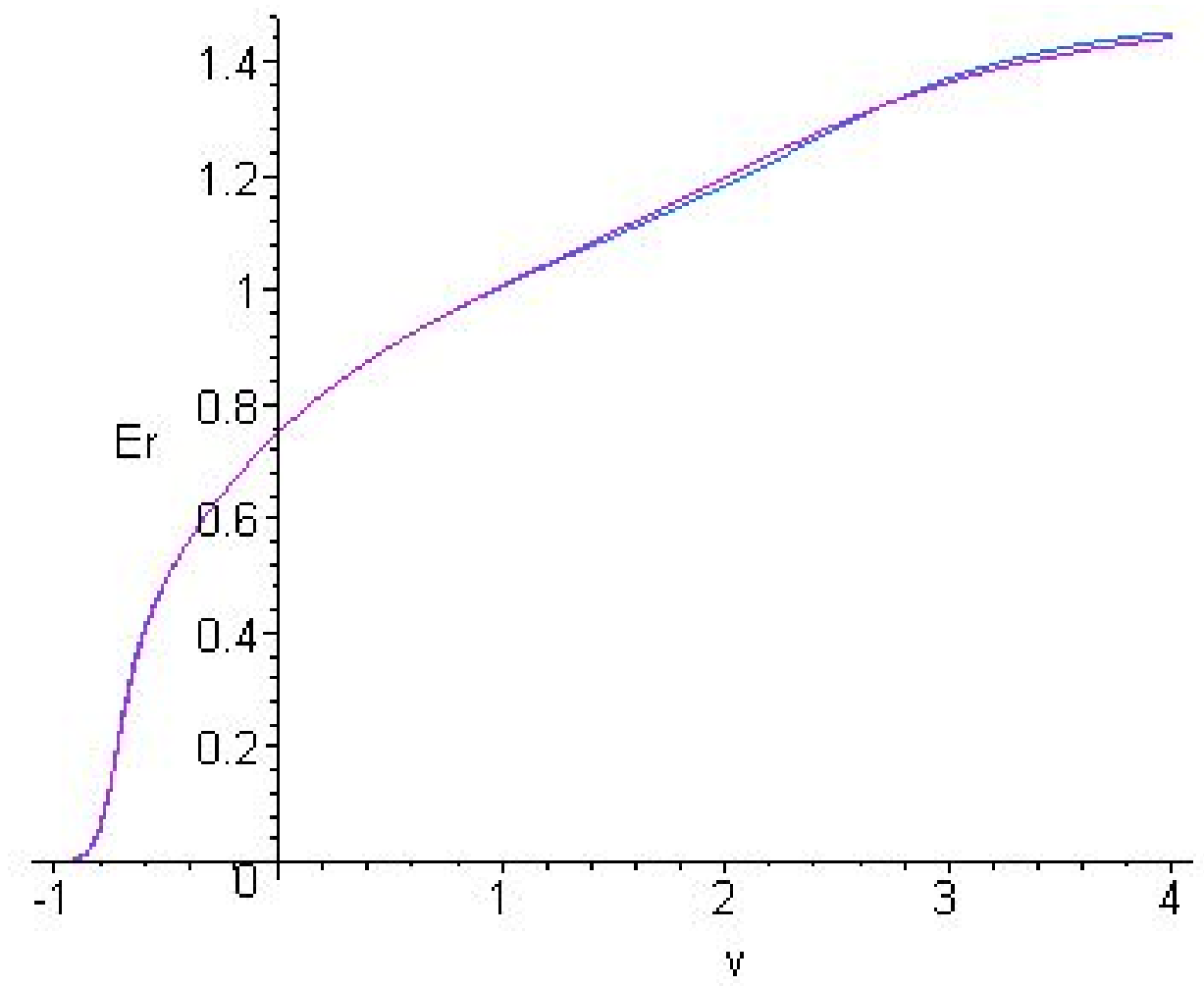} \\
   \phantom{(((a)}(a)    & \phantom{(((a)}(b) \\[5mm]
   \includegraphics[width=170pt]{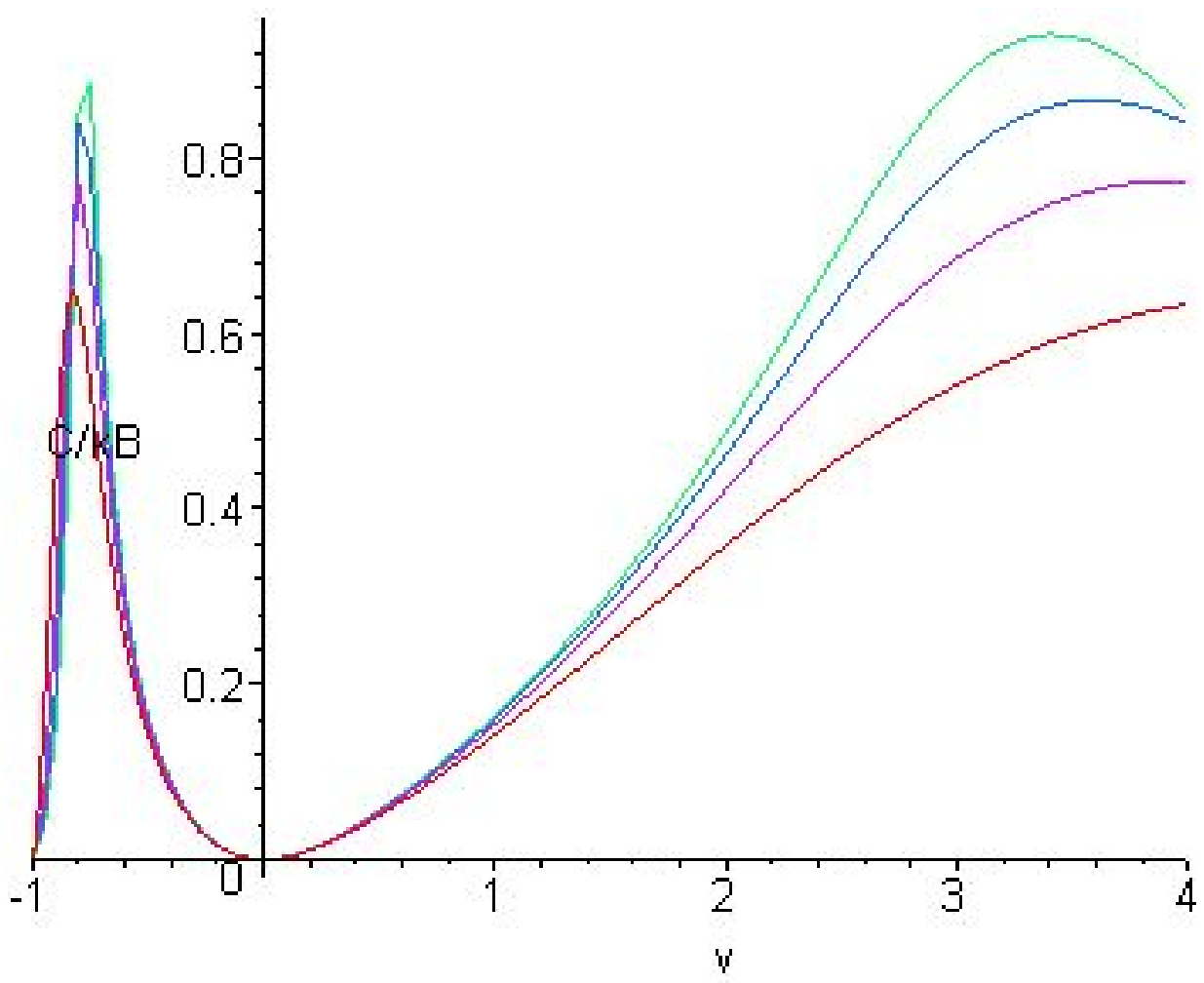} \qquad \qquad & \qquad \qquad 
   \includegraphics[width=170pt]{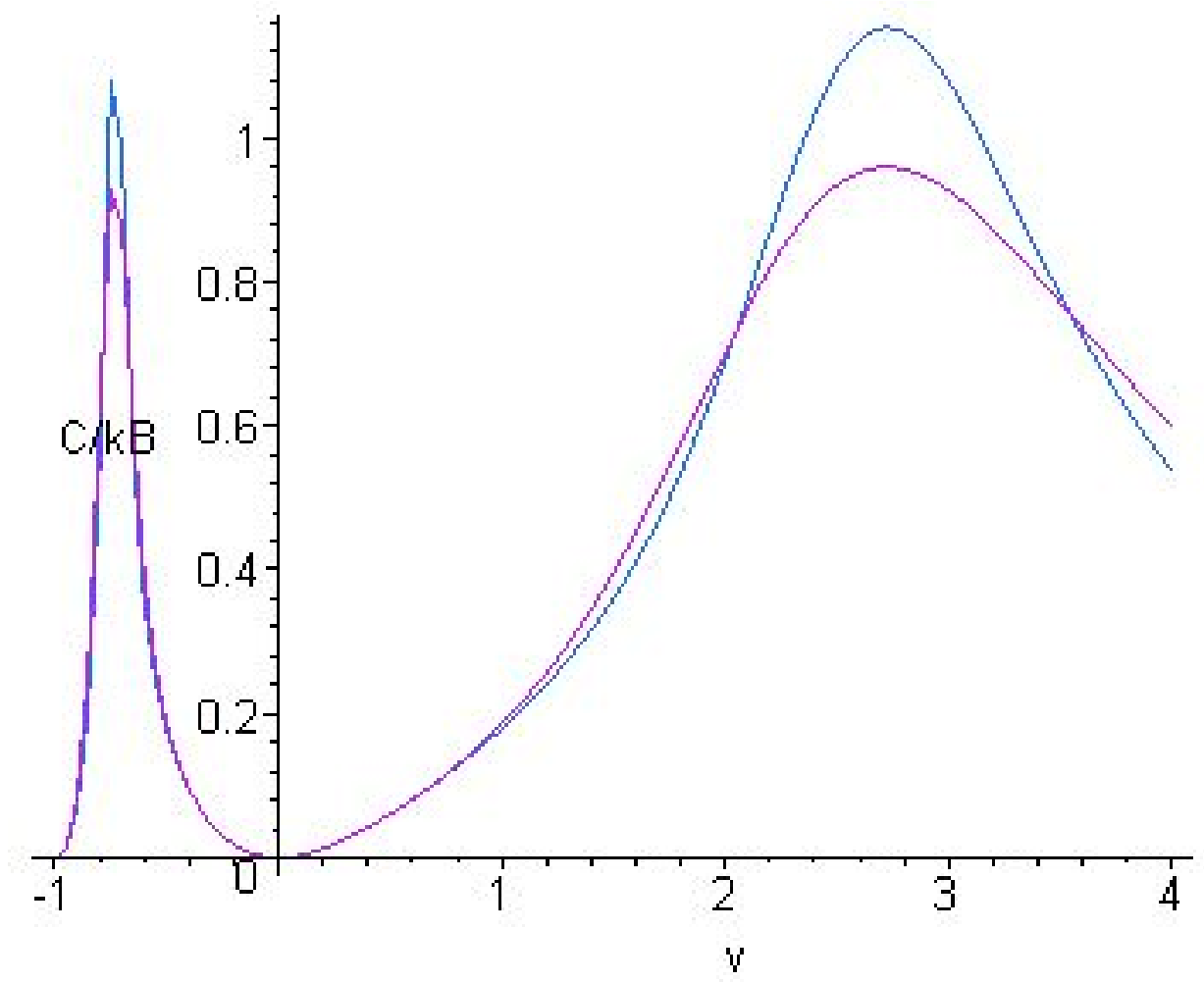} \\
   \phantom{(((a)}(c)    & \phantom{(((a)}(d) \\
\end{tabular}
\caption[a]{\protect\label{EC_q=2} Reduced internal energy $E_r=-E/J$ and
specific heat $C/k_B$ as functions of the temperature-like variable $v$ for the
$q=2$ Potts model on the honeycomb-lattice strips of width $2 \le L_y \le 5$
with free boundary conditions (a) (c) and of width $L_y=4,6$ with cylindrical
boundary conditions (b) (d).  The four curves shown for the case of free
boundary conditions correspond to $L_y=2,3,4,5$ as one moves upward and the two
curves for cylindrical boundary conditions correspond to $L_y=4,6$ as one moves
upward.  The plot includes both the ferromagnetic and antiferromagnetic Potts
models, for which the temperature ranges are $0 \le v \le \infty$ and $-1 \le v
\le 0$, respectively.}
\end{figure}

\begin{figure}[hbtp]
\centering
\begin{tabular}{cc}
   \includegraphics[width=170pt]{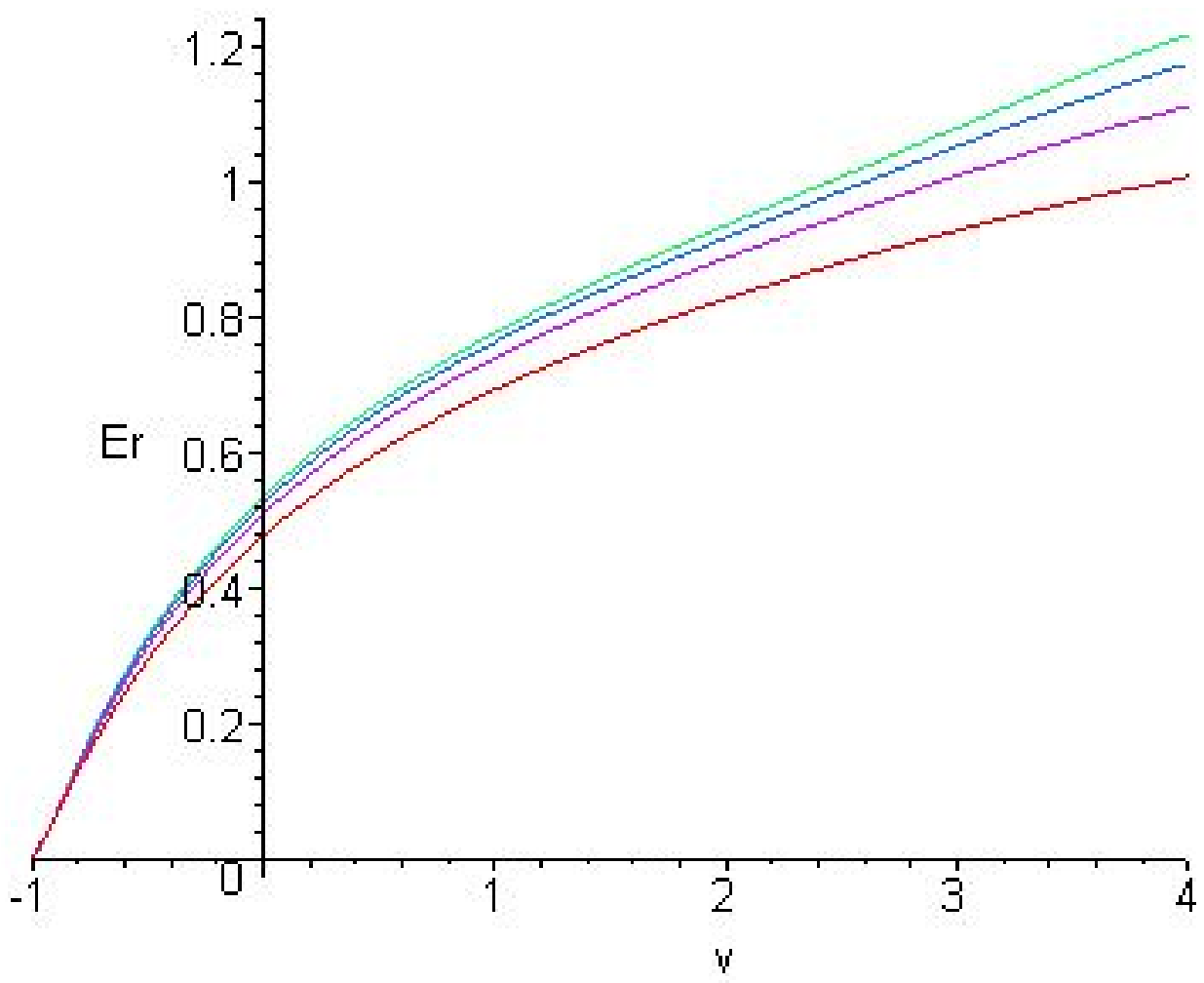} \qquad \qquad & \qquad \qquad 
   \includegraphics[width=170pt]{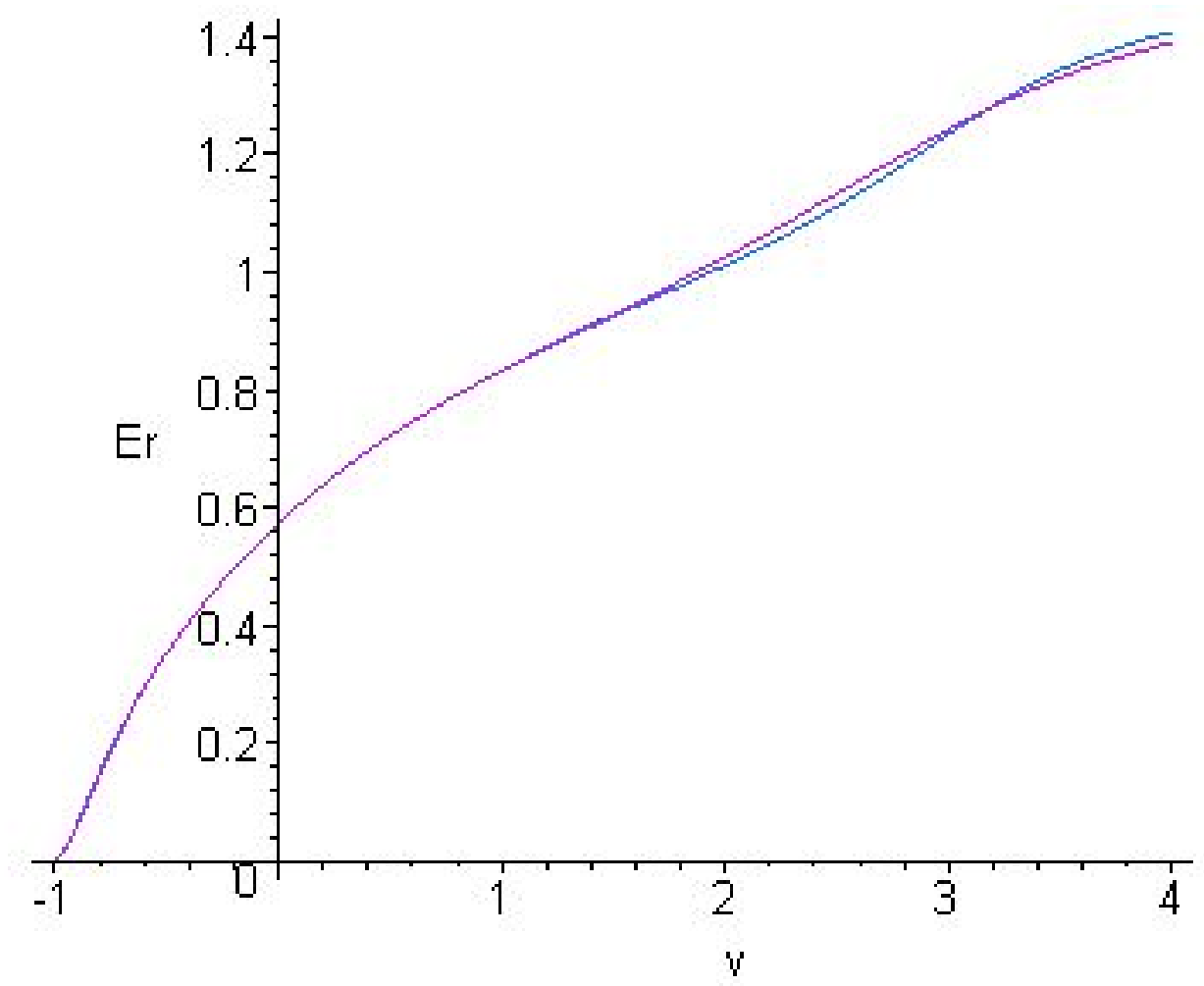} \\
   \phantom{(((a)}(a)    & \phantom{(((a)}(b) \\[5mm]
   \includegraphics[width=170pt]{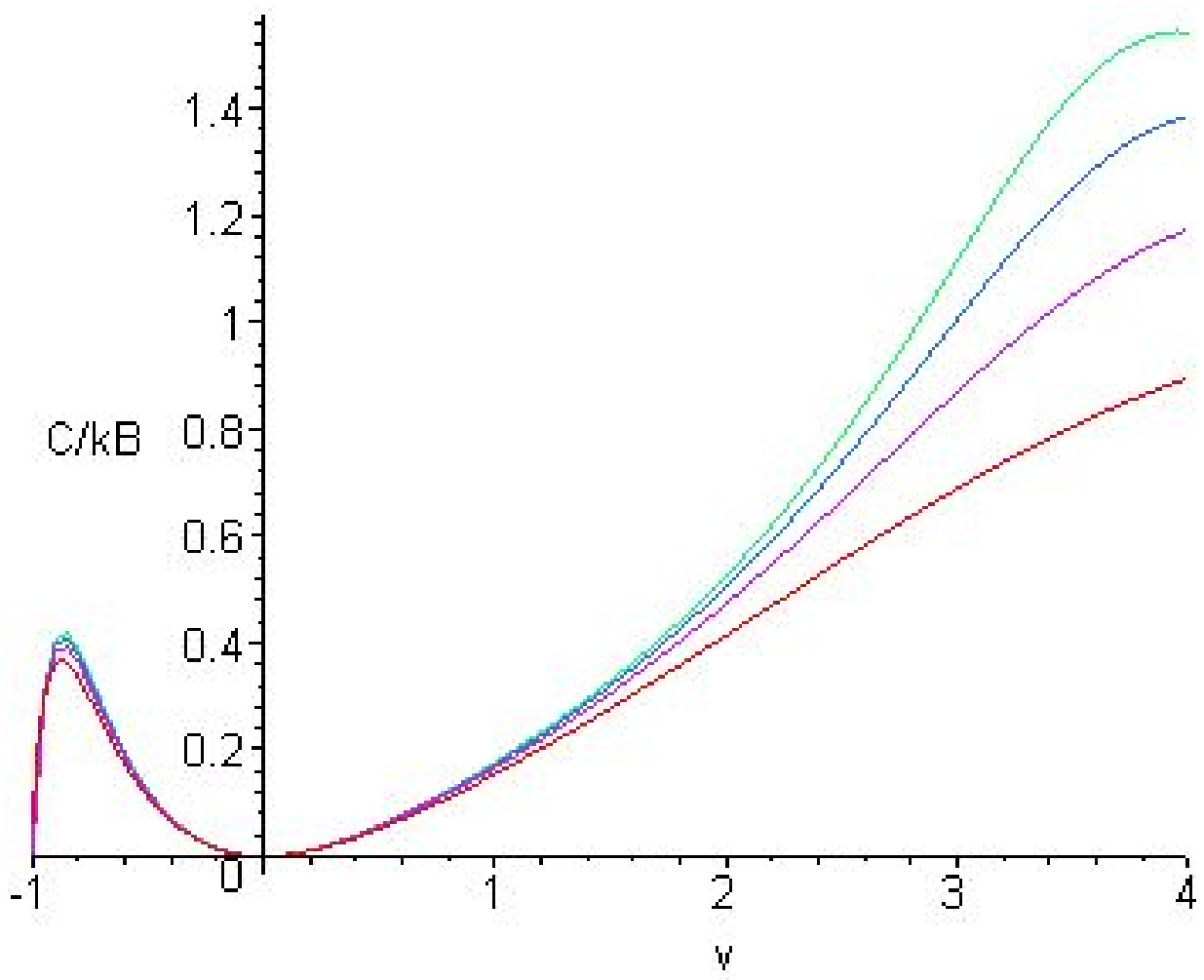} \qquad \qquad & \qquad \qquad 
   \includegraphics[width=170pt]{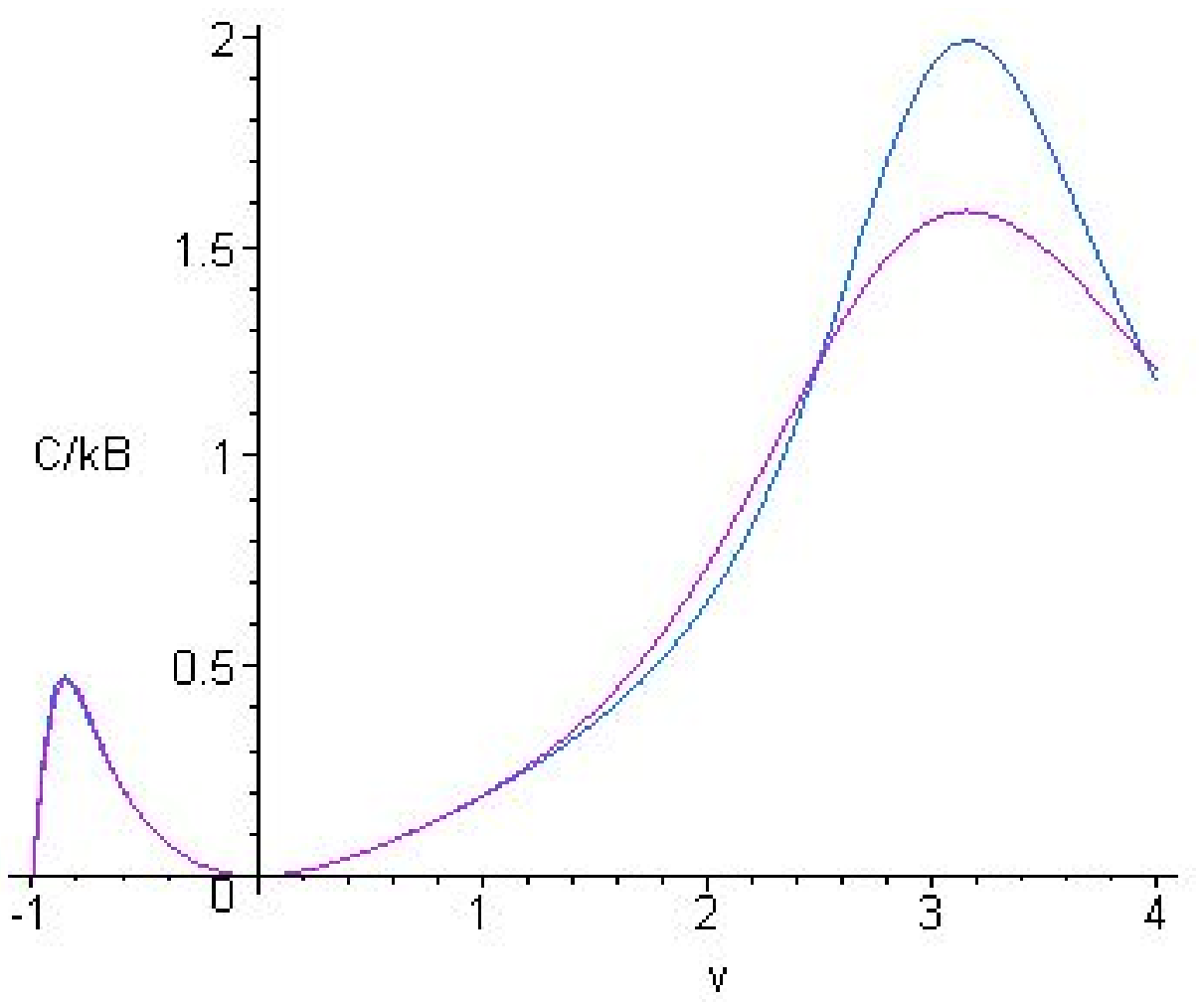} \\
   \phantom{(((a)}(c)    & \phantom{(((a)}(d) \\
\end{tabular}
\caption[a]{\protect\label{EC_q=q5} Reduced internal energy $E_r=-E/J$ and
specific heat $C/k_B$ as functions of the temperature-like variable $v$ for the
Potts model with $q=(3+\sqrt{5})/2$ on the honeycomb-lattice strips of width $2
\le L_y \le 5$ with free boundary conditions (a) (c) and of width $L_y=4,6$
with cylindrical boundary conditions (b) (d).  Ordering of curves is as in
Fig. \ref{EC_q=2}.}
\end{figure}

\begin{figure}[hbtp]
\centering
\begin{tabular}{cc}
   \includegraphics[width=170pt]{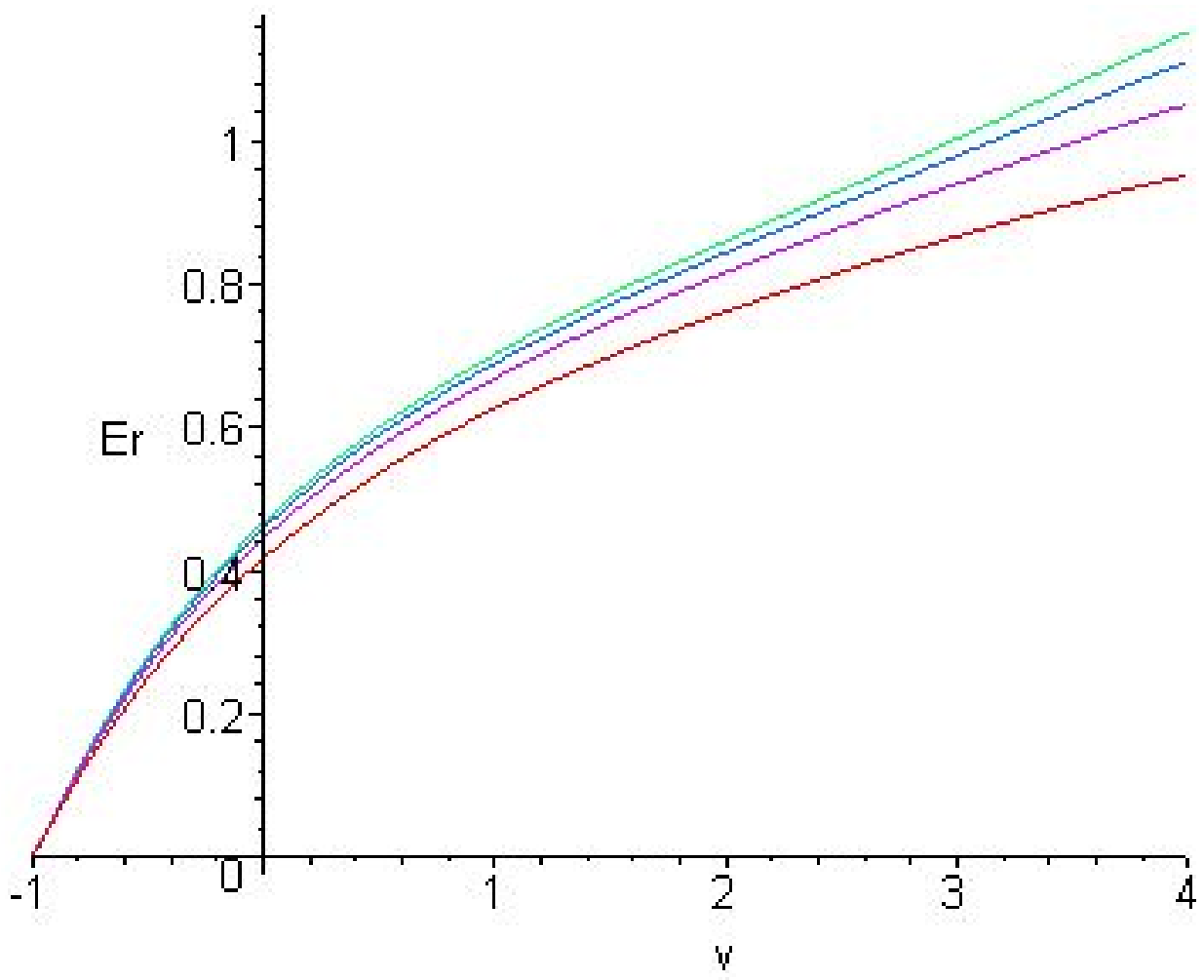} \qquad \qquad & \qquad \qquad 
   \includegraphics[width=170pt]{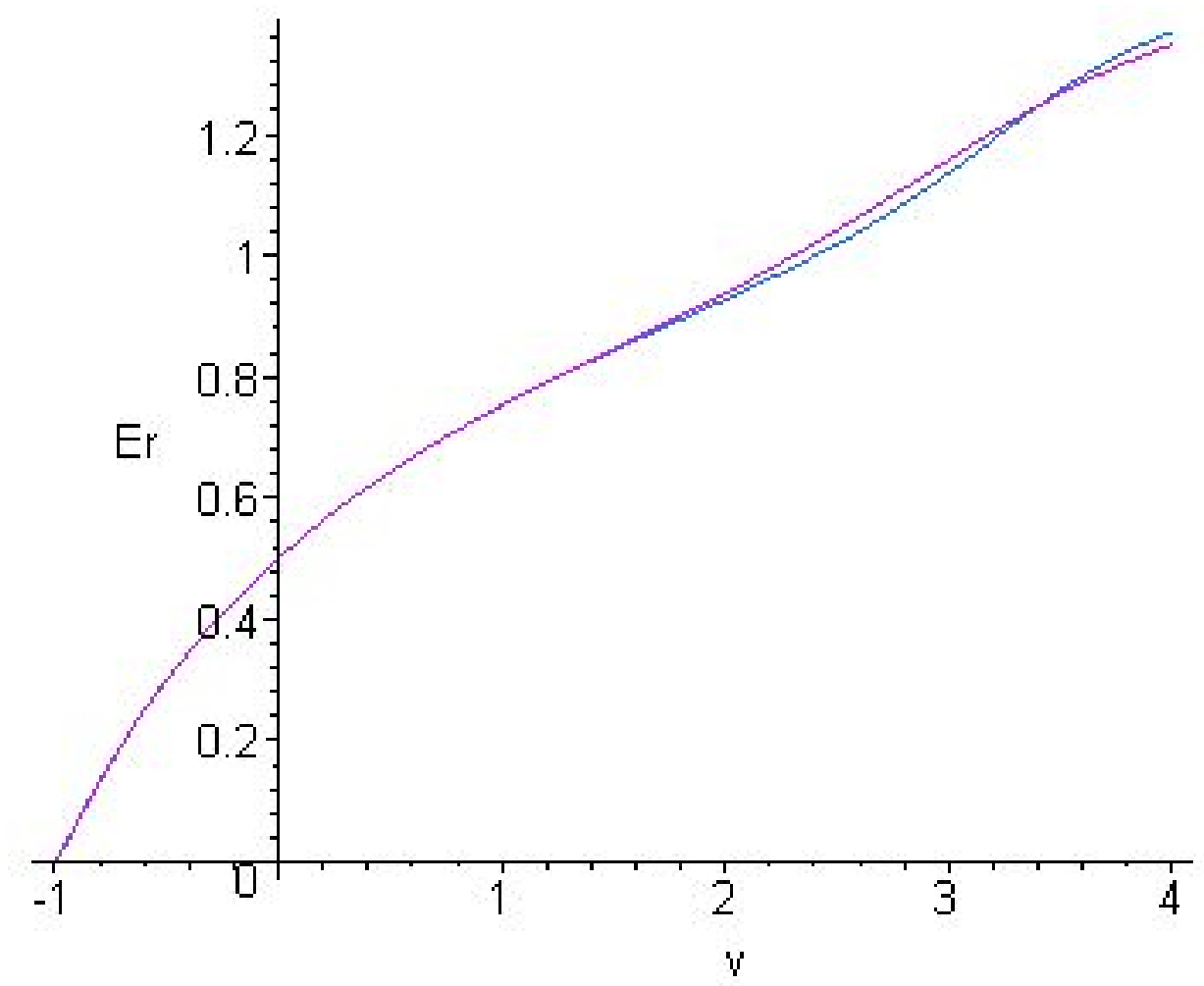} \\
   \phantom{(((a)}(a)    & \phantom{(((a)}(b) \\[5mm]
   \includegraphics[width=170pt]{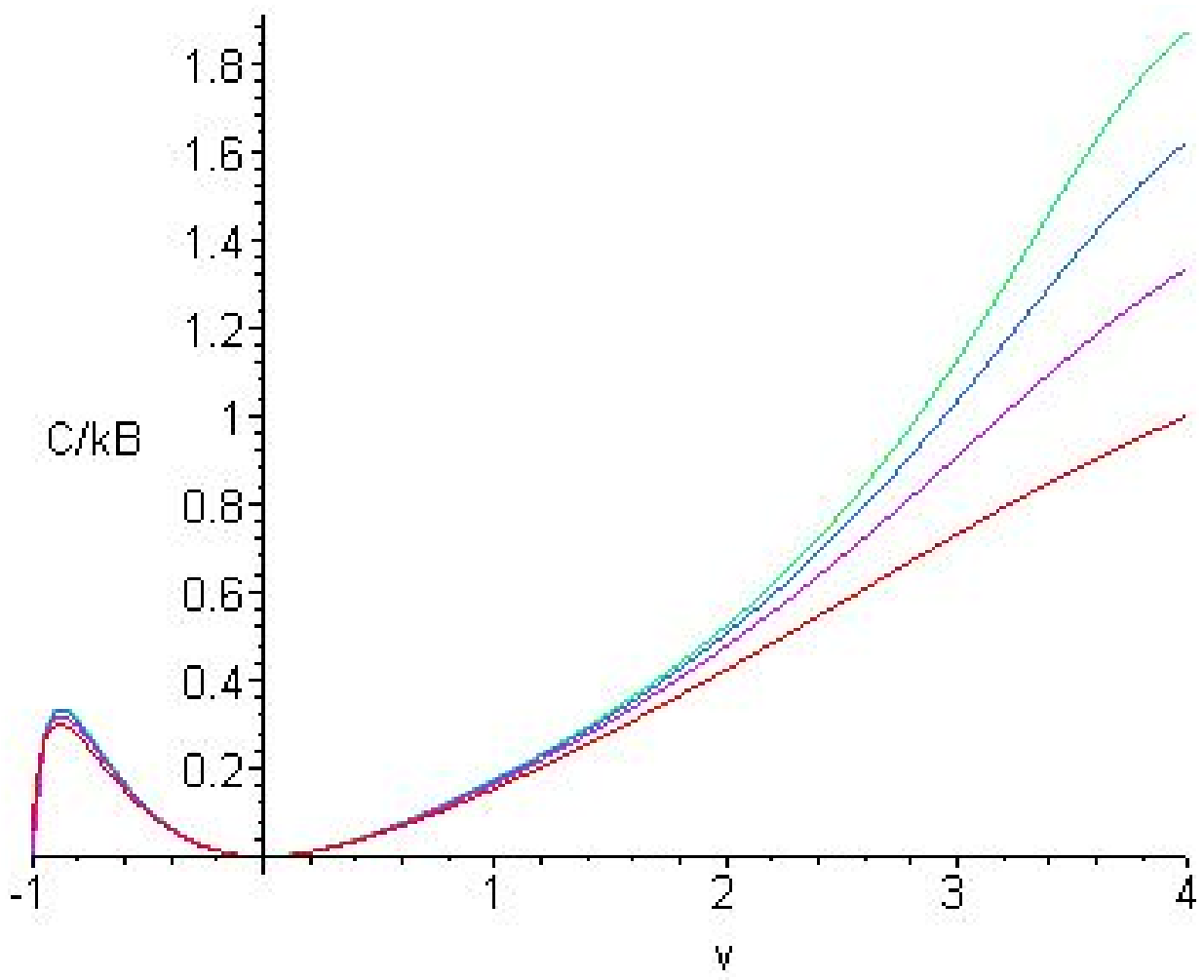} \qquad \qquad & \qquad \qquad 
   \includegraphics[width=170pt]{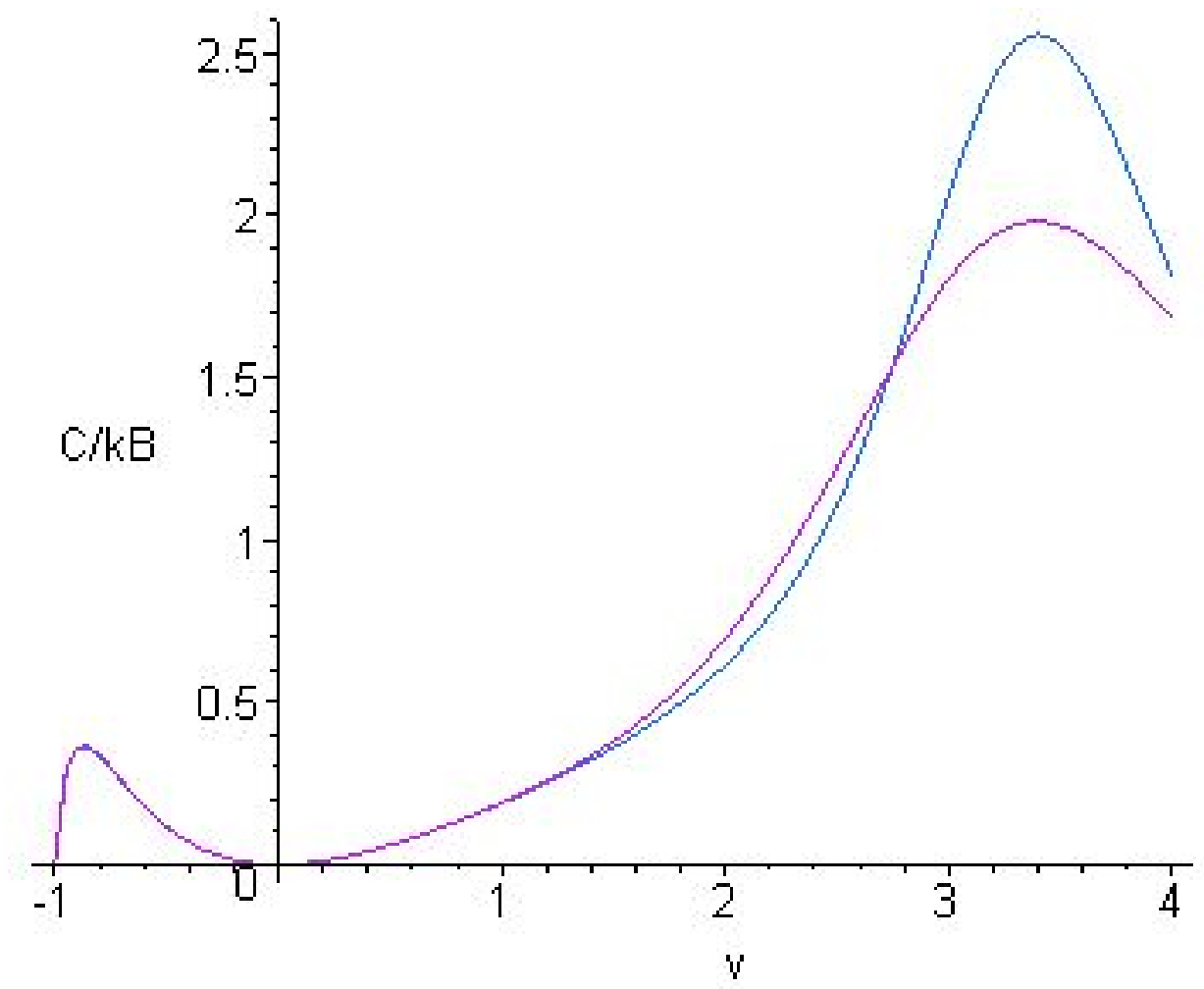} \\
   \phantom{(((a)}(c)    & \phantom{(((a)}(d) \\
\end{tabular}
\caption[a]{\protect\label{EC_q=3} Reduced internal energy $E_r=-E/J$ and
specific heat $C/k_B$ as functions of the temperature-like variable $v$ for the
$q=3$ Potts model on the honeycomb-lattice strips of width $2 \le L_y \le 5$
with free boundary conditions (a) (c) and of width $L_y=4,6$ with cylindrical
boundary conditions (b) (d).  Ordering of curves is as in Fig. \ref{EC_q=2}.}
\end{figure}

We next present plots of the the (reduced) internal energy $E_r$ per site and
the specific heat per site $C$ on infinite-length honeycomb-lattice strips for
three values of $q$ in increasing order, $q=2$, $q=(3+\sqrt{5})/2$, and $q=3$ 
in the respective Figs. \ref{EC_q=2}-\ref{EC_q=3}. Each plot contains
curves for $L_y$ from 2 to 5 for strips with free boundary conditions and for
$L_y=4$ and $L_y=6$ for strips with cylindrical boundary conditions.  The free
energy and its derivatives with respect to the temperature are independent of
the longitudinal boundary conditions in the limit $L_x \to \infty$, although
they depend on the transverse boundary conditions \cite{a}.  As expected, in
the vicinity of the infinite-temperature point $v=0$, the results for the
internal energy are well described by the first few terms of the
high-temperature series expansion given in eq. (\ref{ehightemp}). One can see
the approach of $E_r$ to its zero-temperature limit of 1.5 for the ferromagnet
as $v$ increases through positive values.  This approach is more rapid for the
strips with cylindrical boundary conditions, as is understandable since these
minimize finite-size effects in the transverse direction.  One can also see the
approach of $E_r$ to its zero-temperature limit of zero for the antiferromagnet
as $v$ decreases toward $v=-1$.  

   With regard to the specific heat, the plots show maxima which occur at
values of $v$, denoted $v_m$, that depend on the values of $q$ and $L_y$ and
the transverse boundary conditions.  In the ferromagnetic case, these approach
the critical values for the infinite honeycomb lattice, $v_{PM-FM}$, as $L_y$
increases. For the results shown, this approach is from above (below) in $v$
for the case of free (cylindrical) boundary conditions.  The heights of the
maxima increase as $L_y$ increases, in accordance with the fact that on 2D
lattices, the specific heat diverges at the PM-FM critical point for the values
of $q$ shown. (This divergence is logarithmic for $q=2$ \cite{on}; more
generally, for the interval $0 \le q \le 4$ where the 2D Potts ferromagnet has
a second-order transition, the specific heat exponent is given by
$\alpha=\alpha'=(2/3)(\pi-2\theta)/(\pi-\theta)$ \cite{wurev,baxterbook}, where
$\theta= 2 \, {\rm arccos}(q^{1/2}/2)$ as in eq. (\ref{qt}), so $\alpha=2/9$
for $q=(3+\sqrt{5})/2$ and $\alpha=1/3$ for $q=3$.)  This behavior as a
function of increasing strip width $L_y$ is the analogue, for these
infinite-length strips, of the standard finite-size scaling behavior of the
specific heat on $L \times L$ sections of a regular lattice, for which $|v_m -
v_{PM-FM}| \sim L^{-1/\nu}$ and $C(v=v_m) \sim L^{\alpha/\nu}$, where
$\nu=\nu'$ is the correlation length critical exponent \cite{fss,cft} for the
Potts ferromagnet on a two-dimensional lattice, with $\nu$ being related to
$\alpha$ by the hyperscaling relation $d\nu=2-\alpha$, so that $\nu=1$ for
$q=2$ and $\nu=5/6$ for $q=3$.  Note that $\alpha < 0$ for $0 < q < 2$, so
that, although the Potts ferromagnet has a PM-FM critical point for this range
of $q$, with a divergent correlation length, the specific heat has only a
finite, rather than divergent, nonanalyticity at the critical point.

  We next consider the plots of the specific heat in the case of the Potts
antiferromagnet. For $q=2$, $C$ exhibits a maximum at a value of $v$ that
approaches the value $v_{PM-AFM}=1-\sqrt{3}$ for the 2D lattice as $L_y$
increases, and the height of the maximum increases; these are simply related,
by $K \to -K$, to the behavior for the $q=2$ ferromagnet.  For 
$q=(3+\sqrt{5})/2$ the curves for the
specific heat in the antiferromagnetic region of Fig. \ref{EC_q=q5} exhibit a
maximum not too far from the zero-temperature value $v=-1$.  However, the
height of the maximum does not increase very much as $L_y$ increases, for
either the free or cylindrical boundary conditions.  Finally, we show $E_r$ and
$C$ for $q=3$, in Fig. \ref{EC_q=3}.  For this value of $q$ the Potts
antiferromagnet has no finite-temperature PM-AFM transition and is noncritical
even at $T=0$; consistent with this, although $C$ has a maximum, the height of
this maximum does not increase with increasing $L_y$. 

%
%

%
%

%
%

%
%
\subsection*{Acknowledgment} 

This research was partially supported by the Taiwan NSC grant
NSC-95-2112-M-006-004 and NSC-95-2119-M-002-001 (S.-C.C.) and the U.S. NSF
grant PHY-03-54776 (R.S.).

\newpage

\section{Appendix: Transfer Matrix for $L_y=4$ Strip with Cylindrical Boundary
  Conditions} 
\label{sec.4P}

The number of elements in the basis is eight: ${\bf P} = \{ \delta_{1,2,3,4},
\delta_{1,2,3}+\delta_{1,2,4}+\delta_{1,3,4}+\delta_{2,3,4},
\delta_{1,2}\delta_{3,4}, \delta_{1,2}+\delta_{3,4}, \delta_{1,3}+\delta_{2,4},
\delta_{1,4}\delta_{2,3}, \delta_{2,3}+\delta_{1,4}, 1 \}$.  The transfer
matrix is given by
\footnotesize
\begin{eqnarray}
& & \T = \cr
& &   \left( \begin{array}{cccccccc}
v^8 D_1^2 D_3^2 & 4v^8 D_1 D_3 S_{25} & v^8 D_1^2 T_{13} & 2v^8 D_1 T_{14} &   2v^8 S_{25}^2 & v^9 D_2 D_3^2 & 2v^9 D_3 S_{23} & v^9 T_{18} \\
v^6 D_1^2 D_3 S_{25} & v^6 D_1 T_{22} & v^6 D_1^2 T_{23} & v^6 D_1 T_{24} &   v^6 S_{25} T_{25} & v^7 D_2 D_3 S_{25} & v^7 T_{27} & v^7 T_{28} \\
0 & 0 & v^8 D_1^2 & 2v^9 D_1 & 0 & 0 & 0 & v^{10} \\
v^4 D_1^2 S_{25}^2 & 2v^4 D_1 S_{25} T_{25} & v^4 D_1^2 T_{43} & v^4 D_1 T_{44} & 2v^5 S_{23} S_{33} & v^5 D_2 S_{25}^2 & 2v^5 S_{25} T_{47} & v^5 T_{48} \\
v^4 D_1^2 S_{25}^2 & 2v^4 D_1 S_{25} T_{25} & v^5 D_1^2 T_{53} & 2v^5 D_1 T_{54} & v^4 T_{55} & v^5 D_2 S_{25}^2 & 2v^5 S_{25} T_{47} & v^6 T_{58} \\
v^6 D_1^2 T_{61} & 4 v^6 D_1 T_{62} & v^6 D_1^2 F_2^2 & 2v^6 D_1 F_2 S_{42} &     2v^6 F_2 T_{65} & v^6 T_{66} & 2v^6 T_{67} & v^6 S_{42}^2 \\
v^3 D_1^2 T_{71} & 2v^3 D_1 T_{72} & v^3 D_1^2 T_{73} & 2v^3 D_1 T_{74} & 2v^3 T_{75} & v^3 T_{76} & v^3 T_{77} & v^3 T_{78} \\
D_1^2 T_{23} T_{81} & 4D_1 T_{82} & D_1^2 T_{83} & 2D_1 T_{84} & 2T_{85} T_{85}^\prime & T_{86} & T_{87} & T_{88}
\end{array} \right) \cr
& &
\end{eqnarray}
\normalsize
where the factors $D_k$ and $F_k$ are defined in eqs.
\reff{def_Dk} and \reff{def_Fk}; the $T_{ij}$ are given by
\begin{subeqnarray}
T_{13} &=& 2q + 6v + v^2  \\
T_{14} &=& q^2 + 4qv + 7v^2 + v^3  \\
T_{15} &=& q + 8 v + 4 v^2 \\
T_{18} &=& 2q^2 + 6qv + 8v^2 + v^3  
\end{subeqnarray}
\begin{subeqnarray}
T_{22} &=& 5q^2 + q^2v + 34qv + 10qv^2 + 65v^2 + 33v^3 + 4v^4 \\
T_{23} &=& q^2 + 5qv + 8v^2 + v^3 \\
T_{24} &=& q^3 + 7q^2v + 17qv^2 + 19v^3 + 2v^4 \\
T_{25} &=& q^2 + 6qv + 11v^2 + 2v^3 \\
T_{27} &=& 5q^2 + q^2v + 28qv + 8qv^2 + 41v^2 + 19v^3 + 2v^4 \\
T_{28} &=& q^3 + 6q^2v + 12qv^2 + 11v^3 + v^4
\end{subeqnarray}
\begin{subeqnarray}
T_{43} &=& q^3 + 6q^2v + 14qv^2 + 14v^3 + v^4 \\
T_{44} &=& q^4 + 8q^3v + 26q^2v^2 + 42qv^3 + 33v^4 + 2v^5 \\
T_{47} &=& q^2 + 5qv + 7v^2 + v^3 \\
T_{48} &=& q^4 + 7q^3v + 20q^2v^2 + 28qv^3 + 19v^4 + v^5 
\end{subeqnarray}
\begin{subeqnarray}
T_{53} &=& 2q^2 + 8qv + 10v^2 + v^3 \\
T_{54} &=& q^3 + 6q^2v + 13qv^2 + 12v^3 + v^4 \\
T_{55} &=& q^4 + 8q^3v + 32q^2v^2 + 2q^2v^3 + 68qv^3 + 12qv^4 + 61v^4 + 22v^5 + 2v^6 \\
T_{58} &=& 2q^3 + 10q^2v + 18qv^2 + 14v^3 + v^4 
\end{subeqnarray}
\begin{subeqnarray}
T_{61} &=& q + 6v + 2v^2 \\
T_{62} &=& q^2 + 6qv + qv^2 + 10v^2 + 3v^3 \\
T_{65} &=& q^2 + 4qv + 8v^2 + 2v^3 \\
T_{66} &=& q^2 + 8qv + 3qv^2 + 21v^2 + 16v^3 + 3v^4 \\
T_{67} &=& q^3 + 7q^2v + q^2v^2 + 18qv^2 + 4qv^3 + 17v^3 + 5v^4 
\end{subeqnarray}
\begin{subeqnarray}
T_{71} &=& q^3 + 8q^2v + q^2v^2 + 25qv^2 + 6qv^3 + 32v^3 + 13v^4 + v^5 \\
T_{72} &=& 2q^4 + 17q^3v + q^3v^2 + 62q^2v^2 + 8q^2v^3 + 117qv^3 + 25qv^4 + 98v^4 + 34v^5 + 2v^6 \cr
& & \\
T_{73} &=& q^4 + 7q^3v + 21q^2v^2 + 32qv^3 + 22v^4 + v^5 \\
T_{74} &=& q^5 + 8q^4v + 28q^3v^2 + 54q^2v^3 + 58qv^4 + 30v^5 + v^6 \\
T_{75} &=& q^5 + 9q^4v + 37q^3v^2 + q^3v^3 + 87q^2v^3 + 7q^2v^4 + 118v^4q + 19v^5q + 74v^5 + 21v^6 \cr 
& & + v^7 \\
T_{76} &=& q^4 + 10vq^3 + 2v^2q^3 + 41v^2q^2 + 16v^3q^2 + v^4q^2 + 88v^3q + 52v^4q + 7qv^5 + 88v^4 \cr 
& & + 72v^5 + 17v^6 + v^7 \\
T_{77} &=& 2q^5 + 19q^4v + q^4v^2 + 78q^3v^2 + 8q^3v^3 + 180q^2v^3 + 28q^2v^4 + 240qv^4 + 52qv^5 \cr
& & + 149v^5 + 47v^6 + 2v^7 \\
T_{78} &=& q^6 + 9vq^5 + 36q^4v^2 + 83q^3v^3 + 118q^2v^4 + 99qv^5 + 41v^6 + v^7 
\end{subeqnarray}
\begin{subeqnarray}
T_{81} &=& q^3 + 5q^2v + 12qv^2 + qv^3 + 16v^3 + 4v^4 \\
T_{82} &=& q^6 + 11q^5v + 55q^4v^2 + q^4v^3 + 162q^3v^3 + 9q^3v^4 + 302q^2v^4 + 34q^2v^5 + 345qv^5 \cr
& & + 67qv^6 + qv^7 + 192v^6 + 61v^7 + 4v^8 \\
T_{83} &=& q^6 + 10q^5v + 45q^4v^2 + 118q^3v^3 + 194q^2v^4 + 194qv^5 + qv^6 + 96v^6 + 4v^7 \\
T_{84} &=& q^7 + 11q^6v + 55q^5v^2 + 163q^4v^3 + 313q^3v^4 + 396q^2v^5 + 313qv^6 + qv^7 + 126v^7 \cr
& & + 4v^8 \\
T_{85} &=& q^3 + 6q^2v + 14qv^2 + 13v^3 + v^4 \\
T_{85}^\prime &=& q^4 + 6q^3v + 16q^2v^2 + 25qv^3 + qv^4 + 22v^4 + 4v^5 \\
T_{86} &=& q^6 + 12q^5v + q^5v^2 + 65q^4v^2 + 12q^4v^3 + 206q^3v^3 + 63q^3v^4 + 2q^3v^5 + 412q^2v^4 \cr
& & + 184q^2v^5 + 15q^2v^6 + 512qv^5 + 312qv^6 + 46qv^7 + qv^8 + 320v^6 + 256v^7 \cr & & + 60v^8 + 4v^9 \\
T_{87} &=& 2q^7 + 24q^6v + 132q^5v^2 + 2v^3q^5 + 434q^4v^3 + 18q^4v^4 + 934q^3v^4 + 72q^3v^5 \cr
& & + 1346q^2v^5 + 166q^2v^6 + 1232qv^6 + 230qv^7 + 2qv^8 + 560v^7 + 160v^8 + 8v^9 \cr & & \\
T_{88} &=& q^8 + 12q^7v + 66q^6v^2 + 218q^5v^3 + 477q^4v^4 + 718q^3v^5 + 739q^2v^6 + 486qv^7 \cr
& & + qv^8 + 165v^8 + 4v^9 
\end{subeqnarray}
The vectors $\w$ and $\uu_{\rm id}$ are given by
\begin{subeqnarray}
\w_{\rm odd}^{\rm T} &=&  q \left( D_1^2, 4D_1 F_1, q D_1^2, 2q D_1 F_1, 2F_1^2, q+2v+v^2, 2 F_1^2, q F_1^2 \right) \\ 
\w_{\rm even}^{\rm T} &=&  q \left( D_1^2 X_2^2, 4D_1 X_1 X_2, D_1^2 X_3, 2 D_1 X_4, 2X_1^2, (q+2v+v^2) X_2^2, 2 X_2 X_5, X_6 \right) \cr
& & \\
\uu_{\rm id}^{\rm T} &=&  \left( 1, 0,0,0,0,0,0,0 \right)  
\end{subeqnarray}
where 
\begin{subeqnarray}
X_3 &=& v^6 + 6v^5 + 15v^4q + 20v^3q^2 + 15v^2q^3 + 6q^4v + q^5 \\
X_4 &=& v^7 + 7v^6 + 21v^5q + 35v^4q^2 + 35q^3v^3 + 21q^4v^2 + 7vq^5 + q^6 \\
X_5 &=& v^5 + 5v^4 + 10v^3q + 10v^2q^2 + 5vq^3 + q^4 \\
X_6 &=& v^8 + 8v^7 + 28v^6q + 56v^5q^2 + 70q^3v^4 + 56q^4v^3 + 28v^2q^5 + 8q^6v + q^7 \cr
& &
\end{subeqnarray}

\vfill
\eject
\end{document}